\documentclass[11pt,a4paper]{article}
\usepackage{jheppub}
\usepackage[american]{babel}
\usepackage{bbm}
\usepackage{tikz}
\usetikzlibrary{decorations.pathmorphing}
\usetikzlibrary{decorations.markings}

\newcommand{\dvol}{d\mathrm{vol}}

\newcommand{\Vol}{\mathrm{Vol}}

\newcommand{\parfrac}[2]{\frac{\partial #1}{\partial #2}}

\newcommand{\ud}[2]{^{#1}_{\phantom{#1}#2}}

\newcommand{\ket}[1]{\ensuremath{| #1 \rangle}}

\newcommand{\ts}{\textstyle}

\newcommand{\wb}{\overline}
\newcommand{\wt}{\widetilde}

\newcommand{\smallstrut}{\rule{0pt}{.6em}}

\newcommand{\eg}{\textit{e.g.}}

\newcommand{\ie}{\textit{i.e.}}

\numberwithin{equation}{section}

\newcommand{\mat}[1]{\begin{pmatrix} #1 \end{pmatrix}}

\newcommand{\be}{\begin{equation}} \newcommand{\ee}{\end{equation}}
\newcommand{\bea}{\begin{equation} \begin{aligned}} \newcommand{\eea}{\end{aligned} \end{equation}}

\newcommand{\cC}{\mathcal{C}}
\newcommand{\cD}{\mathcal{D}}

\newcommand{\cG}{\mathcal{G}}

\newcommand{\cI}{\mathcal{I}}

\newcommand{\cK}{\mathcal{K}}
\newcommand{\cL}{\mathcal{L}}

\newcommand{\cN}{\mathcal{N}}
\newcommand{\cO}{\mathcal{O}}

\newcommand{\cQ}{\mathcal{Q}}
\newcommand{\cR}{\mathcal{R}}
\newcommand{\cS}{\mathcal{S}}

\newcommand{\cV}{\mathcal{V}}

\newcommand{\bB}{\mathbb{B}}
\newcommand{\bC}{\mathbb{C}}

\newcommand{\bM}{\mathbb{M}}

\newcommand{\bP}{\mathbb{P}}

\newcommand{\bR}{\mathbb{R}}
\newcommand{\bZ}{\mathbb{Z}}
\newcommand{\fg}{\mathfrak{g}}
\newcommand{\fh}{\mathfrak{h}}
\newcommand{\fm}{\mathfrak{m}}

\newcommand{\fn}{\mathfrak{n}}
\newcommand{\fp}{\mathfrak{p}}

\newcommand{\fR}{\mathfrak{R}}

\newcommand{\ft}{\mathfrak{t}}
\newcommand{\fu}{\mathfrak{u}}

\newcommand{\unit}{\mathbbm{1}}

\def\su{\mathfrak{su}}
\def\sl{\mathfrak{sl}}
\def\so{\mathfrak{so}}

\DeclareMathOperator{\Tr}{Tr}

\DeclareMathOperator{\re}{\mathbb{R}e}
\DeclareMathOperator{\im}{\mathbb{I}m}

\DeclareMathOperator{\diag}{diag}

\DeclareMathOperator{\Li}{Li}

\newcommand{\ketup}{\ensuremath{|\!\!\uparrow\rangle}}
\newcommand{\ketdown}{\ensuremath{|\!\!\downarrow\rangle}}

\title{Black hole microstates in AdS$\boldsymbol{_4}$ \\ from supersymmetric localization}

\author[a,b]{Francesco Benini,}

\author[c]{Kiril Hristov}

\author[d,e]{and Alberto Zaffaroni}

\affiliation[a]{Blackett Laboratory, Imperial College London, \\
South Kensington Campus, London SW7 2AZ, United Kingdom}
\affiliation[b]{International School for Advanced Studies (SISSA), \\
via Bonomea 265, 34136 Trieste, Italy}
\affiliation[c]{Institute for Nuclear Research and Nuclear Energy, Bulgarian Academy of Sciences, \\
Tsarigradsko Chaussee 72, 1784 Sofia, Bulgaria}
\affiliation[d]{Dipartimento di Fisica, Universit\`a di Milano-Bicocca, \\
I-20126 Milano, Italy}
\affiliation[e]{INFN, sezione di Milano-Bicocca, I-20126 Milano, Italy}

\emailAdd{f.benini@imperial.ac.uk}
\emailAdd{khristov@inrne.bas.bg}
\emailAdd{alberto.zaffaroni@mib.infn.it}

\preprint{Imperial/TP/2015/FB/03}

\abstract{This paper addresses a long standing problem, the  counting of the microstates of  supersymmetric asymptotically AdS black
holes in terms of a holographically dual field theory. We focus on a class of asymptotically AdS$_4$ static black holes preserving two real supercharges
which are dual to a topologically twisted deformation of the ABJM theory. We evaluate in the large $N$ limit the topologically twisted index of
the ABJM theory and we show that it correctly reproduces the entropy of the AdS$_4$ black holes. An extremization of the index with respect to a set of chemical potentials
is required. We interpret it as the selection of the exact R-symmetry of the superconformal quantum mechanics describing the horizon of the black hole.}

\begin{document}

\setcounter{tocdepth}{2}
\maketitle

%
%

\section{Introduction}

One of the great successes of string theory is the microscopic explanation of the entropy of a class of asymptotically flat black holes.
An immense literature, which we will not try to refer to here, followed the seminal paper \cite{Strominger:1996sh}.
No similar result exists for asymptotically AdS black holes. This is curious since holography suggests that the microstates
of the black hole should correspond to states in a dual conformal field theory. The AdS/CFT correspondence should be the natural setting
where to explain the black hole entropy in terms of a microscopical theory. Various attempts have been made to
derive the entropy of a class of rotating black holes in AdS$_5$ in terms of states of the dual ${\cal N}=4$ super-Yang-Mills (SYM) theory  \cite{Kinney:2005ej, Grant:2008sk}
but none was completely successful.%
\footnote{They involve counting the $1/16$ BPS states of ${\cal N}=4$ SYM which is still out of reach of our current techniques.}

In this paper we consider the analogous problem for asymptotically  AdS$_4$ black holes. In AdS$_4$ there exist spherically symmetric static BPS black holes,%
\footnote{This is not possible in AdS$_5$.}
preserving at least two real supercharges. The first numeric evidence for these solutions was found in \cite{Cucu:2003yk}, but their analytic construction was discovered in \cite{Cacciatori:2009iz} and further studied by many authors \cite{Dall'Agata:2010gj, Hristov:2010ri, Klemm:2011xw, Donos:2011pn, Donos:2012sy, Halmagyi:2013sla, Halmagyi:2013qoa, Katmadas:2014faa, Halmagyi:2014qza}. They occur in non-minimal ${\cal N}=2$ gauged supergravity in four dimensions and they reduce asymptotically to AdS$_4$ with the addition of magnetic charges for the gauge fields in vector multiplets. This background is sometimes called {\it magnetic} AdS$_4$. The full spacetime can be thought of as interpolating between the asymptotic AdS$_4$ vacuum and the near-horizon AdS$_2 \times S^2$ geometry, leading to a natural holographic interpretation as an RG flow across dimensions. In particular, we have a flow between a CFT$_3$ and a CFT$_1$, from a three-dimensional theory compactified on $S^2$ to a superconformal quantum mechanics (QM).

To be concrete, we focus on a class of supersymmetric black holes that are asymptotic to
AdS$_4\times S^7$. The dual field theory is a topologically twisted ABJM theory \cite{Aharony:2008ug} depending on a choice of magnetic fluxes $\fn_a$ for the
four Abelian gauge fields $U(1)^4 \subset SO(8)$ arising from the reduction on $S^7$. The theory, dimensionally reduced on $S^2$, gives rise to a  supersymmetric quantum mechanics. The holographic picture suggests that it becomes superconformal at low energies.
It also suggests that the original UV R-symmetry of the three-dimensional theory mixes in a non-trivial way with the flavor symmetries along the flow, and that some \emph{extremization principle} is at work to determine the exact linear combination.
The setting is indeed very similar to the one in  \cite{Benini:2012cz, Benini:2013cda}, where the dual to the topologically twisted ${\cal N}=4$ SYM compactified on a Riemann surface $\Sigma$ was studied. The gravity solution interpolates between AdS$_5$ and AdS$_3 \times \Sigma$. In \cite{Benini:2012cz, Benini:2013cda}, the central charge of the dimensionally reduced CFT$_2$ has been computed via c-extremization and successfully compared with the gravity prediction.

Here we focus our attention on the entropy of the black hole. We expect that it can be obtained with a microscopic computation in the dual field theory and we show indeed that this is the case. To this purpose, we evaluate the topologically twisted index introduced in \cite{Benini:2015noa} for the ABJM theory. This is the partition function of the topologically twisted theory on $S^2\times S^1$ and can be computed via localization  \cite{Benini:2015noa}. The result depends on a set of magnetic fluxes $\fn_a$ and chemical potentials $\Delta_a$ for the global symmetries
of the theory. It can be interpreted as the Witten index
$$
Z (\fn_a ,  \Delta_a) =   \Tr \, (-1)^F \, e^{-\beta H} \, e^{i J_a  \Delta_a}
$$
of the dimensionally-reduced quantum mechanics.  The magnetic fluxes $\fn_a$ precisely correspond to the magnetic charges of the black hole. The chemical potentials $\Delta_a$ parametrize the mixing of the R-symmetry with the flavor symmetries. We propose that, in order to find the R-symmetry that sits in the superconformal algebra in the IR, we need to extremize $Z (\fn_a ,  \Delta_a)$ with respect to $\Delta_a$.

The main result of this paper is the evaluation of the topologically twisted index $Z(\fn_a, \Delta_a)$ for ABJM in the large $N$ limit.
We extremize $Z (\fn_a ,  \Delta_a)$ at large $N$ and we show that the extremum exactly reproduces the black hole entropy:
$$
\re\log Z\large |_\text{crit} (\fn_a) = S_\text{BH}(\fn_a) \;.
$$
The critical values of the $\Delta_a$'s coincide with the values of the bulk scalar fields at the horizon of the black hole, which parametrize the bulk dual to the R-symmetry, in perfect agreement with supergravity expectations.

One of the technical challenges of this paper is the evaluation of the topologically twisted index in the large $N$ limit. The index can be expressed as a contour integral,
$$
Z = \sum_{\fm \,\in\, \Gamma_\fh} \oint_\cC \; Z_\text{int}(x,  \fm) \;,
$$
of a meromorphic form $Z_\text{int}$ of Cartan-valued complex variables $x$,  summed over a lattice $\Gamma_\fh$ of magnetic gauge fluxes. The form $Z_\text{int}$ encodes the classical and one-loop contributions to the path-integral, around BPS configurations. We first perform the sum over the magnetic flux lattice. Then we solve, at large $N$, an auxiliary set of equations---which have been dubbed ``Bethe Ansatz Equations'' in a similar context in \cite{Gukov:2015sna}---that give the positions of the poles of the meromorphic integrand. This part of the computation bears many similarities with the large $N$ evaluation of the $S^3$ partition function for ${\cal N}=2$ three-dimensional theories in \cite{Herzog:2010hf, Jafferis:2011zi}, although it is much more complicated. We finally evaluate the partition function $Z$ using the residue theorem.

Our result opens many questions and directions of investigation. Let us mention two of them.

First, it is tempting to speculate that, under certain conditions, the exact R-symmetry in $\cN=2$ superconformal QM can  be found by extremizing the corresponding Witten index. This fact would add to the other extremization theorems valid in higher dimensions. We know that even and odd dimensions work differently. In two and four dimensions, the exact R-symmetry is found by extremizing central charges: $a$-maximization works in four dimensions \cite{Intriligator:2003jj, Barnes:2004jj} and $c$-extremization in two \cite{Benini:2012cz}. In odd dimensions, we have so far the example of three dimensions where the partition function on $S^3$ is extremized \cite{Jafferis:2010un, Jafferis:2011zi, Closset:2012vg}. The natural candidate for an extremization in one dimension is the partition function on $S^1$, which is exactly the Witten index.

Secondly, it would be very interesting to understand better the superconformal quantum mechanics corresponding to the horizon of the black hole. Our computation is done in the topologically twisted three-dimensional theory. The dimensionally reduced QM has infinitely many states corresponding to different gauge fluxes on $S^2$. The topologically twisted index defined in \cite{Benini:2015noa} depends on fugacities $y_a=e^{i \Delta_a- \sigma_a}$ and it counts the supersymmetric ground states, but it necessarily involves a regularization when the fugacities are pure phases, as it is in our case.%
\footnote{The $\sigma_a$ are real masses that make the spectrum of the Hamiltonian discrete; in our case all these masses are zero.}
It would be interesting to understand more precisely how these supersymmetric ground states flow to the microstates of the black hole. This implies
understanding in details the structure of the IR  superconformal quantum mechanics. We leave these very interesting questions for the future.

\

The paper is organized as follows.
In Section \ref{sec: localization} we write the topologically twisted index for the ABJM theory. We evaluate it in the large $N$ limit as a function of the magnetic fluxes $\fn_a$ and the fugacities $y_a = e^{i\Delta_a}$. In particular we show that it scales as $N^{3/2}$. Our computation is valid for $N\gg 1$, which corresponds to  the M-theory limit of ABJM.
In Section \ref{sec: sugra} we review and discuss the general features of the static supersymmetric AdS$_4$ black holes. We emphasize in particular the holographic interpretation.
In Section \ref{sec: comparison} we compare the field theory and supergravity results. We show that the critical value of the index correctly reproduces the black hole entropy. We also show
that the critical values of the chemical potentials $\Delta_a$ match with the horizon values of the scalar fields and we show how this corresponds to the identification of the exact R-symmetry of the problem.
In Section \ref{sec: conclusions} we give a preliminary discussion of  some open issues, like the Witten index extremization and the correct interpretation of the superconformal quantum mechanics.
Finally, in the Appendices we give a derivation of the near horizon black hole metric from the BPS equations of gauged supergravity, we discuss the simplest case of a superconformal quantum mechanics---the free chiral field---and we discuss in details the attractor mechanism for our class of black holes.

\section{The topologically twisted index of ABJM at large $\boldsymbol{N}$}
\label{sec: localization}

A general 3d $\cN=2$ supersymmetric theory with an R-symmetry and integer R-charges, can be placed supersymmetrically on $S^2 \times S^1$ (in fact on $\Sigma_\fg \times S^1$) by performing a partial topological twist on $S^2$. If the theory has also a continuous flavor symmetry, then there is a discrete infinite family of such twists obtained by mixing the R-symmetry with Abelian subgroups of the flavor symmetry, and twisting by these alternative R-symmetries. One can also turn on background flat connections along $S^1$, and real masses. Both can be thought of as a background for the bosonic fields (the connection along $S^1$ and the real scalar) in external vector multiplets coupled to the flavor symmetry; we collectively call them ``complex flat connections''. One can then compute the path-integral of the theory on $S^2 \times S^1$ with such a background: this defines the so-called \emph{topologically twisted index} of the theory \cite{Benini:2015noa}. We briefly review its definition here and then we apply it to the ABJM theory.

We take a metric on $S^2 \times S^1$ and a background for the R-symmetry  given by
\be
\label{background}
ds^2 = R^2 \big( d\theta^2 + \sin^2\theta\, d\varphi^2 \big) + \beta^2 dt^2 \;,\qquad\qquad A^R = \frac12 \cos\theta\, d\varphi \;,
\ee
where $t \cong t+1$ . We take vielbein $e^1 = R\, d\theta$, $e^2 = R\sin\theta\, d\varphi$, $e^3 = \beta\, dt$. We can write  supersymmetric Yang-Mills and Chern-Simons  Lagrangians for a vector multiplet $\cV = (A_\mu, \sigma, \lambda, \lambda^\dag, D)$
\bea
\label{YM and CS}
\cL_\text{YM} & = \Tr\bigg[ \frac14 F_{\mu\nu} F^{\mu\nu} + \frac12 D_\mu\sigma D^\mu\sigma + \frac12 D^2 - \frac i2 \lambda^\dag \gamma^\mu D_\mu\lambda - \frac i2 \lambda^\dag [\sigma,\lambda] \bigg] \;, \\
\cL_\text{CS} & = - \frac{ik}{4\pi} \Tr \bigg[ \epsilon^{\mu\nu\rho} \Big( A_\mu \partial_\nu A_\rho - \frac{2i}3 A_\mu A_\nu A_\rho \Big) + \lambda^\dag \lambda + 2D\sigma \bigg] \;,
\eea
and for matter chiral multiplets  $\Phi = (\phi, \psi, F)$ transforming in a representation $\fR$ of the gauge group
\be
\label{matter}
\cL_\text{mat} = D_\mu \phi^\dag D^\mu\phi + \phi^\dag \Big( \sigma^2 + iD  + \frac{q}{2 R^2} \Big) \phi + F^\dag F + i \psi^\dag ( \gamma^\mu D_\mu -\sigma ) \psi - i \psi^\dag \lambda\phi + i \phi^\dag \lambda^\dag \psi \;,
\ee
where $q$ is the R-charge of the chiral multiplet \cite{Benini:2015noa}. In the previous expression, for example, $\psi^\dag \sigma \psi$ is a shorthand for $\psi_A^\dag \sigma^\alpha (T_\alpha)\ud{A}{B}  \psi^B$, where the indices $A,B$ run over the representation $\fR$ and $\alpha,\beta$ over the Lie algebra. The covariant derivatives in (\ref{YM and CS}) and (\ref{matter}) contain the R-symmetry background (\ref{background}). Supersymmetry is preserved by a constant spinor satisfying $\gamma^3 \epsilon =\epsilon$.

Whenever the theory has flavor symmetries $J^f$, we can turn on supersymmetric backgrounds for the bosonic fields in the  corresponding vector multiplet $\cV^f = (A^f_\mu, \sigma^f, \lambda^f, \lambda^{f \, \dag}, D^f)$. A Cartan-valued magnetic background for the flavor symmetry
\be
\frac1{2\pi} \int_{S^2} F^f = \fn \;,
\ee
is supersymmetric provided that $F^f_{12} = i D^f$.  We can also turn on an arbitrary Cartan-valued vacuum expectation value for  $\sigma^f$ and $A^f_t$. The theory is deformed by
various terms that can be read from the matter Lagrangian (\ref{matter})  where we consider the vector multiplet $\cV = (A_\mu, \sigma, \lambda, \lambda^\dag, D)$ appearing there as running over the gauge as well as flavor symmetries. The flavor gauge background appears in the covariant derivatives of the matter fields and in explicit mass term deformations. The magnetic flux $\fn$ for the flavor symmetry will add up to the magnetic flux for the R-symmetry, providing a family of topological twists. The constant potential $A^f_t$ is a flat connection (or Wilson line) for the flavor symmetry and $\sigma^f$ is a real mass for the three-dimensional theory. The nonvanishing value for $D^f$ induces extra bosonic mass terms in the Lagrangian \cite{Benini:2015noa}.

One can then compute the path-integral of the theory on $S^2 \times S^1$ with such a background using localization techniques \cite{Benini:2015noa}. The path integral is  a  function of the flavor magnetic fluxes $\fn$ and fugacities $y = e^{i(A^f_t + i \beta \sigma^f)}$ for the flavor symmetries and it defines the so-called \emph{topologically twisted index} of the theory \cite{Benini:2015noa}. It is explicitly given by a contour integral of a meromorphic form
\be
\label{path}
Z (\fn, y) = \frac1{|W|} \; \sum_{\fm \,\in\, \Gamma_\fh} \; \oint_\cC \; Z_\text{int}(x, y; \fm, \fn) \;,
\ee
summed over all magnetic fluxes $\fm$ in the co-root lattice $\Gamma_\fh$ of the gauge group and integrated over the zero-mode gauge variables $x=e^{i(A_t + i\beta \sigma)}$, where $A_t$ runs over the maximal torus of the gauge group and $\sigma$ over the corresponding Cartan subalgebra. More precisely, we introduce a variable $u = A_t + i \beta \sigma$ on the complexified Cartan subalgebra $\fg_\bC$ and, given a weight $\rho$, we use a notation where $x^\rho = e^{i\rho(u)}$. The form $Z_\text{int}(x, y ; \fm, \fn)$  receives contributions from the classical action and the one-loop determinants.  The contribution of a chiral multiplet to the one-loop determinant is given by
\be
\label{chiral1-loop}
Z_\text{1-loop}^\text{chiral} = \prod_{\rho \in \fR} \bigg( \frac{x^{\rho/2} \, y^{\rho_f/2}}{1-x^\rho \, y^{\rho_f}} \bigg)^{\rho(\fm)+ \rho_f(\fn) - q +1}
\ee
where $\fR$ is the representation under the  gauge group $G$, $\rho$ are the corresponding weights, $q$ is the R-charge of the field, and $\rho_f$ is the weight of the multiplet under the flavor symmetry group. The contribution of a vector multiplet   to the one-loop determinant is instead given by
\be
\label{vector1-loop}
Z_\text{1-loop}^\text{gauge} =   \prod_{\alpha \in G} (1-x^\alpha) \; (i\, du)^r
\ee
where $\alpha$ are the roots of $G$. The classical action contributes a factor
\be
\label{classical}
Z_\text{class}^\text{CS} = x^{k \fm}
\ee
where $k$ is the Chern-Simons coupling of $G$ (each Abelian and simple factor has its own coupling). A $U(1)$ topological symmetry with holonomy $\xi =e^{i z}$ and flux $\ft$ contributes
\be
\label{topological}
Z_\text{class}^\text{top} = x^\ft \, \xi^\fm \;.
\ee
Supersymmetry selects the contour of integration to be used in (\ref{path}) and determines which poles of  $Z_\text{int}(x, y ; \fm, \fn)$  we have to take. The result can be formulated in terms of the Jeffrey-Kirwan residue \cite{JeffreyKirwan}, and we refer to \cite{Benini:2015noa} for the details.

\subsection{The index of ABJM}
\label{sec: ABJM index}

The low-energy dynamics of $N$ M2-branes on $\bC^4/\bZ_k$ is described by the so-called ABJM theory \cite{Aharony:2008ug}: it is a three-dimensional supersymmetric Chern-Simons-matter theory with gauge group $U(N)_k \times U(N)_{-k}$ (the subscripts are the CS levels) and matter in bifundamental representations. Using standard $\cN=2$ notation, the matter content is described by the quiver diagram
\begin{center}
\begin{tikzpicture}
\draw (-2,0) circle [radius=.4]; \node at (-2,0) {$N$}; \node at (-1.5,-.5) {\footnotesize{$k$}};
\draw (2,0) circle [radius=.4]; \node at (2,0) {$N$}; \node at (2.5,-.5) {\footnotesize{$-k$}};
\draw [decoration={markings, mark=at position 1 with {\arrow[scale=2.5]{>>}}}, postaction={decorate}, shorten >=0.4pt] (-1.5,.5) arc [radius=4.5, start angle = 110, end angle = 70];
\draw [decoration={markings, mark=at position 1 with {\arrow[scale=2.5]{>>}}}, postaction={decorate}, shorten >=0.4pt] (1.5, -.5) arc [radius=4.5, start angle = -70, end angle = -106];
\node at (0,1.2) {$A_i$}; \node at (0,-1.2) {$B_j$};
\end{tikzpicture}
\end{center}
where $i,j=1,2$ and arrows represent bifundamental chiral multiplets, and there is a quartic superpotential
\be
W = \Tr \big( A_1B_1A_2B_2 - A_1B_2A_2B_1 \big) \;.
\ee
For $k=1,2$ the theory has $\cN=8$ superconformal symmetry, while for $k\geq 3$ it has $\cN=6$ superconformal symmetry. In the $\cN=2$ notation, an $SU(2)_A \times SU(2)_B \times U(1)_T \times U(1)_R$ global symmetry is made manifest: the first two factors act on $A_i$ and $B_j$, respectively, as on doublets; $U(1)_T$ is the topological symmetry associated to the topological current $J_T = *\Tr(F-\wt F)$ where $F, \wt F$ are the two field strengths; $U(1)_R$ is an R-symmetry. Working in components, though, one finds the full $SO(6)_R$ symmetry; for $k=1,2$ the R-symmetry is further enhanced to $SO(8)$ quantum mechanically \cite{Aharony:2008ug, Benna:2009xd, Bashkirov:2010kz}.

To relate the symmetries of the theory to the isometries of $\bC^4$, let us consider%
\footnote{For $N=1$, the superpotential vanishes and the manifest global symmetry is enhanced to $SU(2)_A \times SU(2)_B \times U(1)_D \times U(1)_T \times U(1)_R$ of rank $5$, where $U(1)_D$ gives charge 1 to all chiral multiplets. In fact, in this case the theory is four free chiral multiplets describing $\bC^4$, or a NLSM on the orbifold $\bC^4/\bZ_k$, which have a rank-$4$ flavor symmetry \emph{and} a $U(1)_R$ symmetry. In view of the $N>1$ case, we will neglect $U(1)_D$.}
$N=1$ and $k=1$ which describes a single M2-brane moving on $\bC^4$. The theory has gauge symmetry $U(1)_g \times U(1)_{\tilde g}$, and denoting $U(1)_{A/B}$ the Cartans of $SU(2)_{A/B}$, the standard charge assignment is
\be
\begin{array}{c|cc|cccc}
 & U(1)_g & U(1)_{\tilde g} & U(1)_A & U(1)_B & U(1)_T & U(1)_R \\
 \hline
A_1 & 1 & -1 & 1 & 0 & 0 & 1/2 \\
A_2 & 1 & -1 & -1 & 0 & 0 & 1/2 \\
B_1 & -1 & 1 & 0 & 1 & 0 & 1/2 \\
B_2 & -1 & 1 & 0 & -1 & 0 & 1/2 \\
\hline
T & 1 & -1 & 0 & 0 & 1 & 0 \\
\wt T & -1 & 1 & 0 & 0 & -1 & 0
\end{array}
\ee
The monopole $T$ corresponds to the magnetic flux $\fm = (1,-1)$ while $\wt T$ to $\fm = (-1,1)$. These monopoles get their gauge charges from the CS terms. The chosen $U(1)_R$ is the superconformal R-symmetry of an $\cN=2$ superconformal subalgebra. The gauge invariants are $A_i \wt T$ and $B_j T$, which are the coordinates of $\bC^4$ (their R-charge $\frac12$ signals that they are free).

It is convenient to introduce a new basis for the Cartan of global symmetries, where the flavor symmetries $J_{1,2,3}$ act on a copy of $\bC \subset \bC^4$ respectively, $\wt J_4$ is an R-symmetry, and they all have integer charges:
\be
J_1 = \frac{J_B + J_A + J_g - J_T}2 \;,\quad J_2 = \frac{J_B - J_A + J_g - J_T}2 \;,\quad J_3 = J_B \;,\quad \wt J_4 = J_R - J_B + \frac{J_T - J_g}2 \;.
\ee
In terms of charges:
\be
\label{table of charges used}
\begin{array}{c|cccc}
 & U(1)_1 & U(1)_2 & U(1)_3 & \wt{U(1)}_4 \\
A_1 & 1 & 0 & 0 & 0 \\
A_2 & 0 & 1 & 0 & 0 \\
B_1 & 0 & 0 & 1 & 0 \\
B_2 & -1 & -1 & -1 & 2 \\
\hline
T & 0 & 0 & 0 & 0 \\
\wt T & 0 & 0 & 0 & 0
\end{array}
\ee
We will use these symmetries to put the theory on $S^2 \times S^1$ with a topological twist. In particular we call $-\fn_{1,2,3}$ the fluxes and $y_{1,2,3}$ the fugacities associated to $J_{1,2,3}$. To restore the symmetry, we can introduce $\fn_4$ and $y_4$ as well, defined by
\be
\sum\nolimits_a \fn_a = 2 \;,\qquad\qquad \prod\nolimits_a y_a = 1 \;.
\ee
In the main body of the paper, we will not introduce separate parameters $\ft, \xi$ for the topological symmetry, essentially because $J_1 + J_2 - J_3 = J_g - J_T$, \ie{} the topological background is already included up to a gauge background.

For the ABJM theory, the topologically twisted index is computed using the rules discussed above and we find
\begin{multline}
\label{initial Z full}
Z = \frac1{(N!)^2} \sum_{\fm, \wt\fm \in \bZ^N} \int_\cC \; \prod_{i=1}^N \frac{dx_i}{2\pi i x_i} \, \frac{d\tilde x_i}{2\pi i \tilde x_i} \, x_i^{k\fm_i + \ft} \, \tilde x_i^{-k \wt\fm_i + \tilde\ft} \, \xi^{\fm_i} \, \tilde\xi^{-\wt\fm_i} \times
\prod_{i\neq j}^N \Big( 1 - \frac{x_i}{x_j} \Big) \, \Big( 1 - \frac{\tilde x_i}{\tilde x_j} \Big) \times \\
\times \prod_{i,j=1}^N \prod_{a=1,2}
\bigg( \frac{ \sqrt{ \frac{x_i}{\tilde x_j} \, y_a} }{ 1- \frac{x_i}{\tilde x_j} \, y_a } \bigg)^{\fm_i - \wt\fm_j - \fn_a +1}
\prod _{b=3,4} \bigg( \frac{ \sqrt{ \frac{\tilde x_j}{x_i} \, y_b} }{ 1- \frac{\tilde x_j}{x_i} \, y_b } \bigg)^{\wt\fm_j - \fm_i - \fn_b +1} \;.
\end{multline}
For the moment, we introduced all possible parameters including redundant ones. However, the index has a set of symmetries and invariances, some of which correspond to the aforementioned redundancies.

First of all, the index  is actually nonvanishing only if $\ft+\tilde \ft =0 \pmod{k}$. This can be seen by performing the integral over the diagonal $U(1)$. By a change of variables $x_i = z w \hat x_i$ and $\tilde x_i = z \hat{ \tilde x}_i/w$ with $\prod_{i=1}^N \hat x_i = \prod_{i=1}^N \ \hat {\tilde x}_i=1$, we see that  each term in the sum (\ref{initial Z full})  contains an integral
$$
\int \frac{d z}{2 \pi i z} \; z^{k \left( \smallstrut \sum_i \fm_i- \sum_i\wt\fm_i \right) + \ft +\tilde \ft } \;,
$$
which can be non-zero only if $\ft +\tilde \ft =0 \pmod{k}$.

Secondly, the index has nice properties under shift of the arguments:
\bea
\label{shiftfree}
x_i &\to \lambda \, x_i \;;\qquad & \xi &\to \lambda^{-k} \xi \qquad& y_{1,2} &\to \lambda^{-1} y_{1,2} \qquad& y_{3,4} &\to \lambda \, y_{3,4} \qquad& Z &\to \lambda^{N\ft} Z \\
\tilde x_i &\to \tilde \lambda \, \tilde x_i \;;& \tilde\xi &\to \tilde\lambda^k \tilde\xi & y_{1,2} &\to \tilde\lambda \, y_{1,2} & y_{3,4} &\to \tilde\lambda^{-1} y_{3,4} & Z &\to \tilde\lambda^{N \tilde\ft} Z \\
\fm_i &\to \fm_i + \fp \;; & \ft &\to \ft - k\fp & \fn_{1,2} &\to \fn_{1,2} + \fp & \fn_{3,4} &\to \fn_{3,4} - \fp & Z &\to \xi^{N\fp} Z \\
\wt\fm_i &\to \wt\fm_i + \tilde\fp \;;& \tilde\ft &\to \tilde\ft + k \tilde\fp & \fn_{1,2} &\to \fn_{1,2} - \tilde\fp & \fn_{3,4} &\to \fn_{3,4} + \tilde\fp & Z &\to \tilde\xi^{-N\tilde\fp} Z \;,
\eea
where each line represents a different transformation and $\lambda$, $\tilde\lambda$, $\fp$, $\tilde\fp$ (with $\lambda, \tilde\lambda \in \bC^*$ and $\fp,\tilde\fp \in\bZ$) are the parameters.  In the first column we indicated the transformation to be performed on the dummy variables in the expression of $Z$ which gives the transformations reported in the last four columns. The first two transformations can be used to set $\xi = \tilde\xi=1$, and for $k=\pm1$ the last two can be used to set $\ft = \tilde\ft=0$. For larger values of $k$, the best we can do is to set  $\ft+\tilde \ft =0$ since it is a multiple of $k$. However, since we will be mainly interested in the case $k=1$, we will simply take $\ft = \tilde\ft=0$ from the start.%
\footnote{In fact, it is simple to check  that in the large $N$ limit the free energy only depends on $\ft + \tilde\ft$.}

Thirdly, the index is invariant under discrete involutions, which we write for simplicity for $\ft=\tilde \ft=0$ and $\xi=\tilde \xi=1$:
\bea
\label{shiftfreediscrete}
x_i &\leftrightarrow  \tilde  x_i \qquad\quad& \fm_i &\leftrightarrow \tilde \fm_i  \;; \qquad\quad& \{1,2\} &\leftrightarrow \{3,4\} \qquad& & \qquad\qquad\qquad& k &\leftrightarrow -k \\
x_i &\leftrightarrow    1/ x_i & \tilde x_i &\leftrightarrow    1/ \tilde x_i  \;; & y_{a} &\leftrightarrow 1/y_{a} & && k &\leftrightarrow -k \\
\fm_i &\leftrightarrow - \fm_i & \tilde \fm_i &\leftrightarrow - \tilde \fm_i  \;; & \{1,2\} &\leftrightarrow \{3,4\} & y_a &\leftrightarrow 1/y_a & k &\leftrightarrow -k \;.
\eea
In the first two columns we indicated the transformation to be performed on dummy variables in the expression of $Z$. Combining the transformations we see that the index is invariant  under change of sign of $k$ (corresponding to a parity transformation \cite{Aharony:2008ug}) and under inversion of the fugacities.
We will assume then, without loss of generality, that $k>0$.

We thus study the index
\begin{multline}
\label{initial Z}
Z = \frac1{(N!)^2} \sum_{\fm, \wt\fm \in \bZ^N} \int_\cC \; \prod_{i=1}^N \frac{dx_i}{2\pi i x_i} \, \frac{d\tilde x_i}{2\pi i \tilde x_i} \, x_i^{k\fm_i} \, \tilde x_i^{-k \wt\fm_i} \times
\prod_{i\neq j}^N \Big( 1 - \frac{x_i}{x_j} \Big) \, \Big( 1 - \frac{\tilde x_i}{\tilde x_j} \Big) \times \\
\times \prod_{i,j=1}^N \prod_{a=1,2}
\bigg( \frac{ \sqrt{ \frac{x_i}{\tilde x_j} \, y_a} }{ 1- \frac{x_i}{\tilde x_j} \, y_a } \bigg)^{\fm_i - \wt\fm_j - \fn_a +1}
\prod _{b=3,4} \bigg( \frac{ \sqrt{ \frac{\tilde x_j}{x_i} \, y_b} }{ 1- \frac{\tilde x_j}{x_i} \, y_b } \bigg)^{\wt\fm_j - \fm_i - \fn_b +1} \;.
\end{multline}
The Jeffrey-Kirwan residue selects a middle-dimensional contour in $(\bC^*)^{2N}$.  The integrand has no residues in the ``bulk'', and the only residues are at the boundaries $x_i = 0,\infty$, $\tilde x_j=0,\infty$ of the domain. According to the rules discussed in \cite{Benini:2015noa}, we need to choose reference covectors  $\eta$, $\wt\eta$ that, combined with the sign of the Chern-Simons coupling, tell us which residues we have to take. The final result is independent of  $\eta$, $\wt\eta$. We  choose the covectors $-\eta = \wt\eta = (1,\ldots ,1)$ in such a way that we pick all residues at the origin \cite{Benini:2015noa}. Then the range of the sums over $\fm_i$ and $\wt\fm_j$ are bounded above and below, respectively. We can take $\fm_i \leq M -1$ and $\wt\fm_j \geq 1-M$ for some large integer $M$. Performing the summations we get
\be
Z = \frac1{(N!)^2} \int_\cC \; \prod_{i=1}^N \frac{dx_i}{2\pi i x_i} \, \frac{d\tilde x_i}{2\pi i \tilde x_i} \prod_{i\neq j}^N \Big( 1 - \frac{x_i}{x_j} \Big) \, \Big( 1 - \frac{\tilde x_i}{\tilde x_j} \Big) \, A \, \prod_{i=1}^N \frac{ (e^{iB_i})^M}{ e^{iB_i} -1 } \prod_{j=1}^N \frac{ (e^{i\wt B_j})^M}{ e^{i\wt B_j} - 1} \;,
\ee
where we defined the quantities
\be
A = \prod_{i,j=1}^N \, \prod_{a=1,2} \bigg( \frac{ \sqrt{ \frac{x_i}{\tilde x_j} y_a} }{ 1 - \frac{x_i}{\tilde x_j}y_a} \bigg)^{1-\fn_a} \prod_{b=3,4} \bigg( \frac{ \sqrt{ \frac{\tilde x_j}{x_i} y_b} }{ 1 - \frac{\tilde x_j}{x_i}y_b} \bigg)^{1-\fn_b}
\ee
and
\be
\label{BA expressions}
e^{iB_i} = x_i^k \prod_{j=1}^N \frac{ \big( 1- y_3 \frac{\tilde x_j}{x_i} \big) \big( 1- y_4 \frac{\tilde x_j}{x_i} \big) }{ \big( 1- y_1^{-1} \frac{\tilde x_j}{x_i} \big) \big( 1- y_2^{-1} \frac{\tilde x_j}{x_i} \big) } \;,\qquad
e^{i\wt B_j} = \tilde x_j^k \prod_{i=1}^N \frac{ \big( 1- y_3 \frac{\tilde x_j}{x_i} \big) \big( 1- y_4 \frac{\tilde x_j}{x_i} \big) }{ \big( 1- y_1^{-1} \frac{\tilde x_j}{x_i} \big) \big( 1- y_2^{-1} \frac{\tilde x_j}{x_i} \big) } \;.
\ee
After the summation, the contributions from the residues at the origin have moved to the solutions to the ``Bethe Ansatz Equations'' (BAEs)
\be
\label{BAEs}
e^{iB_i} = 1 \;,\qquad\qquad\qquad e^{i\wt B_j} = 1 \;.
\ee
We borrow this terminology from \cite{Gukov:2015sna}, where a similar structure was found. Notice that if we take $|y_a|=1$, then the equations are invariant under the exchange $x_i \leftrightarrow \tilde x_i^*$. Moreover, taking the product of the equations immediately leads to the constraint%
\footnote{In particular we can always find ``obvious'' solutions imposing $x_i = x$, $\tilde x_i = \tilde x$ for all $i$. From the constraint, $\tilde x = \omega_\ell x$ where $\omega_\ell$ is a $kN$-th root of unity. Then
$$
x^{-k} = \frac{(1-y_3 \omega_\ell)^N(1-y_4 \omega_\ell)^N}{ (1-y_1^{-1} \omega_\ell)^N (1-y_2^{-1}\omega_\ell)^N} \;.
$$
These solutions, however, do not contribute to the original integral because they are killed by the vector multiplet determinant.}
\be
\label{constraint from BAEs}
\prod_{i=1}^N x_i^k = \prod_{j=1}^N \tilde x_j^k \;.
\ee
As generically all poles are simple, to take the residues we simply insert a Jacobian and evaluate everything else at the pole, hence we see that the dependence on $M$ disappears. The partition function takes the compact expression
\be
\label{Z resummed}
\boxed{\quad \rule[-1.7em]{0pt}{4em}
Z = \prod_{a=1}^4 y_a^{-\frac{N^2 \fn_a}2} \sum_{I \,\in\, \text{BAE}} \frac1{\det \bB} \; \frac{ \prod_{i=1}^N x_i^N \tilde x_i^N \prod_{i\neq j} \big( 1 - \frac{x_i}{x_j} \big) \big( 1 - \frac{\tilde x_i}{\tilde x_j} \big) }{ \prod_{i,j=1}^N \prod_{a=1,2} (\tilde x_j - y_a x_i)^{1-\fn_a} \prod_{a=3,4} (x_i - y_a \tilde x_j)^{1-\fn_a} } \;.
\quad}
\ee
The sum is over all solutions $I$ to the BAEs, modulo permutations of the $x_i$'s and $\tilde x_j$'s. All instances of $x_i, \tilde x_j$ have to be evaluated on those solutions.

The matrix $\bB$ appearing in the Jacobian is $2N \times 2N$ with block form
\be
\label{Jacobian general}
\bB = \frac{\partial \big( e^{iB_j}, e^{i\wt B_j} \big) }{ \partial( \log x_l, \log \tilde x_l)} = \mat{ x_l \dfrac{\partial e^{iB_j}}{\partial x_l} & \tilde x_l \dfrac{ \partial e^{iB_j}}{\partial \tilde x_l} \\[1em] x_l \dfrac{\partial e^{i\wt B_j}}{\partial x_l} & \tilde x_l \dfrac{ \partial e^{i\wt B_j}}{\partial \tilde x_l} }_{2N\times 2N} \;.
\ee
It is the product of the matrix of derivatives and the diagonal matrix $\diag(x_l, \tilde x_l)$. The two blocks on the diagonal are diagonal matrices, $\partial e^{iB_j}/ \partial x_l = 0$ and $\partial e^{i\wt B_j} / \partial \tilde x_l =0$ for $j \neq l$, while the off-diagonal blocks are more complicated and contain all components. We can introduce the function
\be
D(z) = \frac{ (1-z\, y_3)(1-z\, y_4) }{ (1-z\, y_1^{-1})(1-z\, y_2^{-2}) } \;,
\ee
which allows to write the BAEs in a compact form:
\be
e^{iB_i} = x_i^k \prod\nolimits_{j=1}^N D\Big( \frac{\tilde x_j}{x_i} \Big) \;,\qquad\qquad\qquad e^{i \wt B_j} = \tilde x_j^k \prod\nolimits_{i=1}^N D\Big( \frac{\tilde x_j}{x_i} \Big) \;.
\ee
Then we can introduce the objects
\be
G_{ij} = \parfrac{\log D(z)}{\log z} \Big|_{z = \tilde x_j/ x_i} \;.
\ee
The blocks of $\bB$, imposing $1 = e^{iB_i} = e^{i \wt B_j}$, are
\be
\bB \big|_\text{BAEs} = \mat{ \delta_{jl} \big[ k - \sum_{m=1}^N G_{jm} \big] & G_{jl} \\[.7em] - G_{lj} & \delta_{jl} \big[ k + \sum_{m=1}^N G_{mj} \big] } \;.
\ee
Notice that, because of the relation between the $y_a$'s, $D(0) = D(\infty) = 1$. Moreover the logarithmic derivative $\partial\log D(z)/\partial \log z$ vanishes both at $z\to 0$ and $z \to \infty$. This behavior is sometimes called ``absence of long-range forces'' in the large $N$ matrix model.

For $k>1$, given a solution $\{x_i, \tilde x_j\}$ to the BAEs (\ref{BA expressions})-(\ref{BAEs}), we can obtain more solutions multiplying all $x_i$, $\tilde x_j$ by a common $k$-th root of unity $\omega_k$:
\be
\label{Z_k action}
\bZ_k \,: \qquad \{x_i, \tilde x_j\} \;\to\; \{ \omega_k x_i,\, \omega_k \tilde x_j \} \;.
\ee
Thus, all solutions are ``$k$-fold degenerate''. One can also check that (\ref{Z resummed}) receives the same contribution from those $k$ solutions: both $\det \bB$ and the rest of the expression inside the summation are invariant under (\ref{Z_k action}). Therefore in (\ref{Z resummed}) we could sum over the orbits of (\ref{Z_k action}) and multiply the result by $k$.

\subsection{The Bethe potential}

It is convenient to change variables to  $u_i$, $\tilde u_j$, $\Delta_a$, defined modulo $2\pi$:
\be
x_i = e^{iu_i} \;,\qquad\qquad \tilde x_j = e^{i \tilde u_j} \;,\qquad\qquad y_a = e^{i\Delta_a} \;.
\ee
The relation $\prod_a y_a=1$ becomes $\sum_a \Delta_a = 0 \pmod{2\pi}$.
Then the Bethe ansatz equations become
\bea
\label{log BAEs}
0 &= ku_i + i \sum_{j=1}^N \bigg[ \sum_{a=3,4} \Li_1 \big( e^{i(\tilde u_j - u_i + \Delta_a)} \big) - \sum_{a=1,2} \Li_1 \big( e^{i(\tilde u_j - u_i - \Delta_a)} \big) \bigg] - 2\pi n_i  \\
0 &= k \tilde u_j + i \sum_{i=1}^N \bigg[ \sum_{a=3,4} \Li_1 \big( e^{i( \tilde u_j - u_i + \Delta_a)} \big) - \sum_{a=1,2} \Li_1 \big( e^{i( \tilde u_j - u_i - \Delta_a)} \big) \Big] - 2\pi \tilde n_j \;,
\eea
where $n_i, \tilde n_j$ are integers that parametrize the angular ambiguities. In the following we will take $\Delta_a$ real.

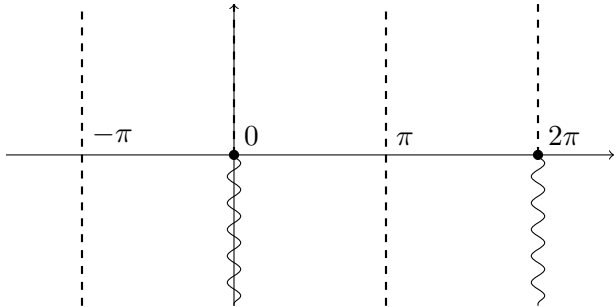
\begin{figure}[t]
\begin{center}\begin{tikzpicture}
\draw [->] (-3,0) -- (5,0); \draw [->] (0,-2) -- (0,2);
\draw [dashed, thick] (-2,-2) -- (-2,2); \draw [dashed, thick] (2,-2) -- (2,2);
\draw [dashed, thick] (0,0) -- (0,2); \draw [dashed, thick] (4,0) -- (4,2);
\draw [fill] (0,0) circle [radius=.06]; \draw [fill] (4,0) circle [radius=.06];
\draw [decorate, decoration=snake] (0,0) -- (0,-2); \draw [decorate, decoration=snake] (4,0) -- (4,-2);
\node [above right] at (0,0) {$0$}; \node [above right] at (-2,0) {$-\pi$}; \node [above right] at (2,0) {$\pi$}; \node [above right] at (4,0) {$2\pi$};
\end{tikzpicture}\end{center}
\caption{Analytic structure of $\Li_n(e^{iu})$.
\label{fig: analytic polylogs}}
\end{figure}

We recall the polylogarithms $\Li_n(z)$ defined by
\be
\Li_n(z) = \sum_{k=1}^\infty \frac{z^k}{k^n}
\ee
for $|z|<1$, and by analytic continuation outside the disk. The first two cases are $\Li_0(z) = z/(1-z)$ and $\Li_1(z) = - \log(1-z)$. For $n\geq 1$, the functions have a branch point at $z=1$ and we shall take the principal determination with a cut $[1,+\infty)$ along the real axis. For different values of $n$, the polylogarithms are related by
\be
\partial_u \Li_n(e^{iu}) = i\, \Li_{n-1}(e^{iu}) \;,\qquad\qquad \Li_n(e^{iu}) = i \int_{+i\infty}^u \Li_{n-1}(e^{iu'})\, du' \;.
\ee
The functions $\Li_n(e^{iu})$ are periodic under $u \to u+2\pi$ and have branch cut discontinuities along the vertical line $[0, -i\infty)$ and its images, as represented in Figure \ref{fig: analytic polylogs}. For us the following inversion formul\ae{} will be important:%
\footnote{The inversion formul\ae{} in the region $-2\pi < \re u < 0$ are simply obtained by sending $u \to -u$.}
\bea
\label{reflection formulae}
\Li_0(e^{iu}) + \Li_0(e^{-iu}) &= -1 \\
\Li_1(e^{iu}) - \Li_1(e^{-iu}) &= -iu + i\pi \\
\Li_2(e^{iu}) + \Li_2(e^{-iu}) &= \frac{u^2}2 - \pi u + \frac{\pi^2}3 \\
\Li_3(e^{iu}) - \Li_3(e^{-iu}) &= \frac i6 u^3 - \frac{i\pi}2 u^2 + \frac{i\pi^2}3 u
\eea
for $0< \re u < 2\pi$. The formul\ae{} in the other regions are obtained by periodicity. Also notice that $\Li_0(z)$ and $\Li_1(z)$ diverge at $z=1$, while $\Li_n(z)$ for $n\geq 2$ have no divergences on the $z$-plane.

All the equations in (\ref{log BAEs}) can be obtained as critical points of the function
\be
\label{function cal V with n}
\cV = \sum_{i=1}^N \Big[ \frac k2 \big(\tilde u_i^2 - u_i^2 \big) -2\pi (\tilde n_i \tilde u_i - n_i u_i) \Big] + \sum_{i,j=1}^N \bigg[ \sum_{a=3,4} \Li_2\big( e^{i(\tilde u_j - u_i + \Delta_a)} \big) - \sum_{a=1,2} \Li_2 \big( e^{i( \tilde u_j - u_i - \Delta_a)} \big) \bigg]
\ee
for some choice of $n_i$, $\tilde n_j$ and up to constants that do not depend on $u_i$, $\tilde u_j$. We call this function the Bethe potential.

\subsection[The BAEs at large $N$]{The BAEs at large $\boldsymbol{N}$}
\label{sec: BAEs large N}

Our goal is to evaluate the twisted index (\ref{Z resummed}) at large $N$. In order to do so, we first seek the dominant solution to the BAEs (\ref{BAEs}) at large $N$, and then evaluate its contribution to (\ref{Z resummed}). A convenient way to solve the BAEs at large $N$ is to first evaluate the functional $\cV$ and then extremize it. Even with this strategy, it is hard to compute all possible large $N$ limits in full generality. Following a similar idea in \cite{Herzog:2010hf}, we first study the BAEs numerically for some large values of $N$, and extract a plausible ansatz for the large $N$ solution; then we extremize the Bethe potential with respect to that ansatz.

For the sake of clarity, in the following we focus on the case $k=1$. Although most of the computations straightforwardly generalize to $k>1$, there are in fact some subtleties related to the identification of the full set of solutions, and we defer the study of those cases to future work. Moreover we are interested in fugacities $|y_a|=1$, \ie{} we will not consider the addition of real masses here.

\begin{figure}[t]
\centering
\includegraphics[scale=.8]{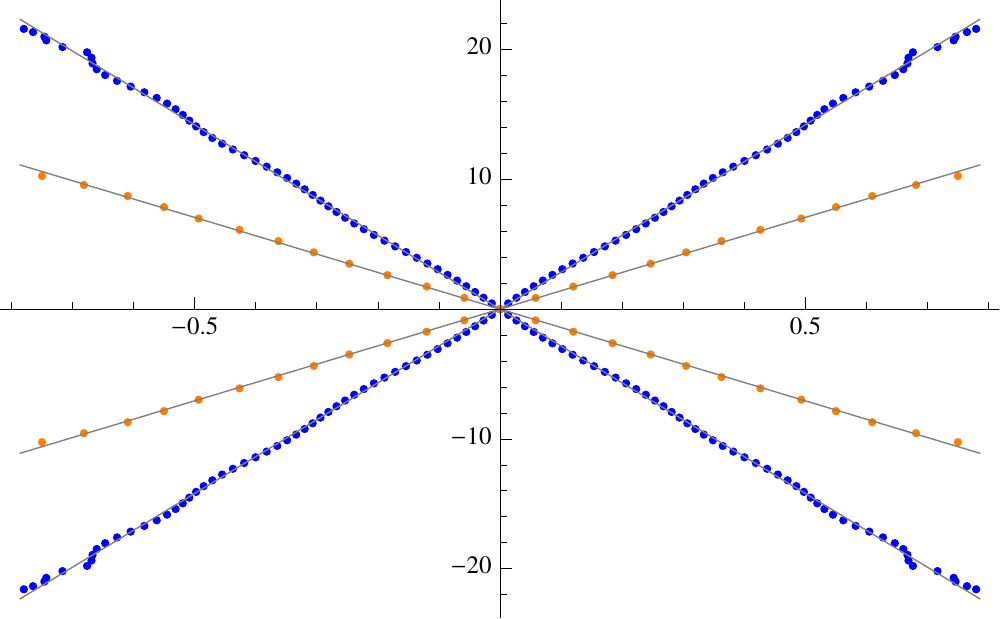}
\caption {Plots of the eigenvalues $u_i$, $\tilde u_j$ for $N=25$ (in orange) and $N=101$ (in blue) for $y_a=i$ and $k=1$. When $N\rightarrow 4N$,  the imaginary parts of  $u_i$, $\tilde u_j$ are approximately doubled---consistently with a scaling $N^\frac12$---while the real parts remain constant. For comparison, we also plot the analytical result.
\label{fig: ploty=i}}
\end{figure}

\begin{figure}[t]
\centering
\includegraphics[scale=.75]{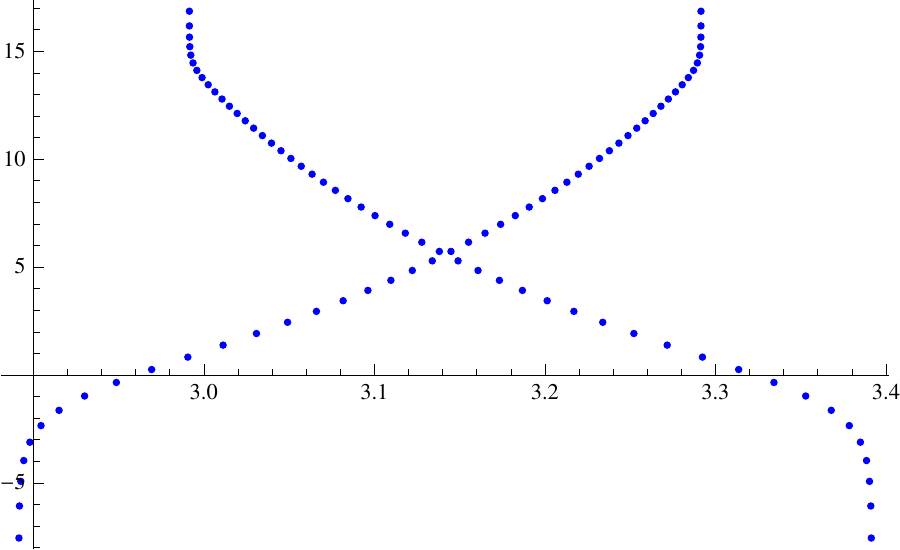}
\hspace{1cm}
\includegraphics[scale=.7]{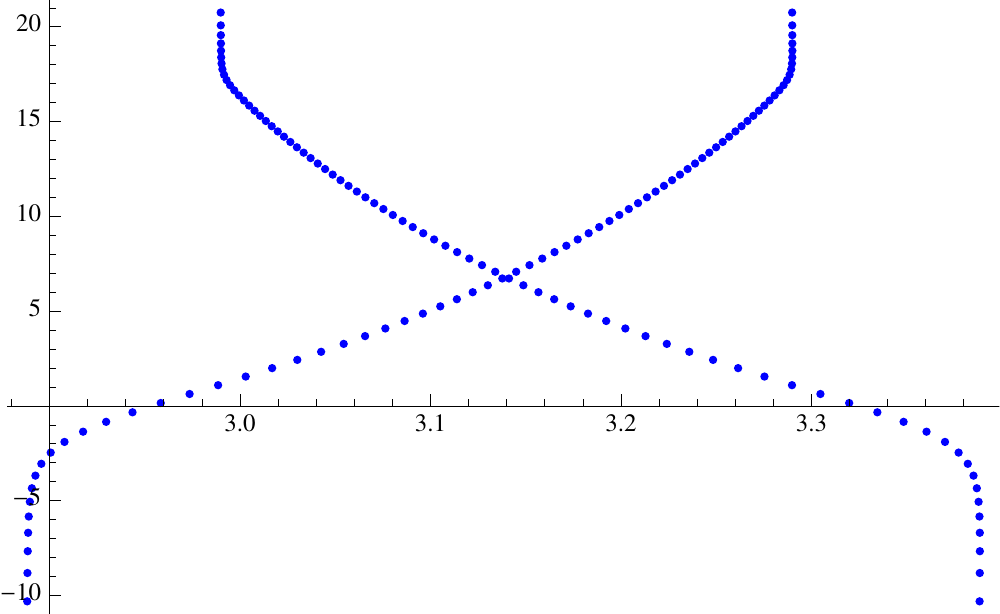}
\caption {Plots of $u_i$, $\tilde u_j$ for $N=50$ (on the left) and $N=75$ (on the right),  $\Delta_1=0.3$, $\Delta_2=0.4$, $\Delta_3=0.5$ with $\sum_a\Delta_a=2\pi$ and $k=1$.
\label{fig: Delta0.3}}
\end{figure}

The numerical analysis can by done with two different methods. The first one involves finding numerical solutions to the system (\ref{BA expressions})-(\ref{BAEs}) by iterating the transformation
\be
\label{iteration}
x_i \to \frac{x_i}{(e^{iB_i})^{1/kC}} \;,\qquad\qquad\qquad \tilde x_j \to \frac{\tilde x_j}{(e^{i\wt B_j})^{1/kC}} \;,
\ee
where $C$ is some large positive integer. If $e^{iB_i}$ is not $1$, then to a first approximation we can move it towards $1$ by rescaling $x_i^k$ (and neglecting the effect on the product).  The second method involves
introducing a time coordinate and setting up a dynamical system
\bea
\label{log BAEsdiff}
\tau \frac{du_i}{dt} &= ku_i + i \sum_{j=1}^N \bigg[ \sum_{a=3,4} \Li_1 \big( e^{i(\tilde u_j - u_i + \Delta_a)} \big) - \sum_{a=1,2} \Li_1 \big( e^{i(\tilde u_j - u_i - \Delta_a)} \big) \bigg] - 2\pi n_i  \\
\bar \tau \frac{d\tilde u_j}{dt}  &= k \tilde u_j + i \sum_{i=1}^N \bigg[ \sum_{a=3,4} \Li_1 \big( e^{i( \tilde u_j - u_i + \Delta_a)} \big) - \sum_{a=1,2} \Li_1 \big( e^{i( \tilde u_j - u_i - \Delta_a)} \big) \Big] - 2\pi \tilde n_j \;,
\eea
whose solutions should approach the equilibrium solution (\ref{log BAEs}) at late times. Here $\tau$ and $\bar \tau$ are complex numbers that have to be chosen so that the equilibrium solution is an attractive fixed point. None of the two methods is really stable and both heavily depend on the choice of constants and initial conditions. However we were lucky enough to find a couple of enlightening examples that we show in Figure  \ref{fig: ploty=i} and \ref{fig: Delta0.3}.

In Figure \ref{fig: ploty=i} we plot the distribution of eigenvalues $u_i$ and $\tilde u_j$ in the symmetric case $y_a=i$ (\ie{} $\Delta_a=\pi/2$) for $N=25$ and $N=101$ (and $k=1$). The distribution has been obtained with the iteration method (\ref{iteration}). We see that the imaginary parts of $u_i$ and $\tilde u_j$ grow with $N$. An analysis for many different values of $N$ reveals that the scaling is consistent with a behaviour $N^\frac12$. On the other hand, the real parts of $u_i$ and $\tilde u_j$ stay bounded when $N$ grows. The difference $\re(\tilde u_i-u_i)$ has minimum value $-\frac\pi2$ and maximum value $\frac\pi2$. For comparison, we also plot the analytical result that we will derive later in this section.

In Figure \ref{fig: Delta0.3} we plot the distribution of $u_i$ and $\tilde u_j$ for the case $\Delta_1=0.3$, $\Delta_2=0.4$, $\Delta_3=0.5$ with $\sum_a\Delta_a=2\pi$ (and $k=1$).  The distribution has been obtained with the dynamical system method (\ref{log BAEsdiff}). The integers $n_i$ and $\tilde n_j$ have been chosen in such a way that the distribution is ``continuous'' on the $u$-plane, as explained below. The plots for $N=50$ and $N=75$ are again consistent with a $N^\frac12$ scaling of the imaginary parts of the eigenvalues. The real parts are not scaling with $N$, and there are two tails of the distribution where they are constant. One can check that the two tails occur when  $\tilde u_i -u_i + \Delta_3=0$ and $\tilde u_i -u_i -\Delta_1=0$. These values correspond to logarithmic singularities in the equations (\ref{log BAEs}) and will play an important role in the following. Notice that both in this case and the previous one, the large $N$ solutions is invariant under the symmetry $x_i \leftrightarrow \tilde x_i^*$.

Thus, we consider an ansatz where the imaginary parts of $u_i$ and $\tilde u_i$ are equal%
\footnote{\label{foo: unequal densities}
One could have considered a more general ansatz where the imaginary parts are unrelated: $u_i = i N^\alpha t_i + v_i$ and $\tilde u_i = i N^\alpha \tilde t_i + \tilde v_i$. In the large $N$ limit this leads to two different density distributions $\rho(t) = di/N dt$ and $\tilde\rho(t) = di/N d\tilde t$. One can take the large $N$ limit of the BAEs in (\ref{log BAEs}) directly, without passing through the Bethe potential, as we do in (\ref{variation delta v inner}). This leads to two copies of (\ref{variation delta v inner}), one containing $\rho(t)$ and one $\tilde\rho(t)$. It follows, for generic values of $\delta v(t)$, that $\rho(t) = \tilde\rho(t)$.}
and scale as $N^\alpha$ for some power $\alpha$ (having in mind $\alpha \sim \frac12$), while the real parts remain of order one:
\be
u_i = i N^\alpha t_i + v_i \;,\qquad\qquad\qquad \tilde u_i = i N^\alpha t_i + \tilde v_i \;.
\ee
We also define
\be
\delta v_i = \tilde v_i - v_i \;.
\ee
Given the permutation symmetry, we can parametrize the points by the variable $t$ instead of the index $i$, by introducing the density
\be
\rho(t) = \frac1N\, \frac{di}{dt} \;.
\ee
At finite $N$ the density is a sum of delta functions, \ie{} $\rho(t) = \frac1N \sum_i \delta(t-t_i)$, while at large $N$ we assume that it becomes a continuous distribution. Summations are replaced by integrals:
$$
\sum_{i=1}^N \quad\to\quad N \int dt\, \rho(t) \;.
$$
The density distribution is normalized:
\be
\sum\nolimits_i 1 = N \qquad\Leftrightarrow\qquad \int dt\, \rho(t) = 1 \;.
\ee

In the large $N$ limit, we seek configurations where $\rho(t)$, $v(t)$, $\tilde v(t)$ and therefore also $\delta v(t)$ are continuous functions. Inspecting the BAEs in (\ref{log BAEs}) we see that they are singular whenever $\delta v(t)$ hits $\Delta_{1,2}$ or $-\Delta_{3,4}$ (or their periodic images), therefore on continuous solutions $\delta v$ does not cross those values. We recall that all angular variables are defined modulo $2\pi$. We can fix part of the ambiguity in $\Delta_a$ by requiring that
\be
\label{inequalities for delta v}
0 < \delta v + \Delta_{3,4} < 2\pi \;,\qquad\qquad\qquad -2\pi < \delta v - \Delta_{1,2} < 0 \;.
\ee
We can fix the remaining ambiguity of simultaneous shifts $\delta v \to \delta v+2\pi$, $\Delta_{1,2} \to \Delta_{1,2} +2\pi$, $\Delta_{3,4} \to \Delta_{3,4}-2\pi$ by requiring that $\delta v(t)$ takes the value $0$ somewhere (led by the numerical analysis, we assume that for $k=1$, $\delta v(t) = 0 \pmod{2\pi}$ is always solved somewhere). Thus, our choice for the angular determination simply corresponds to
\be
\label{general range of Delta_a}
0 < \Delta_a < 2\pi \;.
\ee
Given the symmetry of all functions and equations under the exchange of $a=1 \leftrightarrow 2$ and of $a=3 \leftrightarrow 4$, without loss of generality we can order
\be
\Delta_1 \leq \Delta_2 \;,\qquad\qquad\qquad \Delta_3 \leq \Delta_4 \;.
\ee

Later on we will have to distinguish the cases that $\sum_a \Delta_a = 2\pi$, $4\pi$ or $6\pi$ (while the cases $0$ and $8\pi$ correspond to $y_a=1$ and are singular). Combining (\ref{general range of Delta_a}) with $\sum_a \Delta_a = 2\pi$ one finds $\Delta_1 < 2\pi - \Delta_4$ and $\Delta_2 - 2\pi < -\Delta_3$, therefore the inequalities (\ref{inequalities for delta v}) can be put in the stronger form:
\be
\label{stronger inequalities delta v sum 2pi}
\sum\nolimits_a \Delta_a = 2\pi \qquad\Rightarrow\qquad\qquad - \Delta_3 < \delta v < \Delta_1 \;.
\ee
For $\sum_a \Delta_a = 6\pi$ one finds $2\pi - \Delta_4 < \Delta_1$ and $-\Delta_3 < \Delta_2 - 2\pi$, therefore
\be
\label{stronger inequalities delta v sum 6pi}
\sum\nolimits_a \Delta_a = 6\pi \qquad\Rightarrow\qquad\qquad \Delta_2 - 2\pi < \delta v < 2\pi - \Delta_4 \;.
\ee
For $\sum_a \Delta_a = 4\pi$ one finds that there are two possibilities:
\be
- \Delta_3 < \Delta_2 - 2\pi < \delta v < \Delta_1 < 2\pi - \Delta_4 \qquad\text{or}\qquad \Delta_2 - 2\pi < - \Delta_3 < \delta v < 2\pi - \Delta_4 < \Delta_1 \;.
\ee

At this point we should provide an estimate for what the constants $n_i$, $\tilde n_j$ are on solutions. Let us assume that
\be
0 < \re\big( \tilde u_j - u_i \big) + \Delta_{3,4} < 2\pi \;,\qquad -2\pi < \re\big( \tilde u_j - u_i \big) - \Delta_{1,2} < 0 \;, \qquad \forall\; i,j \;.
\ee
We set $u = \tilde u_j - u_i$ and estimate the function $i \sum_{a=3,4} \Li_1 \big( e^{i(u+\Delta_a)} \big) - i \sum_{a=1,2} \Li_1 \big( e^{i(u - \Delta_a)} \big)$. For large positive imaginary part of $u$, $\Li_1\big( e^{i(u \pm \Delta)} \big) \sim e^{i(u \pm \Delta)} \sim \cO(e^{-N^\alpha})$, thus the function takes extremely small values. For large negative imaginary part of $u$, instead, the function approaches $\sum \Delta_a - 4\pi$. The dependence on $u$ is exponentially suppressed and we observe an ``absence of long-range forces'' as in \cite{Herzog:2010hf}.%
\footnote{There is a difference with respect to \cite{Herzog:2010hf}. In the latter, the matrix model has long-range forces which cancel out if all species of eigenvalues have the same density distribution $\rho(t)$. In our case, the BAEs do not have long-range forces at all, and the condition $\rho = \tilde \rho$ is imposed by the local interactions among the eigenvalues.}
We conclude that the integers $n_i$, $\tilde n_j$ take the values
\be
2\pi n_i = \big( {\ts \sum \Delta_a} - 4\pi \big) \sum_j \Theta \big( \im (u_i - \tilde u_j) \big) \;,\qquad
2\pi \tilde n_j = \big( {\ts \sum \Delta_a} - 4\pi \big) \sum_i \Theta \big( \im (u_i - \tilde u_j) \big) \;.
\ee
Given the ansatz, the Heaviside theta function could be replaced by $\Theta(i>j)$ if the points are ordered by increasing imaginary part. Hence, in the large $N$ limit we will use the function
\be
\label{function cal V}
\cV = \frac12 \sum_{i=1}^N \big(\tilde u_i^2 - u_i^2 \big) + \sum_{i,j=1}^N \; \sum_{\substack{ a=3,4 \;:\; + \\ a=1,2 \;:\; -}} \Big[ \pm \Li_2 \big( e^{i(\tilde u_j - u_i \pm \Delta_a)} \big) \Big] + \sum_{i>j}^N \big( 4\pi - {\textstyle \sum \Delta_a} \big) (\tilde u_j - u_i) \;.
\ee
Notice that from here on we set $k=1$.

The leading contribution from the first term in $\cV$ is easy to compute:
\be
\frac12 \sum_{i=1}^N \big( \tilde u_i^2 - u_i^2 \big) = iN^{1+\alpha} \int dt\, \rho(t) \, t\, \delta v(t) + \cO(N) \;.
\ee
To compute the second term in $\cV$, we break
\be
\label{broken expression}
\sum_{i,j=1}^N \Li_2\big( e^{i(\tilde u_j - u_i + \Delta)} \big) = \sum_{i=1}^N \Li_2\big( e^{i(\tilde u_i - u_i + \Delta)} \big) + \sum_{i>j} \Li_2\big( e^{i(\tilde u_j - u_i + \Delta)} \big) + \sum_{i<j} \Li_2\big( e^{i(\tilde u_j - u_i + \Delta)} \big) \;.
\ee
The first term in this last expression is of $\cO(N)$ and apparently subleading. However it should be kept---as we will see---because its derivative is not subleading on part of the solution when $\delta v$ approaches $\Delta_{1,2}$ or $-\Delta_{3,4}$. Therefore we keep
$$
N \int dt\, \rho(t) \bigg[ \, \sum_{a=3,4} \Li_2 \Big( e^{i \left( \smallstrut \delta v(t) + \Delta_a \right)} \Big) - \sum_{a=1,2} \Li_2 \Big( e^{i \left( \smallstrut \delta v(t) - \Delta_a \right)} \Big) \bigg] \;.
$$
The third term in (\ref{broken expression}) is
\be
\sum_{i<j} \Li_2\big( e^{i(\tilde u_j - u_i + \Delta)} \big) = N^2 \int dt\, \rho(t) \int_t dt'\, \rho(t')\, \Li_2 \big( e^{i \left( \smallstrut \tilde u(t') - u(t) + \Delta \right)} \big) \;.
\ee
We decompose into ``Fourier modes'', $\Li_2(e^{iu}) = \sum_{k=1}^\infty e^{iku}/k^2$. Then we consider the integral
\be
I_k = \int_t dt'\, \rho(t')\, e^{ik \left( \smallstrut \tilde u(t') - u(t) + \Delta \right)} = \int_t dt'\, e^{-kN^\alpha (t'-t)} \sum_{j=0}^\infty \frac{(t'-t)^j}{j!} \partial_x^j \big[ \rho(x)\, e^{ik \left( \smallstrut \tilde v(x) - v(t) + \Delta \right)} \big]_{x=t} \;,
\ee
where in the second equality we have Taylor-expanded the integrand around the lower bound. Performing the integral in $t'$ we see that the leading contribution is for $j=0$, thus
\be
I_k = \frac{\rho(t)\, e^{ik \left( \smallstrut \tilde v(t) - v(t) + \Delta \right)} }{ k N^\alpha} + \cO(N^{-2\alpha}) \;.
\ee
Substituting we find
\be
\label{first summation}
\sum_{i<j} \Li_2 \big( e^{i (\tilde u_j - u_i + \Delta)} \big) = N^{2-\alpha} \int dt\, \Li_3 \big( e^{i \left( \smallstrut \delta v(t) + \Delta \right)} \big)\, \rho(t)^2 + \cO(N^{2-2\alpha}) \;.
\ee
With the second term in (\ref{broken expression}), where the summation is over $i>j$, we should be more careful because, in order to achieve a localization of the integral to the boundary, we should first invert the integrand. Consider first the case that $0<\re (\tilde u_j - u_i + \Delta_{3,4})<2\pi$: the formula to use is
\be
\label{reflection applied 34}
\Li_2 \big( e^{i( \tilde u_j - u_i + \Delta_{3,4})} \big) = - \Li_2 \big( e^{i(u_i - \tilde u_j - \Delta_{3,4})} \big) + \frac{(\tilde u_j - u_i + \Delta_{3,4})^2}2 - \pi (\tilde u_j - u_i + \Delta_{3,4}) + \frac{\pi^2}3 \;.
\ee
Following the same steps as before, the summation $\sum_{i>j}$ of the first term in the latter expression gives something similar to (\ref{first summation}) but with $-\Li_3\big( e^{-i ( \delta v(t) + \Delta_{3,4})} \big)$ in place of $\Li_3$. The two contributions can then be combined using (\ref{reflection formulae}), and result in a cubic polynomial expression. Then consider the case that $-2\pi< \re (\tilde u_j - u_i - \Delta_{1,2}) < 0$: the formula to use is
\be
\label{reflection applied 12}
- \Li_2 \big( e^{i(\tilde u_j - u_i - \Delta_{1,2})} \big) = \Li_2 \big( e^{i(u_i - \tilde u_j + \Delta_{1,2})} \big) - \frac{(\tilde u_j - u_i - \Delta_{1,2})^2}2 - \pi (\tilde u_j - u_i - \Delta_{1,2}) - \frac{\pi^2}3 \;,
\ee
which differs from the previous one by a sign. Again, the result of the summation $\sum_{i>j}$ can be combined with that of $\sum_{i<j}$ to give a cubic polynomial expression. The remaining terms from (\ref{reflection applied 34}) and (\ref{reflection applied 12}), throwing away the constants which do not affect the critical points, are
$$
- \big( 4\pi - {\ts \sum \Delta_a} \big) \sum_{i>j} (\tilde u_j - u_i) \;.
$$
This term is precisely canceled by the last term in (\ref{function cal V}).

To have a competition between the leading terms of order $N^{1+\alpha}$ and $N^{2-\alpha}$, we need
$$
\alpha = \frac12 \;.
$$
Including a Lagrange multiplier $\mu$ to enforce the normalization of $\rho(t)$, the final result is the following large $N$ expression (up to constants independent of $\rho$ and $\delta v$):
\be
\label{function cal V large N}
\boxed{
\begin{aligned}
\;\rule[-1.5em]{0pt}{3.5em} \frac{\cV}{iN^\frac32} &= \int\!\! dt \Bigg[ t\, \rho(t)\, \delta v(t) + \rho(t)^2 \bigg( \, \sum_{a=3,4} g_+ \big( \delta v(t) + \Delta_a \big) - \sum_{a=1,2} g_- \big( \delta v(t) - \Delta_a \big) \bigg) \Bigg] \\
\rule[-1.5em]{0pt}{3.0em} &\quad - \mu \bigg[ \int\!\! dt\, \rho(t) - 1 \bigg] - \frac{i}{N^\frac12} \int dt\, \rho(t) \bigg[ \, \sum_{a=3,4} \Li_2 \big( e^{i \left( \smallstrut \delta v(t) + \Delta_a \right)} \big) - \sum_{a=1,2} \Li_2 \big( e^{i \left( \smallstrut \delta v(t) - \Delta_a \right)} \big) \bigg] \;,
\end{aligned}
}\ee
where we introduced the polynomial functions
\be
g_\pm(u) = \frac{u^3}6 \mp \frac\pi2 u^2 + \frac{\pi^2}3 u \;,\qquad\qquad g_\pm'(u) = \frac{u^2}2 \mp \pi u + \frac{\pi^2}3 \;.
\ee
We remind that the last term can be neglected when computing the value of the functional $\cV$, because $\Li_2$ does not have divergences---however it becomes important when computing the derivatives of $\cV$ because $\Li_1(e^{iu})$ diverges when $u \to 0$.

\paragraph{The special case $\boldsymbol{y_a = i}$.}

This case, corresponding to $\Delta_a = \frac\pi2$, produces a particularly simple numerical solution that we reported in Figure \ref{fig: ploty=i}. The function $\cV$ simplifies to
\be
\frac\cV{iN^\frac32} = \int dt\, \bigg[ t\, \rho(t)\, \delta v(t) + \pi \rho(t)^2 \Big( \frac{\pi^2}4 - \delta v(t)^2 \Big) - \mu\, \rho(t) \bigg] + \mu + \cO\big( N^{-\frac12} \big) \;.
\ee
Setting to zero the variations with respect to $\rho(t)$ and $\delta v(t)$, we get the equations
\be
t\, \delta v(t) + \frac{\pi^3}2 \rho(t) - 2\pi \rho(t) \, \delta v(t)^2 = \mu \;,\qquad\qquad t\, \rho(t) = 2\pi \rho(t)^2 \delta v(t) \;.
\ee
On the support of $\rho(t)$, the solution is $\rho(t) = 2\mu/\pi^3$ and $\delta v(t) = \pi^2t/4\mu$. Calling $[t_-, t_+]$ the support of $\rho$, then $t_+ - t_- = \pi^3/2\mu$ from the normalization. Plugging back into $\cV$ and extremizing with respect to $\mu$ and $t_-$ we obtain $\mu = \pi^2 / 2\sqrt{2}$ and $t_\pm = \pm \pi/\sqrt{2}$. Finally
\be
\rho = \frac1{\sqrt2\, \pi} \;,\qquad\qquad \delta v(t) = \frac t{\sqrt2} \qquad\qquad\text{for } t \in \Big[ -\frac\pi{\sqrt{2}} \,,\, \frac\pi{\sqrt{2}} \Big] \;.
\ee
Since $\delta v(t_\pm) = \pm \pi/2$, in the solution $\delta v(t)$ barely reaches $\Delta_{1,2}$ or $-\Delta_{3,4}$ at the boundaries of the support, and the last term in (\ref{function cal V large N}) of order $N^{-\frac12}$ can be safely neglected, as we did. This solution, corresponding to the solid grey line in Figure \ref{fig: ploty=i}, precisely reproduces the numerical simulation.

\paragraph{The general case.}

To obtain the large $N$ solution to the BAEs in the general case, we again set to zero the variations of $\cV$ in (\ref{function cal V large N}) with respect to $\rho(t)$ and $\delta v(t)$. The latter equation, though, can be obtained as the large $N$ limit of the BAEs directly, and it is instructive to do so first.%
\footnote{As noted in footnote \ref{foo: unequal densities}, such a computation also allows to determine $\rho(t) = \tilde \rho(t)$ if one starts with an ansatz with two independent density distributions.}

Consider the first equation in (\ref{log BAEs}): we manipulate it as we did with the functional $\cV$. In particular, we break the sum $\sum_{j=1}^N$ into $\sum_{j>i} +\sum_{j<i} + \, (j \to i)$. We find:
\begin{multline}
0 = N^\alpha t + N^{1-\alpha} \rho(t) \Big[ \sum\nolimits_{a=3,4} \, g_+' \big( \delta v(t) + \Delta_a \big) - \sum\nolimits_{a=1,2} \, g_-'\big( \delta v(t) - \Delta_a \big) \Big] \\
+ \sum\nolimits_{a=3,4} \, \Li_1\big( e^{i \left( \smallstrut \delta v(t) + \Delta_a \right)} \big) - \sum\nolimits_{a=1,2} \, \Li_1\big( e^{i \left( \smallstrut \delta v(t) - \Delta_a \right)} \big)
+ \cO( N^{1-2\alpha}) + \cO(1) \;.
\end{multline}
The first term comes from $u_i$ with correction $\cO(1)$; the second term comes from the summation $\sum_{j\neq i}$ with corrections $\cO(N^{1-2\alpha})$ and $2\pi n_i$ with corrections $\cO(1)$; the terms on the second line come from $j=i$.

In order to have a competition between the leading terms it must be $\alpha = \frac12$. As long as $\delta v(t) + \Delta_{3,4} \neq 0$ and $\delta v(t) - \Delta_{1,2} \neq 0$, the terms on the second line are $\cO(1)$ and can be neglected. One finds the equation
\be
\label{variation delta v inner}
0 = t + \rho(t) \bigg[ \, \sum_{a=3,4} g_+' \big( \delta v(v) + \Delta_a \big) - \sum_{a=1,2} g_-' \big( \delta v(t) - \Delta_a \big) \bigg] \;.
\ee
However, when $\delta v(t)$ approaches $\Delta_{1,2}$ or $-\Delta_{3,4}$ the terms on the second line blow up (in particular, smooth solutions never cross those values) and may compete with those on the first line. In order to have a competition it must be
\be
\delta v(t) = \varepsilon_a \Big( \Delta_a - e^{- N^\frac12 Y_a(t)} \Big) \qquad\qquad \varepsilon_a = (1,1,-1,-1)
\ee
for some value of $a=1,2,3,4$ and with $Y_a(t)>0$ of order one. Then the second line contributes $- N^\frac12 \varepsilon_a Y_a(t) + \frac\pi 2 + \cO(e^{-N^{1/2}})$, which competes with the other leading terms. One finds the equation
\be
\label{variation delta v tails}
\varepsilon_a Y_a(t) = t + \rho(t) \bigg[ \, \sum_{b=3,4} g_+' \big( \varepsilon_a \Delta_a + \Delta_b \big) - \sum_{b=1,2} g_-' \big( \varepsilon_a \Delta_a - \Delta_b \big) \bigg] \;.
\ee
The equations (\ref{variation delta v inner}) and (\ref{variation delta v tails}) correspond to $\partial\cV/\partial \, \delta v(t) = 0$.

The variation of $\cV$ with respect to $\rho(t)$ is not affected by the terms suppressed by $N^{-1/2}$ in (\ref{function cal V large N}), because $\Li_2(e^{iu})$ has no divergences. Thus we find that the large $N$ limit of the BAEs is the system of equations:
\be
\label{BAEs large N}
\boxed{
\begin{aligned}
\rule{0pt}{1.5em} \mu &= t\, \delta v + 2\rho \sum\nolimits^*_b \Big[ \pm g_\pm(\delta v \pm \Delta_b) \Big] && \hspace{8em} \sum\nolimits^*_b = \sum_{\substack{b = 3,4 \,:\, + \\ b=1,2 \,:\, -}} \quad \\[-1em]
0 &= t + \rho \sum\nolimits^*_b \Big[ \pm g_\pm'( \delta v \pm \Delta_b) \Big] \qquad & \text{if }& \delta v \not\approx \varepsilon_a \Delta_a  \\[.5em]
\;\rule[-1em]{0pt}{1em} \varepsilon_a Y_a &= t + \rho \sum\nolimits^*_b \Big[ \pm g_\pm'( \varepsilon_a \Delta_a \pm \Delta_b) \Big] \quad & \text{if }& \delta v = \varepsilon_a \Big( \Delta_a - e^{- N^\frac12 Y_a} \Big)
\end{aligned}
}
\ee
as well as $1 = \int dt\, \rho$, $\rho >0$ on its support and $Y_a > 0$.

\paragraph{The solution for $\boldsymbol{\sum \Delta_a = 2\pi}$.}

We then proceed to solve the equations. First we solve the system (\ref{BAEs large N}) for generic values of $\delta v$, which we call the ``inner interval''. It turns out that $\rho(t)$ is a linear function, while $\delta v(t)$ is the ratio of two linear functions and the sign of its derivative equals the sign of $\mu$. This solution is reliable until one of the conditions $-2\pi < \delta v - \Delta_{1,2}<0$ or one of $0<\delta v + \Delta_{3,4}<2\pi$ is saturated. This defines the ``inner interval'' $[t_<, t_>]$: one saturation happens on one side and one on the other side. The inequalities (\ref{stronger inequalities delta v sum 2pi}) fix that in the inner interval $\delta v(t)$ goes from $-\Delta_3$ to $\Delta_1$. Imposing that $\rho>0$ at the extrema fixes $\mu>0$, therefore $\delta v(t)$ is increasing. Outside the inner interval, in two regions that we call the ``left and right tails'' respectively, $\delta v$ remains frozen at its limiting values $-\Delta_3$ and $\Delta_1$ up to exponentially small corrections, and the equations determine $\rho(t)$ and the correction $Y_{1,3}(t)$. The end of the tails is where $\rho(t)=0$. Reassuringly, $\rho(t)$ turns out to be increasing in the left tail and decreasing in the right tail.

Summarizing, the inner interval is $[t_<, t_>]$ with
\be
t_< \text{ s.t. } \delta v(t_<) = - \Delta_3 \;,\qquad\qquad t_> \text{ s.t. } \delta v(t_>) = \Delta_1 \;.
\ee
The points $t_<$ and $t_>$ are also those where $Y_{3,1}=0$. Then we define $t_\ll$ and $t_\gg$ as the values where $\rho=0$ and those bound the left and right tails. Schematically:
\begin{center}
\begin{tikzpicture}[scale=2]
\draw (-2,0) -- (2,0);
\draw (-1.5,-.05) -- (-1.5, .05); \draw (-0.5,-.05) -- (-0.5, .05); \draw (0.5,-.05) -- (0.5, .05); \draw (1.5,-.05) -- (1.5, .05);
\node [below] at (-1.5,0) {$t_\ll$}; \node [below] at (-1.5,-.3) {$\rho=0$};
\node [below] at (-0.5,0) {$t_<$}; \node [below] at (-0.5,-.3) {$\delta v = -\Delta_3$}; \node [below] at (-0.5,-.6) {$Y_3 = 0$};
\node [below] at (0.5,0) {$t_>$}; \node [below] at (0.5,-.3) {$\delta v=\Delta_1$}; \node [below] at (0.5,-.6) {$Y_1 = 0$};
\node [below] at (1.5,0) {$t_\gg$}; \node [below] at (1.5,-.3) {$\rho=0$};
\end{tikzpicture}
\end{center}
Finally we fix $\mu$ by requiring that $\int dt\, \rho(t) = 1$.

The solution is as follows. The transition points are at
\be
\label{solution sum 2pi -- init}
t_\ll = - \frac\mu{\Delta_3} \;,\qquad\quad t_< = - \frac\mu{\Delta_4} \;,\qquad\quad t_> = \frac\mu{\Delta_2} \;,\qquad\quad t_\gg = \frac\mu{\Delta_1} \;.
\ee
In the \emph{left tail} we have
\be
\begin{aligned}
\rho &= \frac{\mu + t\Delta_3}{(\Delta_1 + \Delta_3)(\Delta_2 + \Delta_3)(\Delta_4 - \Delta_3)} \\[.5em]
\delta v &= - \Delta_3 \;,\qquad\qquad Y_3 = \frac{- t\Delta_4 -\mu}{\Delta_4 - \Delta_3}
\end{aligned}
\qquad\qquad\qquad t_\ll < t < t_< \;.
\ee
In the \emph{inner interval} we have
\be
\begin{aligned}
\rho &= \frac{2\pi \mu + t(\Delta_3 \Delta_4 - \Delta_1 \Delta_2)}{(\Delta_1 + \Delta_3)(\Delta_2 + \Delta_3)(\Delta_1 + \Delta_4)(\Delta_2 + \Delta_4)} \\[.5em]
\delta v &= \frac{\mu(\Delta_1 \Delta_2 - \Delta_3 \Delta_4) + t \sum_{a<b<c} \Delta_a \Delta_b \Delta_c }{ 2\pi \mu + t ( \Delta_3 \Delta_4 - \Delta_1 \Delta_2) }
\end{aligned}
\qquad\qquad t_< < t < t_>
\ee
and $\delta v'>0$. In the \emph{right tail} we have
\be
\begin{aligned}
\rho &= \frac{\mu - t \Delta_1}{(\Delta_1 + \Delta_3)(\Delta_1 + \Delta_4)(\Delta_2 - \Delta_1)} \\[.5em]
\delta v &= \Delta_1 \;,\qquad\qquad Y_1 = \frac{t\Delta_2 - \mu}{\Delta_2 - \Delta_1}
\end{aligned}
\qquad\qquad\qquad t_> < t < t_\gg \;.
\ee
Finally, the normalization fixes
\be
\label{solution sum 2pi -- end}
\mu = \sqrt{ 2 \Delta_1 \Delta_2 \Delta_3 \Delta_4} \;.
\ee
The solution satisfies $\int dt\, \rho(t) \, \delta v(t) = 0$.

\begin{figure}[t]
\centering
\includegraphics[scale=.8]{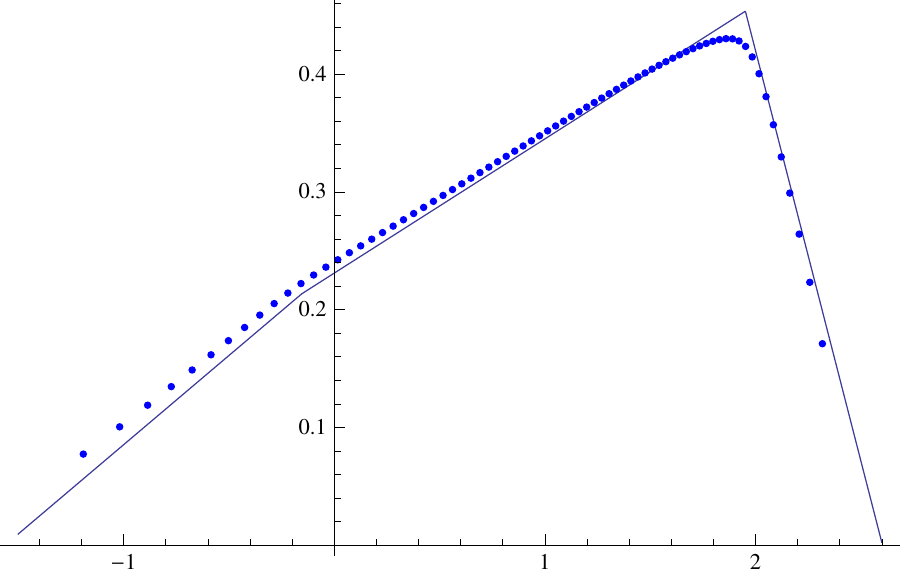}
\hspace{1cm}
\includegraphics[scale=.8]{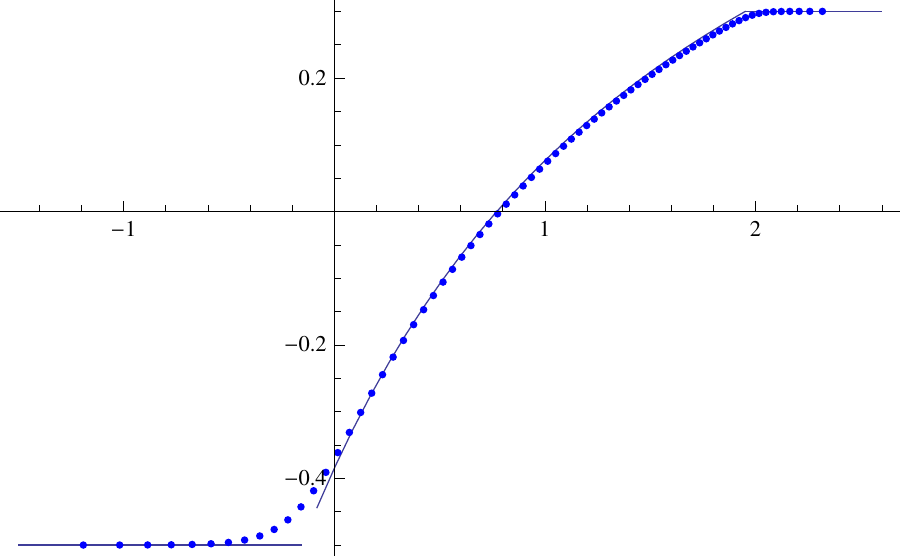}
\caption {Plots of the density of eigenvalues $\rho(t)$ and the function $\delta v(t)$ for $N=75$, $\Delta_1=0.3$, $\Delta_2=0.4$, $\Delta_3=0.5$ with $\sum_a\Delta_a=2\pi$ and $k=1$. The blue dots represent the numerical simulation, while the solid grey line is the analytical result.
\label{fig: comparisonplots}}
\end{figure}

In Figure \ref{fig: comparisonplots} we consider a case with generic $\Delta_a$'s---the same case considered in Figure \ref{fig: Delta0.3}---and compare the numerical simulation of the large $N$ solution to the BAEs, with the analytical result: we plot the density of eigenvalues $\rho(t)$ and the function $\delta v(t)$.

\paragraph{The solution in the other ranges.}

For $\sum\Delta_a = 4\pi$, it turns out that there are no consistent solutions to the large $N$ BAEs. One can run an argument similar to the one we had before, concluding that it is not possible to construct a solution with an inner interval where $\delta v(t)$ transits between two singular values, and two tails where $\delta v$ is frozen while $\rho(t)$ dies off to zero. This implies that, for such a range of parameters, the order of the index $Z(\Delta_a)$ is smaller than for the other ranges.

The solution for $\sum\Delta_a = 6\pi$ is very similar to the one in (\ref{solution sum 2pi -- init})-(\ref{solution sum 2pi -- end}). The function $\delta v(t)$ is decreasing from $2\pi - \Delta_4$ to $\Delta_2 - 2\pi$, as prescribed by (\ref{stronger inequalities delta v sum 6pi}), and $\mu<0$. The solution is obtained from (\ref{solution sum 2pi -- init})-(\ref{solution sum 2pi -- end}) by performing the substitutions
$$
\Delta_{3,4} \,\to\, \wt\Delta_{4,3} \;,\qquad\qquad \Delta_{1,2} \,\to\, \wt\Delta_{2,1} \;,\qquad\qquad \delta v \,\to\, - \delta v \;,\qquad\qquad \mu \,\to\, -\mu
$$
where $\wt\Delta_a = 2\pi - \Delta_a$.

In fact, notice that $\sum \Delta_a = 6\pi$ is equivalent to $\sum \wt\Delta_a = 2\pi$, therefore there is a pairing between points in the two ranges of the parameter space, and a corresponding map between BAE solutions. It turns out that, when evaluated on paired solutions, the twisted index $Z$ takes the same value. This can be understood by the following argument. The matrix model for $Z$ in (\ref{initial Z}) is invariant---possibly up to a sign---under the three involutions in (\ref{shiftfreediscrete}). These transformations can be combined to show invariance of $Z$ under each of the three operations:
\be
k \,\leftrightarrow\, -k \;,\qquad\qquad (12) \,\leftrightarrow\, (34) \;,\qquad\qquad y_a \,\leftrightarrow\, \frac1{y_a} \;.
\ee
The last one, in particular, corresponds to $\Delta_a \leftrightarrow 2\pi - \Delta_a$ and allows to map every solution for $\sum\Delta_a = 2\pi$ to a solution for $\sum\Delta_a = 6\pi$, which produces the same value of the index $Z$.

\subsection[The entropy at large $N$]{The entropy at large $\boldsymbol{N}$}

We are interested in the large $N$ limit of the twisted index, or partition function, (\ref{Z resummed}) and more precisely of its logarithm---the entropy. With the dominant solution to the BAEs at large $N$ in hand, we can compute the large $N$ limit of the expression in (\ref{Z resummed}) and plug the solution in. After various manipulations, we can recast the twisted index in a particularly convenient form:
\bea
\label{Z for large N}
Z &= (-1)^{N(\fn_1 + \fn_2)} \bigg( \frac{y_1^{\fn_1-1} y_2^{\fn_2-1} }{ y_3^{\fn_3-1} y_4^{\fn_4-1} } \bigg)^\frac N2 \; \sum_{I \,\in\, \text{BAE}} \; \frac1{\det \bB} \; \prod_{j>i} \Big( 1 - \frac{x_j}{x_i} \Big)^2 \Big( 1 - \frac{\tilde x_j}{\tilde x_i} \Big)^2 \\
&\quad \times \; \prod_i \frac{\tilde x_i}{x_i} \prod_{a=3,4} \Big( 1 - y_a \frac{\tilde x_i}{x_i} \Big)^{\fn_a - 1} \prod_{a=1,2} \Big( 1- y_a^{-1} \frac{ \tilde x_i}{x_i} \Big)^{\fn_a-1} \\
&\quad \times \; \prod_{j>i} \prod_{a=3,4} \Big( 1 - y_a \frac{\tilde x_j}{x_i} \Big)^{\fn_a - 1} \Big( 1 - y_a^{-1} \frac{ x_j}{\tilde x_i} \Big)^{\fn_a - 1}\prod_{a=1,2} \Big( 1- y_a^{-1} \frac{ \tilde x_j}{x_i} \Big)^{\fn_a-1} \Big( 1- y_a \frac{ x_j}{\tilde x_i} \Big)^{\fn_a-1} \;.
\eea
This time we have already reorganized the products $\prod_{i,j}$ into the diagonal parts $\prod_i$ and the off-diagonal parts, the latter written in terms of $\prod_{j>i}$ solely. Notice that the first two factors are just phases that can be neglected, as we will be interested in $\log |Z|$.

We start with the products $\prod_{j>i}$. The terms on the third line are treated as in Section \ref{sec: BAEs large N}. For $a=3,4$ using $0 < \delta v + \Delta_{3,4} < 2\pi$ we find
\bea
K_{a=3,4} &= \log \prod_{j>i} \Big( 1 - y_a \frac{\tilde x_j}{x_i} \Big)^{\fn_a-1} \Big( 1 - y_a^{-1} \frac{x_j}{\tilde x_i} \Big)^{\fn_a-1} \\
&= - N^\frac32 (\fn_a-1) \int dt\, \rho(t)^2 \Big[ \Li_2\big( e^{i(\delta v+\Delta_a)} \big) + \Li_2 \big( e^{-i(\delta v + \Delta_a)} \big) \Big] + \cO(N) \\
&= - N^\frac32 (\fn_a-1) \int dt\, \rho(t)^2 g_+'\big( \delta v(t) + \Delta_a \big) + \cO(N) \;.
\eea
Instead, for $a=1,2$ using $-2\pi < \delta v - \Delta_{1,2} < 0$ we find
\bea
K_{a=1,2} &= \log \prod_{j>i} \Big( 1 - y_a^{-1} \frac{\tilde x_j}{x_i} \Big)^{\fn_a-1} \Big( 1 - y_a \frac{x_j}{\tilde x_i} \Big)^{\fn_a-1} \\
&= - N^\frac32 (\fn_a-1) \int dt\, \rho(t)^2 g_-' \big( \delta v(t) - \Delta_a \big) + \cO(N) \;.
\eea
The contribution of the Vandermonde determinant is similar:
\be
\log \prod_{j>i} \Big( 1 - \frac{x_j}{x_i} \Big)^2 \Big( 1 - \frac{\tilde x_j}{\tilde x_i} \Big)^2 = - N^\frac32 \; \frac{2\pi^2}3 \int dt\, \rho(t)^2 + \cO(N) \;.
\ee
Then we consider the products $\prod_i$. The term
\be
\log \prod_{i=1}^N \frac{\tilde x_i}{x_i} = iN \int dt\, \rho(t) \, \delta v(t) = \cO(N)
\ee
is subleading. The term
\bea
J_{a=3,4} &= \log \prod_i \Big( 1 - y_a \frac{\tilde x_i}{x_i} \Big)^{\fn_a-1} = N (\fn_a-1) \int dt\, \rho(t) \log \big( 1-e^{i(\delta v + \Delta_a)} \big) \\
&= - N^\frac32 \; (\fn_a-1) \int_{\delta v \,\approx\, - \Delta_{3,4}} \hspace{-3em} dt\, \rho(t) \, Y_a(t) + \cO(N)
\eea
only contributes in the tail where $\delta v \,\approx\, -\Delta_{3,4}$, and the term
\bea
J_{a=1,2} &= \log \prod_i \Big( 1 - y_a^{-1} \frac{\tilde x_i}{x_i} \Big)^{\fn_a-1} = N (\fn_a-1) \int dt\, \rho(t) \log \big( 1-e^{i(\delta v-\Delta_a)} \big) \\
&= - N^\frac32 \; (\fn_a-1) \int_{\delta v \,\approx\,\Delta_{1,2}} \hspace{-2em} dt\, \rho(t) \, Y_a(t) + \cO(N)
\eea
only contributes in the tail where $\delta v \,\approx\, \Delta_{1,2}$.

The last term to evaluate is $-\log\det\bB$. Suppose that all entries of the matrix $\bB$ are of order one and bounded by some constant $c$. Then $\det\bB = \sum_{\text{perm }\sigma} B_{1,\sigma(1)} \dots B_{2N,\sigma(2N)}$ and then
$$
\log \det \bB \,\sim\, \log \big[ (2N)! \, c^{2N} \big] = \cO(N \log N) \;,
$$
which is subleading. Therefore we only get a contribution if some entries diverge with $N$. If we decompose $\bB = \wt\bB + \bB_0$ where the entries of $\wt\bB$ diverge, while those of $\bB_0$ are bounded, we have
$$
\log \det \bB = \log\det \wt\bB + \log \det (\unit + \wt\bB^{-1} \bB_0) \;.
$$
Therefore, provided that $\wt\bB^{-1}$ exists and its entries are bounded, the leading term is $\log\det \wt\bB$. Following the discussion after (\ref{Jacobian general}), the matrix $\bB$ evaluated on the solutions to the BAEs takes the form
\be
\bB \big|_\text{BAEs} = \mat{ \delta_{jl} \big[ k - \sum_m G_{jm} \big] & G_{jl} \\[.2em] - G_{lj} & \delta_{jl} \big[ k + \sum_m G_{mj} \big] }
\ee
with
\be
G_{ij} = \frac z{y_1-z} + \frac z{y_2-z} - \frac z{y_3^{-1}-z} - \frac z{y_4^{-1}-z} \; \Big|_{z = \tilde x_j / x_i} \;.
\ee
The function $G(z)$ diverges at $z=y_{1,2}$ and $z = y_{3,4}^{-1}$ which are phases, therefore the only terms that can diverge are the diagonal ones $G_{ii}$. We see that we can choose $\wt \bB$ to have diagonal matrices in all four blocks. Reorganizing the indices, $\wt\bB$ can be rewritten as a block-diagonal matrix made of $2\times 2$ blocks $\bM_i$: $\wt \bB = \text{diag}(\bM_i)$ with
\be
\bM_i = \mat{ 1 - G_{ii} & G_{ii} \\ -G_{ii} & G_{ii} } \qquad\qquad\Rightarrow\qquad \det \bM_i = G_{ii}
\ee
when $G_{ii}$ diverges, and $\bM_i = \unit_2$ when $G_{ii}$ does not. We made a choice of the $\cO(1)$ terms such that $\bM_i$ is invertible and the inverse has bounded entries. We then compute
\bea
-\log \det \bB &= - \log \prod_i G_{ii} + \cO(N\log N) = -N \int_{G(t) \,\approx\, \infty} \hspace{-2em} dt\, \rho(t) \log G(t) + \cO(N\log N) \\
&= - N^\frac32 \int_{\delta v \,\approx\, \varepsilon_a \Delta_a} \hspace{-2em} dt\, \rho(t)\, Y_a(t) + \cO(N\log N)
\eea
using the behavior $\delta v = \varepsilon_a \big( \Delta_a - e^{-N^{1/2} Y_a} \big)$ in the tails.

Putting everything together we find the following functional for the entropy at large $N$:
\be
\label{Z large N functional}
\boxed{\begin{aligned}
\quad\rule{0pt}{2em} \re \log Z &= - N^\frac32 \int dt\, \rho(t)^2 \bigg[ \frac{2\pi^2}3 + \sum_{\substack{ a=3,4 \;:\; + \\ a=1,2 \;:\; -}} (\fn_a-1) g_\pm'\big( \delta v(t) \pm \Delta_a \big) \bigg] \quad \\
\rule[-2em]{0pt}{1em} &\quad - N^\frac32 \sum_{a=1}^4 \fn_a \int_{\delta v \,\approx\, \varepsilon_a \Delta_a} \hspace{-2em} dt\, \rho(t) \, Y_a(t) \quad \;,
\end{aligned}}
\ee
up to corrections of order $N \log N$. We took the real part to get rid of irrelevant phases in $Z$.

\

Finally we should take the solution to the BAEs, plug it in the functional (\ref{Z large N functional}) and compute the integral. From the solution for $\sum \Delta_a = 2\pi$, we obtain the following surprisingly simple expression for the entropy:
\be
\label{Z large N}
\boxed{\quad \rule[-1.4em]{0pt}{3.4em}
\re\log Z = - \frac{N^\frac32}3 \, \sqrt{2\Delta_1 \Delta_2 \Delta_3 \Delta_4} \; \sum\nolimits_a \frac{\fn_a}{\Delta_a} \;.
\quad}
\ee
Notice that this expression is symmetric under permutations of the indices $a=1,2,3,4$. Such a symmetry is expected for $k=1$, because the index parametrizes the four complex factors in the $\bC^4$ fiber of the normal bundle to the M2-branes.

\section{AdS$\boldsymbol{_4}$ black holes in $\boldsymbol{\cN=2}$ supergravity}
\label{sec: sugra}

We now move to discuss a class of supersymmetric static asymptotically AdS$_4$ black holes, holographically dual to the ABJM theory twisted on $S^2$ that we have discussed so far. We first present the general features of this class of black holes, and then we depict their holographic interpretation, focusing on the asymptotic AdS$_4$ region and the AdS$_2\times S^2$ horizon.

The BPS black-hole solutions in AdS$_4$---similarly to many higher dimensional solutions \textit{\`a la} Maldacena-Nu\~nez \cite{Maldacena:2000mw, Nieder:2000kc, Gauntlett:2001qs, Naka:2002jz, Cucu:2003yk, Gauntlett:2003di, Kim:2005ez, Gauntlett:2007ph, Benini:2013cda}---preserve supersymmetry due to the topological twist on the internal space $S^2$ (or more generally on any Riemann surface $\Sigma$). The noteworthy feature in four dimensions is the existence of full analytic solutions for a completely general set of parameters, as first discovered in \cite{Cacciatori:2009iz}, elaborated upon in \cite{Dall'Agata:2010gj, Hristov:2010ri} and further generalized in various directions in \cite{Klemm:2011xw, Halmagyi:2013sla, Halmagyi:2013qoa, Katmadas:2014faa, Halmagyi:2014qza} and references therein. The complete spacetime can be thought of as interpolating between the asymptotic AdS$_4$ vacuum and the near-horizon AdS$_2 \times \Sigma$ geometry, leading to a natural holographic interpretation of those black holes as RG flows across dimensions.

Here we are specifically interested in solutions to the maximal $D=4$ ${\cal N}=8$ gauged supergravity, which can in turn be embedded in eleven-dimensional supergravity with an M-theory interpretation as wrapped M2-branes. In particular we focus on black holes that are asymptotic to AdS$_4\times S^7$. The topological twist on the internal two-dimensional space requires a background $SO(2)$ gauge field turned on, and therefore without loss of generality we can restrict our attention to the ${\cal N}=2$ truncation of the maximal supergravity \cite{Cvetic:1999xp, Duff:1999gh}.%
\footnote{See also \cite{Katmadas:2015ima} for the embedding of these black holes in 11D.}
We follow the standard conventions of \cite{Andrianopoli:1996cm} and consider the so-called magnetic STU model with electric FI gaugings that arises exactly as a truncation of ${\cal N}=8$ supergravity. It consists of three vector multiplets (in addition to the gravity multiplet) with the prepotential
\be
\label{prepotential}
F = -2 i \sqrt{X^0 X^1 X^2 X^3} \;,
\ee
and can be seen from the 11D point of view as a Kaluza-Klein reduction on S$^7$ (the $X^\Lambda$ are the holomorphic sections of the underlying special K\"{a}hler manifold). In addition, the gravitino R-symmetry is electrically gauged as specified by the FI parameters
\be
\label{FIparameters}
 \xi_0 = \xi_1 = \xi_2 = \xi_3 = \frac12 \;,
\ee
which complete the ${\cal N}=2$ data necessary for the unique definition of the Lagrangian and BPS variations. Further details about the supergravity model can be found in \cite{Cacciatori:2009iz, Hristov:2010ri} and in Appendix \ref{app: horizon}, where for completeness we present an explicit derivation of the BPS equations and the near-horizon geometry that eventually leads to the crucial entropy formula.

Before presenting the black hole solution, a word on notation is in order.  The ${\cal N}=2$ STU model has four gauge fields that correspond to the Cartan subalgebra of the $SO(8)$ isometry of $S^7$. The standard ${\cal N}=2$ supergravity symplectic index $\Lambda = \{0,1,2,3 \}$ used above is actually somewhat unnatural from the point of view of maximal supergravity and the field theory side, where the four gauge fields appear symmetrically.
Therefore, with an abuse of notation we will introduce the index $a = \{1,2,3,4 \}$, and identify the original $\Lambda =\{ 0,1,2,3 \}$ with $a=\{ 4,1,2,3 \}$ in this order. The index $a$ is the same as that used in Section \ref{sec: localization} and it allows to write all formul\ae{} in a manifestly permutation-invariant way. We do not distinguish between the upper and lower position of the index $a$.

The 4D black hole metric%
\footnote{Here we only consider the case of spherical horizon, mostly following the notation of \cite{Hristov:2010ri}. The case of higher-genus Riemann surfaces is analogous, and it is discussed together with the spherical case in Appendix \ref{app: horizon}.}
is compactly written as
\be
\label{BH_metric}
ds^2 = - e^{\mathcal{K} (X)} \Big(g r - \frac c{2 g r} \Big)^2 dt^2 + \frac{ e^{-\mathcal{K} (X)} \, dr^2}{\big( g r - \frac c{2 g r} \big)^2} + 2 e^{-\mathcal{K} (X)} \, r^2 \big( d\theta^2 + \sin^2 \theta\, d\phi^2 \big) \;,
\ee
where $g$ and $c$ are parameters while the K\"{a}hler potential is
\be
\label{Kaehler}
e^{- \mathcal{K} (X)} = i \left( \bar{X}^{\Lambda} F_{\Lambda} - X^{\Lambda} \bar{F}_{\Lambda} \right) = 8 \sqrt{X^0 X^1 X^2 X^3} = 8 \sqrt{X_1 X_2 X_3 X_4} \;.
\ee
The real sections $X_a$ are constrained in the range $0 < X_a < 1$ and satisfy $\sum_a X_a = 1$. They are given by
\be
\label{sections}
X_a = \frac{1}{4} - \frac{\beta_a}{r} \;,\qquad\qquad\qquad \sum\nolimits_a \beta_a = 0 \;,
\ee
in terms of parameters $\beta_a$ subject to the above constraint and further ones spelled below. The solution for the sections above defines also the background values for the physical scalar fields, which are typically chosen as
\be
\label{scalars}
z_1 \equiv \frac{X_1}{X_4} = \frac{r - 4 \beta_1}{r - 4 \beta_4} \;,\qquad z_2 \equiv \frac{X_2}{X_4} = \frac{r - 4 \beta_2}{r - 4 \beta_4} \;,\qquad z_3 \equiv \frac{X_3}{X_4} = \frac{r - 4 \beta_3}{r - 4 \beta_4} \;.
\ee
The parameters $\beta_a$ also specify the constant $c$, which is related to the value $r_h$ of the radial coordinate at the horizon:
\be
\label{r_h}
r^2_h = c = 4 \big( \beta_1^2 + \beta_2^2 + \beta_3^2 + \beta_4^2 \big) - \frac12 \;.
\ee
We already set the unit of the AdS$_4$ curvature $g = 1/\sqrt{2}$, \ie{} the parameters $\beta_a$ have been rescaled in the appropriate units more suitable for holographic use. The black hole has a regular horizon only for a restricted region in the parameter space of $\beta_a$ that ensures that $r_h$ is real and the scalars $X_a$ are positive.

Another crucial element of the solution is given by the background fluxes that carry magnetic charges through the sphere:
\be
\label{field_strengths}
F^{a}_{t r} = 0 \;,\qquad\qquad\qquad F^a_{\theta \phi} = - \frac{\fn_a}{\sqrt{2}} \, \sin \theta \;.
\ee
The four magnetic charges of the black hole $\fn_a$ are integer and fulfil the twisting relation
 \be
 \label{twisting}
 \sum_{a=1,2,3,4} \fn_a = 2 \;,
 \ee
which ensures that two out of the original eight supercharges are preserved by the black hole solution. Supersymmetry further relates the magnetic charges to the parameters $\beta_a$ that specify how the scalars run along the RG flow:
\be
\label{beta to n}
\fn_a - \frac12 = 16 \beta_a^2 - 4 \sum\nolimits_b \beta_b^2 \;.
\ee
Let us define the following quantities:%
\footnote{The signs in $\Pi$ are chosen in such a way that each term contains two positive and two negative signs, and there is one $\fn_a$ which always enters with positive sign.}
\bea
\label{def Pi, F2, Theta}
\Pi &= \frac18 \big( \fn_1 + \fn_2-\fn_3-\fn_4 \big) \big( \fn_1-\fn_2+\fn_3 - \fn_4 \big)\big( \fn_1 - \fn_2 - \fn_3 + \fn_4) \\
F_2 &= \frac12 \sum_{a < b} \fn_a \fn_b - \frac14 \sum_a \fn_a^2 \;,\qquad\qquad\qquad \Theta = \big( F_2 \big)^2 - 4 \fn_1 \fn_2 \fn_3 \fn_4 \;.
\eea
It is easy to check that
\be
\Pi = \big( 1 - \fn_1 - \fn_2 \big) \big( 1-\fn_1 - \fn_3 \big) \big(1- \fn_2 - \fn_3 \big) = 2^{12} \big( \beta_1 + \beta_2 \big)^2 \big( \beta_1 + \beta_3 \big)^2 \big( \beta_2 + \beta_3 \big)^2 >0 \;.
\ee
We can then invert the relations in (\ref{beta to n}), up to a common sign:
\be
\label{n to beta}
\beta_a = \mp \frac{ 4 \big( \fn_a - \frac12 \big)^2 +1 - \sum_b \fn_b^2}{16\sqrt\Pi} \;.
\ee
Here the sign equals the sign of $-(\beta_1+\beta_2)(\beta_1+\beta_3)(\beta_2+\beta_3)$, in other words the sign is correlated with that of
\be
\sqrt\Pi = \pm 64 (\beta_1 + \beta_2) (\beta_1 + \beta_3) ( \beta_2 + \beta_3) \;.
\ee
With a little bit of algebra, we find
\be
\label{rh and K(rh)}
r_h^2 = \frac{\Theta}{4\Pi} \;,\qquad\qquad\qquad\qquad e^{-2\cK(r_h)} = \frac{2\Pi^2}{\Theta^2} \, \big( F_2 \pm \sqrt\Theta \big) \;.
\ee
One can also write the first relation as
\be
\fn_a = 16 \beta_a^2 - r_h^2 = - (r_h + 4 \beta_a) (r_h - 4 \beta_a) \qquad \Rightarrow\qquad r_h^2 = -\frac{1}{16} \sum_a \frac{\fn_a}{X_a (r_h)} \;.
\ee
Although both signs in the formul\ae{} above are compatible with supersymmetry, it turns out (see Appendix \ref{app: horizon}) that smooth solutions exist only if three of the $\fn_a$ are negative, and in that case one should take the upper sign.

The black hole above preserves two supercharges, packaged in the corresponding Killing spinor solution,
\be
\label{Killing_spinor_bh}
\varepsilon_A = e^{\mathcal{K}/4} \sqrt{r - \frac cr } \, \varepsilon_A^0 \;,
\ee
written in terms of a constant spinor $\varepsilon^0$ obeying the following relations:
\be
\label{KSprojections}
\varepsilon_A^0 = \epsilon_{A B} \, \gamma^{\hat{t}} \, \varepsilon^{B, 0} \;,\qquad\qquad\qquad  \varepsilon_A^0 = \sigma^{3 \; B}_A \,  \gamma^{\hat{\theta} \hat{\phi}} \,\varepsilon_B^0 \;,
\ee
where the hatted indices are flat. Note that the Killing spinors are constant in time and on the sphere and therefore the group of rotations on the sphere commutes with the fermionic symmetries, leading to the corresponding symmetry algebra $U(1|1) \times SO(3)$.

This is the general black hole solution we want to describe holographically, and in the following we analyze separately the asymptotic region that defines our UV theory, and the near-horizon IR region related to a 1D superconformal quantum mechanics. Afterwards we discuss the definitions of the black hole entropy and the R-symmetry from the ${\cal N}=2$ supergravity point of view.

\subsection[The asymptotic AdS$_4$ vacuum]{The asymptotic AdS$\boldsymbol{_4}$ vacuum}
\label{sec: sugra1}

It is easy to take the limit $r \rightarrow \infty$ of the full black hole solution \eqref{BH_metric}-\eqref{field_strengths}: one gets the metric
\be
\label{UV intermediate metric}
ds_2 \;\simeq\; -r^2 dt^2 + \frac{dr^2}{r^2} + r^2 \big( d\theta^2 + \sin^2\theta\, d\phi^2 \big) \;,
\ee
constant scalars $z_1 = z_2 = z_3 = 1$ and non-vanishing magnetic field strengths as in \eqref{field_strengths}. This background was dubbed ``magnetic AdS$_4$'' in \cite{Hristov:2011ye}: not all Killing vectors of AdS$_4$ are preserved by the magnetic fluxes, the usual supersymmetry enhancement does not take place, and the corresponding symmetry group remains $U(1|1) \times SO(3)$ as explained in detail in the reference. As standard in cases of twisting, the isometries of the internal manifold $S^2$ commute with the supersymmetries and therefore the fermions effectively become scalars under rotation. Of course, as we further go in the UV the background asymptotes to standard AdS$_4$ and the field strengths (which in vielbein coordinates read $F^a_{\hat\theta\hat\phi} = -\fn_a \sin\theta/\sqrt2\, r^2$) go to zero, since magnetic AdS$_4$ is a non-normalizable deformation of AdS$_4$.

The complementary boundary picture is also clear: the dual boundary theory is a relevant deformation of the maximally supersymmetric ABJM theory, semi-topologically twisted by the presence of the magnetic charges. The fluxes $\fn_a$ give a family of twisted ABJM theories whose Euclidean version is precisely the one discussed
in Section \ref{sec: localization}. The holographic dictionary can be made precise, as discussed in details in \cite{Hristov:2013spa}. The boundary values of the gauge fields and the scalar fields $z_i$ correspond to relevant deformations of the ABJM Lagrangian: the gauge fields introduce a magnetic background for the R- and global symmetries, while the scalars $z_i$ induce mass deformations for the boundary scalar fields. In the Euclidean version the latter precisely correspond to the terms induced in the matter Lagrangian (\ref{matter}) by a constant  auxiliary $D^f$.  Finally, the bulk spinor (\ref{Killing_spinor_bh}) restricts to a constant boundary spinor, as appropriate for a topological twist.

\subsection[The near-horizon geometry AdS$_2 \times S^2$]{The near-horizon geometry AdS$\boldsymbol{_2 \times S^2}$}
\label{sec: sugra2}

Taking the opposite limit, $r \rightarrow r_h$, leads instead to%
\footnote{One performs the standard change of variables $r = r_h + \epsilon$ and expands at leading order in $\epsilon$.}
the AdS$_2 \times S^2$ metric
\be
\label{ads2s2}
ds^2 = \frac{e^{-\mathcal{K} (r_h)}}2 \, ds^2_{\text{AdS}_2} + 2 e^{-\mathcal{K} (r_h)} \, r_h^2 \, ds^2_{S^2} \;,
\ee
with $e^{-\mathcal{K} (r_h)} = 8 \sqrt{X_1 (r_h) X_2 (r_h) X_3 (r_h) X_4 (r_h)},$ and the same magnetic charges $\fn_a$ as before. We defined the unit-radius spaces $ds^2_{\text{AdS}_2} = (-dt^2 + dz^2)/z^2$ and $ds^2_{S^2} = d\theta^2 + \sin^2\theta\, d\phi^2$. All isometries of AdS$_2$ are preserved by the background gauge field. This in turn leads to the appearance of new fermionic symmetries, and the full symmetry group becomes $SU(1,1|1) \times SO(3)$ as discussed in \cite{Hristov:2013spa}. The Killing spinors in this case are full Killing spinors on AdS$_2$ and are obtained from the general ones by dropping the first relation in \eqref{KSprojections}, still keeping them constant on the sphere. We can therefore talk about a genuine superconformal symmetry in the IR, leading to a dual superconformal quantum mechanics.

Making use of the relations \eqref{beta to n}-\eqref{rh and K(rh)}, we can express the near-horizon metric in terms of the magnetic charges $\fn_a$ (see also \cite{Katmadas:2014faa,Halmagyi:2014qza} for similar expressions in the literature). Recalling that smooth solutions are obtained only with the upper sign in those expressions, we find the IR metric
\be
\label{ads2s2v2}
ds^2 = R^2_{\text{AdS}_2} \, ds^2_{\text{AdS}_2} + R^2_{S^2} \, ds^2_{S^2} \;,
\ee
with
\be
\label{radiusSofA}
R^2_{\text{AdS}_2} = \frac{\Pi}{\sqrt2 \, \Theta} \, \big( F_2 + \sqrt{\Theta} \big)^{1/2} \;,\qquad\qquad R^2_{S^2} = \frac1{\sqrt2} \big( F_2 + \sqrt{\Theta} \big)^{1/2} \;,
\ee
where the quantities $\Pi$, $F_2$, $\Theta$ are defined in (\ref{def Pi, F2, Theta}). The physical scalars are given by
\bea
\label{z of n}
z_1 &=  \frac{2(\fn_2 + \fn_3)(\fn_1-\fn_4)^2 - (\fn_1 + \fn_4) \big[ (\fn_2-\fn_3)^2 + (\fn_1-\fn_4)^2 \big] + 4 (\fn_4 - \fn_1) \sqrt{\Theta}}{2\fn_4 (\fn_4 - \fn_1 + \fn_2 - \fn_3)( \fn_4 - \fn_1 - \fn_2 + \fn_3)} \\
z_2 &= \frac{2(\fn_1 + \fn_3)(\fn_2-\fn_4)^2 - (\fn_2 + \fn_4) \big[ (\fn_1 - \fn_3)^2 + (\fn_2 - \fn_4)^2 \big] + 4 (\fn_4 - \fn_2) \sqrt{\Theta}}{2\fn_4 (\fn_4 + \fn_1 - \fn_2 - \fn_3)( \fn_4 - \fn_1 - \fn_2 + \fn_3)} \\
z_3 &=  \frac{2 (\fn_1+\fn_2)(\fn_3-\fn_4)^2 - (\fn_3+\fn_4) \big[ (\fn_1-\fn_2)^2 + (\fn_3-\fn_4)^2 \big] + 4 (\fn_4 - \fn_3) \sqrt{\Theta}}{2 \fn_4 (\fn_4 + \fn_1 -\fn_2 -\fn_3)( \fn_4 - \fn_1 + \fn_2 - \fn_3) } \;.
\eea
The sections at the horizon are then obtained from
\be
\label{neweq}
X_{1,2,3} = \frac{z_{1,2,3}}{1+z_1+z_2+z_3} \;,\qquad\qquad\qquad X_4 = \frac{1}{1+z_1+z_2+z_3} \;.
\ee
Smooth solutions are found if exactly three of the $\fn_a$ are negative. More details are given in Appendix \ref{app: horizon}.

These expressions can be further related to the different quartic invariants of the symplectic group and can be justified by the implicit electromagnetic duality of 4D ${\cal N}=2$ supergravity, see \cite{Katmadas:2014faa,Halmagyi:2014qza} for more details. Electromagnetic duality will likely play a more important role for generalizing our results to solutions with electric charges on top of the magnetic ones we consider.

\subsection{The entropy and R-symmetry}
\label{sec: sugra3}

At leading order, the entropy of the black hole is given by the area of the horizon via the Bekenstein-Hawking formula
\be
\label{entropy_gravity}
S_\text{BH} = \frac{\text{Area}}{4 G_\text{4D}} = \frac{\pi R_{S^2}^2}{G_\text{4D}} = \frac{\sqrt2\, \pi g^2}{G_\text{4D}}  \big( F_2 + \sqrt{\Theta} \big)^{1/2} \;,
\ee
where $G_\text{4D}$ is the four-dimensional Newton constant and we reinstated $g$ for dimensional reasons. We can also write the entropy in a more suggestive form using the symplectic sections $X_a$ to compare more directly with the field theory expression \eqref{Z large N},
\be
\label{entropy_gravity2}
S_\text{BH} = - \frac{2\pi g^2}{G_\text{4D}} \, \sqrt{X_1 (r_h) X_2 (r_h) X_3 (r_h) X_4 (r_h)} \; \sum_a \frac{\fn_a}{X_a (r_h)} \;.
\ee
Let us stress that this is only the leading contribution to the gravitational entropy, which should be supplemented by the higher-derivative corrections following the Wald formalism, and possibly by other quantum corrections. The leading answer for the entropy was confirmed by verifying the first law of thermodynamics in the canonical and grand-canonical ensembles for black holes in AdS$_4$ \cite{Hristov:2013sya}. Here we will not consider any corrections to the above formula, in accordance to the fact that we focused only on the leading $N^{3/2}$ contribution to the index on the field theory side.

As a last important remark about the supergravity solutions, let us note that the theory under consideration has four $U(1)$ gauge fields, which can be thought of as the four Cartan generators of the original $SO(8)$ R-symmetry in the maximal gauged supergravity in 4D. The $U(1)$ R-symmetry of ${\cal N}=2$ supergravity is gauged by a particular combination of those four $U(1)$'s, called the graviphoton. As shown in \cite{Hristov:2011qr} for general matter-coupled ${\cal N}=2$ supergravities, in asymptotically AdS spacetimes the graviphoton field strength $F_{\mu\nu}^\text{gp}$ is given by
\be
\label{graviphoton}
F_{\mu \nu}^\text{gp} = e^{\mathcal{K}/2} \, X^{\Lambda} F_{\Lambda, \mu \nu} \;,
\ee
where $F_{\Lambda}$ are the field strengths of the four gauge fields. This formula is correct only in the case of purely real (or purely imaginary, depending on conventions) sections $X^{\Lambda}$, which is the case here. In the context of the AdS/CFT correspondence, such a formula allows us to extract the exact R-symmetry from supergravity and it tells us how it changes from the boundary, where $X^{\Lambda} = 1/4$, to the horizon, where we find $X^{\Lambda} (r_h)$.

The notion of R-symmetry defined in \eqref{graviphoton} exists everywhere in the bulk, however it gets a clear holographic meaning only in the UV and the IR, where there is a corresponding exact R-symmetry for the superconformal 3D QFT and quantum mechanics, respectively. In the next section we will compare the field theory parameters $\Delta_a$ with the sections $X_a (r_h)$ at the horizon.

\subsection{The attractor mechanism}
\label{sec: sugra4}

The notion of attractor mechanism in black hole solutions refers to the way the expectation values of the scalars are fixed at the horizon in terms of the black hole charges. This has been explored carefully in the literature and we elaborate on it in Appendix \ref{sec: attractorsAPP}, while here we present a shortened version for the black holes we consider.

Let us first notice that there is a simple quantity that exists at generic points in spacetime,
\be
\label{eq:R}
{\cal R} = \sum\nolimits_a F_a \fn_a \;,
\ee
which is properly defined in an electromagnetic invariant way in Appendix \ref{sec: attractorsAPP} for more general black holes. The sections $F_a \equiv \partial F/\partial X_a$ are derived from the prepotential \eqref{prepotential}:
\be
F_a = - \frac{i}{X_a} \sqrt{X_1 X_2 X_3 X_4} \;.
\ee
It is therefore easy to see that $|{\cal R}|$ at the black hole horizon gives the entropy \eqref{entropy_gravity2}, up to a numerical prefactor.

Unlike the entropy, ${\cal R}$ is defined for all values of the sections $X_a$ at any point in spacetime, and for a static geometry it is a function of the radial coordinate $r$ only. It is therefore a natural measure of the holographic RG flow between the asymptotic AdS$_4$ and the near-horizon AdS$_2 \times S^2$ geometry. We observe that ${\cal R}$ matches functionally the index \eqref{Z large N}, if we assume a proportionality between $X_a$ and $\Delta_a$ (see Section \ref{sec: comparison}).

The quantity ${\cal R}$ is interesting for the attractor mechanism since it provides a function that the scalars extremize at the horizon,
\be
\frac{\partial {\cal R}}{\partial X_a} \Big|_\text{horizon} = 0 \;,
\ee
under the constraint $\sum_a X_a = 1$, and this determines the sections $X_a (r_h)$ and correspondingly the physical scalars $z_i (r_h)$ in terms of the charges $\fn_a$. We refer to Appendix \ref{sec: attractorsAPP} for the derivation of the above formula in the general context of half-BPS attractors in $\cN=2$ gauged supergravity.

\section{Comparison of index and entropy}
\label{sec: comparison}

We can finally compare the field theory and gravity results. We show that the topologically twisted index $|Z|$ in the large $N$ limit is extremized at a value of $\Delta_a$ which is proportional to the value of the sections $X^\Lambda$ at the horizon, and that the value of  $\log |Z|$ at the critical point precisely reproduces the entropy of the black hole.

The topologically twisted index is a function of the magnetic fluxes $\fn_a$ and the chemical potentials $\Delta_a$, while the black hole entropy only depends on $\fn_a$. The physical interpretation of the $\Delta_a$ is the following. The path integral of the topologically twisted theory can be interpreted as the Witten index
\be
\label{index}
Z (\fn_a ,  \Delta_a) =   \Tr \, (-1)^F \, e^{-\beta H} \, e^{i \sum_{a=1}^3 J_a  \Delta_a}
\ee
of the supersymmetric quantum mechanics obtained by reducing the theory on $S^2$ in the presence of the magnetic fluxes $\fn_a$ \cite{Benini:2015noa}. Here $J_a$ denote the currents associated with the global symmetries, as defined in Section \ref{sec: localization},  and the Hamiltonian depends explicitly on the fluxes $\fn_a$. The ${\cal N}=2$ quantum mechanics has supersymmetry algebra $\fu(1|1)$:
\be
\cQ^2 = \wb \cQ^2 = 0 \;,\qquad\{\cQ, \wb\cQ\} = 2H \;, \qquad [R,\cQ] = \cQ \;,\qquad [R,\wb\cQ] = - \wb\cQ \;,
\ee
where $\wb \cQ = \cQ^\dagger$ and $R$ is the R-symmetry generator. The R-symmetry $R$  is not unique, however. The generators $J_a$ of flavor symmetries, by definition, commute with $H$, $\cQ$, $\wb\cQ$, $R$---therefore any other symmetry $R' = R + \sum_a c_a J_a$ is an equally good R-symmetry. In particular, the fermion number $(-1)^F$ is a discrete R-symmetry transformation, which often is part of the continuous family of R-symmetries. In ABJM, the fermion number can be written in terms of the 3D superconformal R-symmetry $R_0$ that assigns charge $\frac12$ to the chiral multiplets $A_i$ and $B_j$:
\be
(-1)^F = e^{i\pi R_0} \, e^{- \frac{i\pi}{2} \sum_{a=1}^3 J_a} \;.
\ee
The topologically twisted index can then be written as%
\footnote{By the notation $(-1)^{R(\Delta_a)}$ we mean $e^{i\pi R(\Delta_a)}$.}
\be
\label{indexR}
Z (\fn_a ,  \Delta_a) =   \Tr \, (-1)^{R(\Delta_a)} \; e^{-\beta H}
\ee
as a function of the trial R-symmetry
\be
\label{Rsymm}
R(\Delta_a) = R_0 +  \frac{1}{\pi} \sum_{a=1}^3 \Big( \Delta_a - \frac{\pi}{2} \Big) \, J_a \;.
\ee
Thus, the fugacities $\Delta_a$  parametrize the mixing of the R-symmetry with the flavor symmetries, \ie{} the space of trial R-symmetries. Given the AdS$_2$ factor at the horizon, we expect that our quantum mechanics becomes superconformal at low energies. The IR superconformal algebra will single out a particular R-symmetry---the one sitting in the algebra---and a particular value for $\Delta_a$. It is natural to ask how to find the exact IR superconformal R-symmetry.

We can probe the mixing of the R-symmetry with the flavor symmetries using the dual supergravity solution. As already discussed, the graviphoton field strength $F^\text{gp}_{\mu \nu} = e^{\cK/2} X^{\Lambda} F_{\Lambda, \mu \nu}$ in (\ref{graviphoton}) depends on the radial coordinate through the sections $X^\Lambda$ and it is different at the boundary and at the horizon. Its expression suggests the identification
\be
\frac{\Delta_1}{\Delta_4} =  \frac{X_1}{X_4} \;,\qquad\qquad  \frac{\Delta_2}{\Delta_4} =  \frac{X_2}{X_4} \;,\qquad\qquad \frac{\Delta_3}{\Delta_4} =  \frac{X_3}{X_4} \;.
\ee
The constraint $\sum_a \Delta_a= 2 \pi n$ is compatible with $\sum_a X_a=1$ valid everywhere in the bulk.  Let us assume to be in the range $\sum_a \Delta_a=2\pi$. At the boundary, where the solution asymptotes to AdS$_4\times S^7$, the scalar fields $X_a$ are all equal and we find $\Delta_a= \pi/2$. This reproduces the UV superconformal R-symmetry of ABJM. At the horizon, on the other hand, the values of the scalars depend on the charges $\fn_a$ and, using (\ref{neweq}), we find
\be
\frac{\bar\Delta_{1,2,3}}{2\pi} =  X_{1,2,3}(r_h) = \frac{z_{1,2,3}}{1+z_1+z_2+z_3} \;,\qquad\qquad
\frac{\bar\Delta_4}{2\pi} = X_4(r_h) = \frac{1}{1+z_1+z_2+z_3}
\ee
in terms of the horizon values of the scalars in \eqref{z of n}. We can argue that $\bar\Delta_a$  determine, through (\ref{Rsymm}), the exact R-symmetry of the IR superconformal quantum mechanics.

\

Here comes the main result of our paper. First, with an explicit computation one can check that $\bar\Delta_a$ is a critical point of the function $|Z|$:
\be
\boxed{\quad\rule[-1.3em]{0pt}{3.1em}
\parfrac{\re\log Z}{\Delta_{1,2,3}} \Big|_{\sum_a \Delta_a = 2\pi} (\bar\Delta_a) = 0 \;.
\quad}
\ee
In fact, $\bar\Delta_a$ is the only critical point of $\log|Z|$ in the range $0< \Delta_a <2\pi$ (with $\sum_a\Delta_a=2\pi$). Setting to zero the derivatives of (\ref{Z large N}) with respect of $\Delta_{1,2,3}$ and expressing them in terms of $z_{1,2,3}$, one precisely obtains the equations (\ref{eq WYZ 1})-(\ref{eq WYZ 3}) that are solved in Appendix \ref{app: horizon}:  they lead to the two solutions in (\ref{solution z's}), but only the one with upper signs can possibly satisfy $z_{1,2,3} > 0$.

Second, we can then compare the value of $\log |Z|$ at the critical point $\bar\Delta_a$ with the black hole entropy. Using  \eqref{entropy_gravity2} and the relation%
\footnote{See for example \cite{Marino:2011nm}.}
\be
\frac{2g^2}{G_\text{4D}} = \frac{2 \sqrt{2}}{3} \, k^{1/2} \, N^{3/2} \;,
\ee
we find
\be
\boxed{\quad\rule[-.7em]{0pt}{2em}
\re\log Z\large \big|_\text{crit} (\fn_a) = \text{BH Entropy} \, (\fn_a) \;.
\quad}
\ee
Thus, we have reproduced the black hole entropy with a microscopic counting of ground states in a dual field theory, at the leading order $N^{3/2}$.

Let us notice that $\bar\Delta_a$ is a critical point of the function $\re\log Z$, but it is not a maximum. The Hessian of $\re\log Z$ has one negative and two positive eigenvalues, therefore the critical point is a saddle point. In fact, we should have expected this from the general large $N$ expression (\ref{Z large N}) of $\re\log Z$:  since, generically, at least one of the integers $\fn_a$ is negative (and in fact three of them should be negative to have regular black hole solutions), it follows that $\re\log Z$ diverges to positive infinity when the corresponding $\Delta_a$ goes to zero.

\subsection{The case with three equal fluxes}

To give a concrete example, we consider the simple case where
\be
\fn_1 = \fn_2 = \fn_3 \equiv \fn \;,\qquad\qquad \fn_4 = 2-3\fn \;.
\ee
From \eqref{def Pi, F2, Theta} we have
\be
F_2 = - (6\fn^2 - 6\fn + 1) \;,\qquad\qquad\qquad \Theta = (1-6\fn)(1-2\fn)^3 \;,
\ee
which lead to smooth supergravity solutions with regular horizon for $\fn<0$.

Consider the field theory expression in (\ref{Z large N}). For our particular choice of fluxes, we expect the critical point to lie along the submanifold $\Delta_1 = \Delta_2 = \Delta_3 \equiv \Delta$, $\Delta_4 = 2\pi - 3\Delta$, with $0 \leq \Delta \leq \frac{2\pi}3$.  We can therefore restrict $Z$ to such a submanifold:
\be
\re\log Z (\Delta) = -\frac{2N^\frac32}{3} \, \sqrt{\frac{2\Delta}{2 \pi - 3\Delta}} \, \big( 3 \pi \fn + (1-6\fn) \Delta \big) \;.
\ee
In the range $0 \leq \Delta \leq \frac{2\pi}3$ and for $\fn<0$, which is the region in the flux parameter space where a black hole with regular horizon exists, the function has a critical point at
\be
\bar\Delta = \frac\pi2 \bigg( 1 - \sqrt{ \frac{1-2\fn}{1-6\fn}} \, \bigg) \;,
\ee
that is also a positive maximum.%
\footnote{For $\fn>0$, instead, the function has a negative minimum in the range for $\Delta$.}
At the maximum the function takes the value
\be
\re\log Z(\bar\Delta) = \frac{2\pi}3 N^\frac32  \sqrt{ F_2 + \sqrt{\Theta}} \;, \ee
which precisely matches the entropy of the black hole  \eqref{entropy_gravity}.

Let us stress that, while restricted to the symmetric locus $\Delta_1 = \Delta_2 = \Delta_3 \equiv \Delta$ the index has a maximum, in the full parameter space spanned by the three independent parameters $\Delta_1$, $\Delta_2$ and $\Delta_3$ the critical point is a saddle point.

\section{Discussion and Conclusions}
\label{sec: conclusions}

In this paper we have computed the large $N$ limit of the topologically twisted index of the 3D ABJM theory, which counts (with phases) the ground states of the theory compactified on $S^2$ with R- and flavor magnetic fluxes. We have argued that this is relevant for understanding the physics of magnetically charged BPS black holes in AdS$_4$, arising in 4D maximal $\cN=8$ gauged supergravity. Each black hole can be given a holographic interpretation as the RG flow from the 3D ABJM theory twisted by the corresponding magnetic fluxes to a 1D superconformal quantum mechanics, whose ground states are counted by the index. Indeed, the leading $N^{3/2}$ contribution to the index precisely reproduces the leading Bekenstein-Hawking entropy of the black hole.

The matching proceeds in two steps. First, the index $Z(\fn_a, \Delta_a)$ is a function of fugacities $e^{i\Delta_a}$ as well as of magnetic fluxes $\fn_a$ for the flavor symmetries, and one has to \emph{extremize} $Z$ with respect to the $\Delta_a$'s. Comparing with supergravity, we observe that this procedure selects the exact superconformal R-symmetry in the IR $\su(1,1|1)$ superconformal algebra. Second, we observe that the index at the critical point, $Z\big( \fn_a, \bar\Delta_a(\fn_a) \big)$, precisely reproduces the black hole entropy $S_\text{BH}(\fn_a)$.

A possible interpretation could be the following.  We are evaluating a partition function with chemical potentials $\Delta_a$ for the flavor symmetries. The vanishing of the derivative with respect to $\Delta_a$  is equivalent to the vanishing of  the electric charge of the system,  which must be zero since the black hole is electrically neutral.  It is then conceivable that we get the entropy by extremization. However this argument is not completely satisfactory. The partition function we are computing is supersymmetric and treats bosons and fermions with different sign. Moreover the argument makes no use of the exact superconformal R-symmetry, whose role in the game is strongly suggested by the supergravity analysis.

It would be more interesting to have a clear mapping of the states counted by the topologically twisted index of the 3D ABJM theory to the black hole microstates. Although we do not yet have a clear understanding of this point, let us make some general observations.

\subsection*{A naive argument}

Let us first give a superficial argument that originally motivated our investigation. Suppose that the quantum mechanics describing the modes on $S^2$ is gapped with a finite number of ground states. Then the index reduces to
\be
\label{index in conclusions}
Z(\Delta_a) = \Tr_{H=0} \, (-1)^F e^{i \sum J_a \Delta_a} = \Tr_{H=0} \, (-1)^{R(\Delta_a)} \;,
\ee
where the Hamiltonian $H$ is a function of $\fn_a$. In the last expression we have written the index as a function on the space of R-symmetries of the theory (assuming that all IR R-symmetries are visible in the UV, \ie{} there are no accidental ones).
Then further suppose that, at low energies, the system develops 1D $\cN=2$ superconformal symmetry and the ground states are invariant under $\sl(2,\bR)$ conformal transformations: these assumptions follow from the fact that the supergravity solution develops an AdS$_2$ factor at the horizon. Then the $\su(1,1|1)$ algebra implies that the ground states have $R_c=0$, where $R_c \in \su(1,1|1)$ is the \emph{superconformal} \mbox{R-symmetry}. In other words, we conclude that in the space of all possible R-symmetries, there is one that assigns $(-1)^{R_c}=1$ to all ground states. But then, since (\ref{index in conclusions}) is a finite sum of phases, it is clear that it is \emph{maximized} when all phases are $1$. Since, as stressed in \cite{Benini:2015noa}, the overall phase of the index defined through the path-integral is ambiguous because of fermionic Fock space quantizations, we conclude that $|Z|$ is maximized:
$$
\max_{\Delta_a} \, \big| Z(\Delta_a) \big| = \max_{\Delta_a} \, \big| \Tr_{H=0} \, (-1)^{R(\Delta_a)} \big| = \big| \Tr_{H=0} \, (-1)^{R_c} \big| = \Tr_{H=0} 1 \;.
$$
Thus, an argument  of this kind ``would prove'' two statements: (1) that the index function $\big| Z(\Delta_a) \big| = \big| \Tr_{H=0}\, (-1)^{R(\Delta_a)} \big|$ has a maximum at the point $\bar\Delta_a$ where the trial R-symmetry equals the IR superconformal \mbox{R-symmetry}, $R(\bar\Delta_a) = R_c$; (2) that the index evaluated at the maximum, $\big| Z(\bar\Delta_a) \big|$, computes the number of ground states (as opposed to a weighted sum).

Unfortunately, this argument is too superficial and it does not apply to the black holes. First of all, if at low energies we just have a finite number of zero-energy ground states separated from the rest by a gap, then the low-energy theory is just $H=0$: a bunch of states with no dynamics. An example is a collection of $|\fn|$ 1D free Fermi multiplets (which can be obtained from a 3D free chiral multiplet on $S^2$, with negative magnetic flux $\fn$): the index is
$$
\big| Z_\text{chiral}(\fn, \Delta) \big| = \bigg| \Big( \frac{y^{1/2}}{1-y} \Big)^\fn \bigg| \qquad\qquad\text{with $y=e^{i\Delta}$}
$$
which, for $\fn<0$, is maximized at $y=-1$ with $|Z_\text{chiral}(\fn,\pi)| = 2^{|\fn|}$ (correct number of states in the fermionic Fock space). On such theories $\su(1,1|1)$ simply does not act, and therefore it is hard to understand how this trivial superconformal quantum mechanics can be dual to AdS$_2$ (although compare with \cite{Maldacena:1998uz}).

A non-trivial superconformal quantum mechanics with states with $H>0$ necessarily has a continuous spectrum that spans $\bR_+$, just because the spectrum must be invariant under dilations. Then the states are necessarily non-normalizable, and computing an index (for instance of $L^2$-normalizable states as in \cite{Sethi:1997pa}) is in general very difficult. In such cases, our index---which is an equivariant index as opposed to an $L^2$ index---is defined by first deforming the Hamiltonian with real masses $\sigma_a$ (that make the spectrum discrete), and then performing analytic continuation to $\sigma_a=0$ exploiting holomorphy in $\Delta_a + i\beta \sigma_a$. In this setup the argument above does not apply.

Indeed, the ABJM index in (\ref{Z large N}) diverges when some $\Delta_a$ vanish.%
\footnote{Some divergence had to be expected. The BPS black holes are the near-horizon geometry of $N$ M2-branes wrapping the $S^2$ in the Calabi-Yau geometry $\bigotimes_{a=1}^4 \cL_a(-\fn_a) \, \bP^1$, which is the total space of four line bundles over $\bP^1$ with first Chern classes $-\fn_a$. When some $\fn_a<0$, there are non-trivial holomorphic sections and the M2-branes can be well separated, giving rise to flat directions. This, however, only explains $\cO(N)$ divergences, not $\cO(N^{3/2})$.}
This excludes the possibility of a finite Hilbert space of normalizable ground states gapped from the rest, and so the superficial argument does not apply. In fact, the index has a saddle---not a maximum---at the point that corresponds to the superconformal R-symmetry and that reproduces the BH entropy.

\subsection*{The $\boldsymbol{I}$-extremization principle}

We would like to propose that the $I$-extremization principle, stating that
\begin{enumerate}
\item the index is extremized at the superconformal R-symmetry, and
\item the value of the index at the extremum is the regularized number of ground states,
\end{enumerate}
has a general validity in $\cN=2$ superconformal quantum mechanics, under certain assumptions suitable for the black holes. Obviously, it would be desirable to precisely understand what assumptions are necessary, and to have a rigorous proof.

A better understanding of all these issues necessarily involves a better understanding of the superconformal quantum mechanics with $\su(1,1|1)$ symmetry.
Here  we just notice that a simple example of superconformal quantum mechanics with continuous spectrum is provided by a free chiral multiplet (this can be obtained from a 3D free chiral multiplet on $S^2$ with $\fn>0$). We study this example in some details in Appendix \ref{app: freefield}. It turns out that the index diverges at $\Delta=0$, it has a minimum at the superconformal R-symmetry and its value gives the zeta-regularized number of states: $\frac12$. In this case, extremization can be proven from time-reversal invariance and integrality of the R-charge spectrum.

\subsection*{Relations with the literature and future directions}

Let us briefly comment about the connection between our results and several other streams of ideas in the literature. The 3D topologically twisted index considered in this paper becomes an equivariant Witten index for the dimensionally reduced quantum mechanics. We should notice that there exist another chiral index in $\cN=2$ superconformal quantum mechanics---the superconformal index---which makes use of $L_0$ that has discrete spectrum \cite{Michelson:1999zf, BrittoPacumio:2000sv, Britto-Pacumio:2001zma}, as reviewed in Appendix \ref{sec: superconformal index}. The relation between the equivariant and the superconformal indices is not obvious and deserves investigation.

It would be interesting to better understand the relation of our procedure with other extremization mechanisms that appear in the physics of black holes. As we showed in Section \ref{sec: sugra} and Appendix \ref{sec: attractorsAPP}, the entropy can be obtained by extremizing with respect to the value of the scalar fields at the horizon. This has a natural interpretation in terms of an attractor mechanism \cite{Ferrara:1996dd}, which plays an important role in asymptotically flat black holes. We also recognize many similarities with Sen's entropy function formalism \cite{Sen:2005wa}, of which we might provide a supersymmetric version. In this context one could investigate the relation between the twisted index before extremization and Sen's entropy function.

If the $I$-extremization principle turned out to be correct, it should be added to the list of well-established theorems in other dimensions: $a$-maximization in 4D \cite{Intriligator:2003jj, Barnes:2004jj}, $F$-maximization in 3D \cite{Jafferis:2010un, Jafferis:2011zi, Closset:2012vg} and $c$-extremization in 2D \cite{Benini:2012cz, Benini:2013cda}.

To provide tests of the proposed $I$-extremization principle, one could study more general black holes in the same supergravity model, but with both magnetic and electric charges: we are currently investigating this direction. Other obvious generalizations are to look at the twisted index for CS level $k>1$, and on higher-genus Riemann surfaces. In fact, as discussed in Appendix \ref{app: horizon}, there are analogous families of BPS black holes with toroidal and higher-genus horizons. It would also be interesting to generalize our computations to other less symmetric theories, from the 11D point of view. For instance, starting with the geometries AdS$_4 \times \text{SE}_7$ and their field theory duals (possibly considering toric Sasaki-Einstein cones as in \cite{Hanany:2008cd, Hanany:2008fj, Martelli:2008si, Aganagic:2009zk, Benini:2009qs, Benini:2011cma}) and placing them on a Riemann surface, one can obtain $\frac14$-BPS black holes in broad families of 4D $\cN=2$ gauged supergravities.

A very important question is whether the index provides the \emph{exact} number of black hole microstates, beyond the leading contribution in $N$. It is known that in some examples (\eg{} \cite{Dabholkar:2010rm}) the black hole represents only part of the conformally-invariant states, while other ones are represented by graviton waves or other modes. It would be interesting to compute $1/N$ corrections, both in supergravity and in the large $N$ expansion of the index, to clarify the issue.

On a different note, let us also emphasize that the integral expression for the topologically twisted index found in \cite{Benini:2015noa}, as the one for the elliptic genus in \cite{Benini:2013nda, Benini:2013xpa}, provides a novel type of large $N$ ``matrix models'': the integrands are standard, but they are integrated along non-trivial contours. These models probably have a rich mathematical structure deserving its own attention.

\section*{Acknowledgements}

We would like to thank Alessandro Tomasiello for initial collaboration on this project and many insightful discussions. We would also like to thank Nikolay Bobev, Jan de Boer, Diego Hofman, Stefanos Katmadas, Sungjay Lee, Chiara Toldo, Stijn van Tongeren and Toby Wiseman for useful discussions.
AZ is  grateful to the Ecole Normale Superieure (LPTENS), Paris VI - Jussieu (LPTHE), the Benasque Center for Science, the Mainz Institute for Theoretical Physics (MITP) for hospitality and its partial support during the completion of this work.
FB is supported by the Royal Society as a Royal Society University Research Fellowship holder, and by the MIUR-SIR grant RBSI1471GJ ``Quantum Field Theories at Strong Coupling: Exact Computations and Applications''.
AZ is supported by the INFN and the MIUR-FIRB grant RBFR10QS5J ``String Theory and Fundamental Interactions''.

\appendix

\section{Supergravity solutions}
\label{app: horizon}

In this appendix we derive the black hole horizon solutions, in order to study in what region of the parameter space the solutions are smooth with regular horizon. For completeness we consider the general case with AdS$_2\times \Sigma_\fg$ horizon, where $\Sigma_\fg$ is a Riemann surface of arbitrary genus $\fg$.

\subsection[4D $\cN=2$ gauged supergravity from $\cN=8$]{4D $\boldsymbol{\cN=2}$ gauged supergravity from $\boldsymbol{\cN=8}$}

We use the Lagrangian and BPS equations given in  \cite{Gauntlett:2001qs}, which conveniently summarizes  the results in \cite{Duff:1999gh,Cvetic:1999xp}. Note that this is not  the standard ${\cal N}=2$ gauged supergravity notation, but rather the natural notation imposed from the reduction of 11D supergravity on $S^7$. For the bosonic fields we use the normalization and index structure from the main text, and make explicit comments about the relation with the conventions in \cite{Gauntlett:2001qs} when needed.

The $S^7$ reduction of 11D supergravity gives the 4D ${\cal N}=8$ $SO(8)$ gauged supergravity. Using the reduction ansatz of \cite{Cvetic:1999xp} one finds a consistent reduction to $U(1)^4$ gauged supergravity:
\bea
\label{app embedding}
ds^2 &= \Delta^\frac23 \, ds_4^2 + \frac2{g^2 \Delta^\frac13} \sum_a \frac1{L_a} \big( d\mu_a^2 + \mu_a^2 ( d\varphi_a + g A_a)^2 \big) \\
G_4 &= \sqrt2\, g \sum_a \big( L_a^2 \mu_a^2 - \Delta \, L_a \big) \, \epsilon_4 - \frac1{\sqrt2\, g} \sum_a L_a^{-1} ( \ast \,  dL_a ) \wedge d\mu_a^2 \\
&\qquad - \frac{\sqrt{2}}{g^2} \sum_a L_a^{-2} d\mu_a^2 \wedge (d\varphi_a + g A_a) \wedge \ast F_a \;.
\eea
Here $a = 1,\dots,4$, the $L_a$ satisfy $L_1L_2L_3L_4 = 1$ and parametrize the scalars, $A_a$ are 1-forms with field strengths $F_a = dA_a$, $\Delta = \sum_a L_a \mu_a^2$ is the warp factor, $\sum_a \mu_a^2 = 1$ and $0 \leq \varphi_a < 2\pi$ parametrize $S^7$, $U(1)^4 \subset SO(8)$ is parametrized by $\varphi_a$, $\ast$ is the Hodge operator on $ds_4^2$ and $\epsilon_4$ is its volume form.%
\footnote{The $A_a$ here are the same from the main text, related to the $A_{\alpha}$ in  \cite{Gauntlett:2001qs} by $A_a = 2 A_{\alpha}$ and $g=e$.}

The reduction gives a 4D theory with bosonic action
\be
\label{4d N=2 sugra bosonic action}
\cL = \frac1{2\kappa^2} \, \bigg[ R - \frac12 (\partial \vec\phi)^2 - \frac12  \sum\nolimits_a e^{\vec a_a \cdot \vec\phi} F_a^2 - V(\phi) \bigg]
\ee
where
\be
V = -4 g^2 \big( \cosh \phi_{12} + \cosh \phi_{13} + \cosh \phi_{14} \big) \;.
\ee
In this Lagrangian we have parametrized the constrained scalar fields $L_a$ with%
\footnote{This is yet another parametrization of the physical scalars. To compare with the main text, the functions $L_a$ are proportional to the section $X_a$ so that we can write $z_{1,2,3} = L_{1,2,3}/L_4$.}
\be
\vec\phi = (\phi_{12}, \phi_{13}, \phi_{14}) \;.
\ee
We can combine them into a symmetric tensor $\phi_{a b}$, which is self-dual ($\phi_{34} = \phi_{12}$, $\phi_{24} = \phi_{13}$ and $\phi_{23} = \phi_{14}$) and zero on the diagonal, $\phi_{aa} = 0$. The $L_a$ are then given by
\be
L_a = e^{- \vec a_a \cdot \vec\phi/2}
\ee
with
\be
\vec a_1 = (1,1,1) \;,\qquad \vec a_2 = (1, -1, -1) \;,\qquad \vec a_3 = (-1, 1, -1) \;,\qquad \vec a_4 = (-1, -1, 1) \;.
\ee
In fact (\ref{4d N=2 sugra bosonic action}) is the bosonic action of 4D $\cN=2$ $U(1)^4$ gauged supergravity with the three axions set to zero \cite{Cvetic:1999xp}. We stress that (\ref{4d N=2 sugra bosonic action}) is \emph{not} a consistent reduction without the three axions  \cite{Cvetic:1999xp}. They are sourced by $F \wedge F$, so it is consistent to set them to zero only if $F \wedge F = 0$. We can still consider either electric or magnetic charges.

The fermionic fields of the ${\cal N}=8$ $SO(8)$ gauged supergravity are the gravitini $\psi_\mu^I$ and the spin-$\frac12$ fields $\chi^{[IJK]}$, where $I,J,K$ are $SO(8)$ indices. We can decompose $I$ in the pair $(a,i)$ with $a=1,\dots,4$ and $i=1,2$. The gravitini variations are (see (2.15) in \cite{Duff:1999gh})
\be
\delta \psi_\mu^{a i} = \nabla_\mu \epsilon^{a i} - g \sum_{b j} \Omega_{a b} A^{b}_\mu \varepsilon^{ij} \epsilon^{a j} + \frac g{4\sqrt2} \sum_b e^{-\vec a_b \cdot \vec\phi/2} \gamma_\mu \epsilon^{a i} + \frac1{4 \sqrt{2}} \sum_{b \nu\lambda j} \Omega_{a b} e^{\vec a_b \cdot \vec\phi/2} F^{b}_{\nu\lambda} \gamma^{\nu\lambda} \gamma_\mu \varepsilon^{ij} \epsilon^{a j}
\ee
where $\varepsilon^{ij}$ is the antisymmetric tensor, $\epsilon^{a i}$ are the Killing spinors and
\be
\label{def Omega}
\Omega = \frac12 \mat{ 1 & 1 & 1 & 1 \\ 1 & 1 & -1 & -1 \\ 1 & -1 & 1 & -1 \\ 1 & -1 & -1 & 1} \;.
\ee
The spin-$\frac12$ fermions $\chi^{[IJK]}$ are totally antisymmetric. It turns out \cite{Duff:1999gh} that $\delta\chi^{[IJK]} = 0$ unless at least two indices have the same $a$ (then different $i$ because of antisymmmetry), but they cannot all three have the same $a$ because of antisymmetry. One can then write
\be
\label{delta chi}
\delta \chi^{a i\, b j\, c k} = \delta\underline\chi^{a c k} \delta^{ab} \varepsilon^{ij} + \delta\underline\chi^{ba i} \delta^{b c} \varepsilon^{jk} + \delta\underline\chi^{c b j} \delta^{c a} \varepsilon^{ki}
\ee
which is automatically antisymmetric in the pairs $(a,i)$ \textit{etc.}, where
\be
\delta \underline\chi^{a b i} = - \frac1{\sqrt2} \sum_{\mu j} \gamma^\mu \partial_\mu \phi_{a b} \varepsilon^{ij} \epsilon^{b j} - g \sum_{c d  j} \Sigma_{a b c} \Omega_{c d} e^{-\vec a_d \cdot \vec\phi/2} \varepsilon^{ij} \epsilon^{b j}
+ \frac1{2} \sum_{d \mu\nu} \Omega_{a d} e^{\vec a_d \cdot \vec\phi /2} F^{d}_{\mu\nu} \gamma^{\mu\nu} \epsilon^{b i}
\ee
if $a \neq b$ while $\delta\underline\chi^{a a i} \equiv 0$. Clearly $\delta\chi$ in (\ref{delta chi}) vanishes if $a \neq b \neq d$, while if $a = b$ then $\delta\chi = \delta\underline\chi^{a c k} \varepsilon^{ij}$. The BPS equations then reduce to $\delta\underline\chi^{a b i} = 0$.
In the formula, $\phi_{a b}$ is defined above and
\be
\Sigma_{a b c} = \begin{cases} |\varepsilon_{a b c}| &\text{for } a, b, c \neq 1 \\ \delta_{b c} &\text{for } a = 1 \\ \delta_{a c} &\text{for } b = 1 \\ 0 &\text{otherwise.} \end{cases}
\ee
At this point we can choose the gauge coupling constant
\be
g = 1/\sqrt2 \;,
\ee
such that the UV metric is the unit-radius AdS$_4$ as in the main text; the coupling constant $g$ can be reinstated at the end by sending $L_a \to \sqrt2\, g L_a$.

\subsection{Wrapped M2-branes}

The black-hole solutions can be thought of as the near-horizon geometry of a large number of M2-branes wrapping a Riemann surface $\Sigma_\fg$. To construct them, we consider the metric ansatz
\be
\label{near horizon metric ansatz}
ds^2 = e^{2f_1}(-dt^2 + dr^2) + e^{2f_2+2h}(dx^2 + dy^2)
\ee
where $f_{1,2}$ are functions of $r$ and $h$ is a function of $x,y$. We choose  vielbein $e_{\hat t} = e^{f_1} dt$, $e_{\hat r} = e^{f_1} dr$, $e_{\hat x} = e^{h+f_2} dx$, $e_{\hat y} = e^{h+f_2} dy$. We fix
\be
e^{2h} =
\begin{cases} \frac4{(1+x^2+y^2)^2} &\text{for } S^2 \\ 2\pi & \text{for } T^2 \\ \frac1{y^2} &\text{for } H^2 \end{cases}
\ee
so that $ds_\Sigma^2 = e^{2h}(dx^2 + dy^2)$ is a constant curvature metric on the Riemann surface with
\be
R^\Sigma_{ab} = \kappa\, g^\Sigma_{ab} \;,\qquad\qquad\qquad\quad R^\Sigma = 2\kappa \;,
\ee
and $\kappa = 1$ for $S^2$, $\kappa = 0$ for $T^2$, and $\kappa = -1$ for $H^2$. The range of coordinates are $(x,y) \in \bR^2$ for $S^2$, $(x,y) \in [0,1)^2$ for $T^2$, and $(x,y) \in \bR \times \bR_{>0}$ for $H^2$. In the $H^2$ case the upper half-plane has to be quotiented by a suitable Fuchsian group to get a compact Riemann surface $\Sigma_{\fg>1}$. The ranges are chosen in such a way that
\be
\Vol(\Sigma_\fg) = \int e^{2h} dx\,dy = 2\pi \eta \;,\qquad\qquad \eta \equiv \begin{cases} 2|\fg-1| &\text{for } \fg \neq 1 \\ 1 & \text{for } \fg = 1 \end{cases}
\ee
where we defined the positive number $\eta$. The case of genus $\fg>1$ follows from the Gauss-Bonnet theorem $\frac12 \int R^\Sigma \, \dvol_\Sigma = 4\pi(1-\fg)$.

The field strengths are taken as
\be
F^{a} = -\frac{\fn_a}{\sqrt{2}} \, e^{2h} dx \wedge dy = -\frac{\fn_a}{\sqrt{2}} \, \dvol_\Sigma \;.
\ee
On curved Riemann surfaces, we can choose a gauge connection proportional to the spin connection: defining $\tilde \omega_\mu = \frac12 \omega_\mu^{ab} \varepsilon_{ab}$ on $\Sigma_\fg$, we have
\be
d\tilde\omega = \frac{R^\Sigma}2 \, \dvol_\Sigma \;.
\ee
The parameters $\fn_a$ will be quantized later. Notice that the ansatz considered here contains, for $\kappa=1$, the supergravity solution presented in Section \ref{sec: sugra}, however the radial coordinate used here is not the same as the one used in (\ref{BH_metric}), as it is obvious by comparing with (\ref{near horizon metric ansatz}).

We choose the following projectors on spinors:
\be
\label{projectors on spinors}
\gamma_{\hat r} \epsilon^{a i} = \epsilon^{a i} \;,\qquad 0 = \partial_{t,x,y} \epsilon^{a i} \;,\qquad \gamma_{\hat x \hat y} \epsilon^{a i} = - \varepsilon^{ij} \epsilon^{a j} \;,\qquad \epsilon^{a i} = 0 \text{ for } a = 2,3,4 \;.
\ee
The first two conditions generically select the Poincar\'e supercharges (versus possible conformal supercharges on AdS); the second is a symplectic reduction for M2-branes on $\Sigma_\fg$; the fourth one---to be compared with $\Omega$ in (\ref{def Omega})---means that we only keep the diagonal supercharge coupled to all fluxes with charge $+1$ (additional supercharges arise if some fluxes are zero and so other rows of $\Omega$ vanish), as in \cite{Gauntlett:2001qs}.

\

Let us start with the gravitino variation. If $a\neq 1$ then $\delta\psi_\mu^{a i} = 0$ automatically. We then define $\epsilon^i = \epsilon^{1i}$, and get
\be
0 = \delta\psi_\mu^{1i} = \nabla_\mu \epsilon^i - \frac{1}{2\sqrt{2} } \sum_b A_\mu^{b} \varepsilon^{ij} \epsilon^j + \frac1{8} \sum_b L_b \gamma_\mu \epsilon^i + \frac1{8\sqrt{2}} \sum_b L_b^{-1} F^{b}_{\nu\lambda} \gamma^{\nu\lambda} \gamma_\mu \varepsilon^{ij} \epsilon^j \;.
\ee
From $\mu = \hat t$ we get
\be
0 = \Big[ e^{-f_1} f_1' + \frac14 {\ts \sum_b L_b} - \frac{e^{-2f_2}}4 {\ts \sum_b \fn_b L_b^{-1}} \Big]\, \epsilon^i \;.
\ee
From $\mu = \hat r$, and using $\partial_{\hat r} = e^{-f_1}\partial_r$, we get
\be
0 = \Big[ 2 e^{-f_1} \partial_r + \frac14 {\ts \sum_b L_b} - \frac{e^{-2f_2}}4 {\ts \sum_b \fn_b L_b^{-1}} \Big]\, \epsilon^i \;.
\ee
Combining the two we get $\partial_r \epsilon^i = \frac12 f_1' \epsilon^i$, \ie{}
\be
\epsilon^i(r) = e^{f_1(r)/2} \, \epsilon^i_0 \;,\qquad\qquad\qquad \epsilon_0 = \text{const} \;.
\ee
From $\mu = \hat x$ we get
\be
0 = - \Big[ \frac12 e^{-f_2-h} \partial_y h + \frac1{2\sqrt2} \, {\ts \sum_b A_{\hat x}^b} \Big] \varepsilon^{ij} \epsilon^j
+ \Big[ \frac12 e^{-f_1} f_2' + \frac18 {\ts \sum_b L_b} + \frac{e^{-2f_2}}8 {\ts \sum_b \fn_b L_b^{-1}} \Big] \gamma_{\hat x} \epsilon^i
\ee
which gives two equations. We have an analogous equation for $\mu = \hat y$. Combining the two we find
\be
\sum\nolimits_a \fn_a = 2 \kappa \;,
\ee
and an equation for $f_2^\prime$.

Now let us look at the gaugino variation $\delta\underline\chi^{ab i}$. Given our ansatz for $\epsilon^{bi}$, it follows that we obtain non-trivial equations only for $b=1$ and therefore for $a\neq1$. We get
\be
0 = - \frac1{\sqrt2} \bigg[ e^{-f_1} \partial_r \phi_{a1} + \sum_d \Omega_{a d} L_d - e^{-2 f_2} \sum_d \Omega_{a d} \fn_d L_d^{-1} \bigg] \, \varepsilon^{ij} \epsilon^j \;.
\ee
These are three equations for $a = 2,3,4$.

\

The final full set of BPS equations is:
\bea
\label{BPS equations full set}
e^{-f_1}f_1' &= - \frac 1{4}\big(L_1 + L_2 + L_3 + L_4 \big) + \frac{e^{-2f_2}}{4} \big( \fn_1 L_1^{-1} + \fn_2 L_2^{-1} + \fn_3 L_3^{-1} + \fn_4 L_4^{-1} \big) \\
e^{-f_1}f_2' &= - \frac 1{4}\big(L_1 + L_2 + L_3 + L_4 \big) - \frac{e^{-2f_2}}{4} \big( \fn_1 L_1^{-1} + \fn_2 L_2^{-1} + \fn_3 L_3^{-1} + \fn_4 L_4^{-1} \big) \\
e^{-f_1} \vec\phi_1' &= - \frac 1{2}\big(L_1 + L_2 - L_3 - L_4 \big) + \frac{e^{-2f_2}}{2} \big( \fn_1 L_1^{-1} + \fn_2 L_2^{-1} - \fn_3 L_3^{-1} - \fn_4 L_4^{-1} \big) \\
e^{-f_1} \vec\phi_2' &= - \frac 1{2}\big(L_1 - L_2 + L_3 - L_4 \big) + \frac{e^{-2f_2}}{2} \big( \fn_1 L_1^{-1} - \fn_2 L_2^{-1} + \fn_3 L_3^{-1} -\fn_4 L_4^{-1} \big) \\
e^{-f_1} \vec\phi_3' &= - \frac 1{2}\big(L_1 - L_2 - L_3 + L_4 \big) + \frac{e^{-2f_2}}{2} \big(\fn_1 L_1^{-1} - \fn_2 L_2^{-1} - \fn_3 L_3^{-1} + \fn_4 L_4^{-1} \big) \\
\sum\nolimits_a \fn_a &= 2 \kappa \;.
\eea

To understand the quantization condition, consider the case of M2-branes on $T^*\Sigma_\fg$ \ie{} take $\fn_{2,3,4}=0$ and $\fn_1 = 2 \kappa$. In this case we know that on $T^*S^2 \simeq \bC^2/\bZ_2$ there are two (negative) units of flux, and on $T^*\Sigma_{\fg>1}$ there are $2(\fg-1)$ units of flux. We conclude that the quantization condition is
\be
\fn_a \in \frac2\eta\, \bZ \;.
\ee
In the case of $S^2$ considered in the main text, the $\fn_a$ are integers. On a higher genus Riemann surface, a more refined quantization is possible.

\subsection[AdS$_2 \times \Sigma_\fg$ solutions]{AdS$\boldsymbol{_2 \times \Sigma_\fg}$ solutions}

We could solve the BPS equations in (\ref{BPS equations full set}), which are a system of coupled ODEs, to find the complete black hole solutions discussed in the main text and their generalization with $\Sigma_\fg$ horizon. Instead, we will here analyze only the near-horizon geometry AdS$_2 \times \Sigma_\fg$, for which the equations become algebraic. This will be enough to study the region in parameter space where smooth solutions with regular horizon exist.

We set $e^{2f_1(r)} = e^{2f}/r^2$ and all other functions constant. We get the algebraic system:
\begin{align}
\label{eq f}
\frac4{e^f} &= \Big(L_1 + L_2 + L_3 + \frac1{L_1L_2L_3} \Big) - e^{-2f_2} \big( \fn_1 L_1^{-1} + \fn_2 L_2^{-1} + \fn_3 L_3^{-1} + \fn_4 L_1L_2L_3 \big) \\
\label{eq g}
0 &= \Big(L_1 + L_2 + L_3 + \frac1{L_1L_2L_3} \Big) +e^{-2f_2} \big( \fn_1 L_1^{-1} + \fn_2 L_2^{-1} + \fn_3 L_3^{-1} + \fn_4 L_1L_2L_3 \big) \\
\label{eq phi 1}
0 &= \Big(L_1 + L_2 - L_3 - \frac1{L_1L_2L_3} \Big) - e^{-2f_2} \big( \fn_1 L_1^{-1} + \fn_2 L_2^{-1} - \fn_3 L_3^{-1} - \fn_4 L_1L_2L_3 \big) \\
\label{eq phi 2}
0 &=  \Big(L_1 - L_2 + L_3 - \frac1{L_1L_2L_3} \Big) - e^{-2f_2} \big( \fn_1 L_1^{-1} - \fn_2 L_2^{-1} + \fn_3 L_3^{-1} - \fn_4 L_1L_2L_3 \big) \\
\label{eq phi 3}
0 &= \Big(L_1 - L_2 - L_3 + \frac1{L_1L_2L_3} \Big) - e^{-2f_2} \big( \fn_1 L_1^{-1} - \fn_2 L_2^{-1} - \fn_3 L_3^{-1} + \fn_4 L_1L_2L_3 \big)
\end{align}
together with $\sum \fn_a = 2 \kappa$. We have substituted $L_1L_2L_3L_4=1$.

First notice that it must be $L_a > 0$ for all $a$. Then consider a linear combination of (\ref{eq g})-(\ref{eq phi 3}) with coefficients equal to the last row of $\Omega$ (\ref{def Omega}): it gives
\be
2 e^{2f_2} = \fn_4 L_1^2 L_2^2 L_3^2 - \fn_1 L_2L_3 - \fn_2 L_3L_1 - \fn_3 L_1L_2 \;.
\ee
The combination $(\text{\ref{eq f}}) + (\text{\ref{eq g}})$ gives
\be
e^f = \frac{2L_1L_2L_3}{1 + L_1^2L_2L_3 + L_1L_2^2L_3 + L_1L_2L_3^2} \;.
\ee
We can define the positive non-vanishing variables
\be
z_1 = L_1^2L_2L_3 \;,\qquad z_2 = L_1L_2^2L_3 \;,\qquad z_3 = L_1L_2L_3^2 \;,
\ee
which correspond to the physical scalars used in the main text. In fact, they are simply given by $z_{1,2,3} = L_{1,2,3}/L_4$.%
\footnote{The $L_a$ are proportional to the $X_a$.}
The relations above are inverted by
\be
L_1^4 = \frac{z_1^3}{z_2z_3} \;,\qquad L_2^4 = \frac{z_2^3}{z_1z_3} \;,\qquad L_3^4 = \frac{z_3^3}{z_1z_2} \;,\qquad L_4^4 = \frac1{z_1z_2z_3} \;.
\ee

Taking three linear combinations of (\ref{eq g})-(\ref{eq phi 3}), with coefficients equal to the first three rows of $\Omega$ (\ref{def Omega}), we get
\begin{align}
\label{eq WYZ 1}
0 &= (\fn_1 z_2 + \fn_2 z_1) z_3(z_3-1) + (\fn_3-\fn_4 z_3) z_1 z_2 (z_3+1) \\
\label{eq WYZ 2}
0 &= (\fn_2 z_3 + \fn_3 z_2) z_1(z_1-1) + (\fn_1 - \fn_4 z_1) z_2 z_3 (z_1 +1) \\
\label{eq WYZ 3}
0 &= (\fn_1 z_3 + \fn_3 z_1) z_2(z_2-1) + (\fn_2 - \fn_4 z_2) z_1 z_3 (z_2+1) \;.
\end{align}
Solving the first or the second equation for $z_2$, we get
\be
\label{expr Y}
z_2 = - \frac{\fn_2 z_1 z_3 (z_3 - 1)}{\fn_1 z_3 (z_3 -1) + (\fn_3 - \fn_4 z_3) z_1 (z_3 + 1)} = - \frac{\fn_2 z_1 z_3( z_1 -1)}{\fn_3 z_1( z_1 -1) + (\fn_1-\fn_4 z_1) z_3 (z_1 + 1)} \;.
\ee
Each of the two expressions is valid if its numerator and denominator are both non-vanishing. Unless $\fn_2=0$ or $z_1=z_3=1$, at least one of the two expressions is valid; we can then substitute in (\ref{eq WYZ 1}) or (\ref{eq WYZ 2}), respectively, obtaining
\be
\label{eq WZ}
0 = \fn_1 z_3(z_3-1) - \fn_3 z_1(z_1-1) + \fn_4 z_1 z_3 (z_1-z_3) \;.
\ee
If we substitute the first expression of $z_2$ (\ref{expr Y}) in (\ref{eq WYZ 3}) we obtain a complicated equation:
\be
\label{eq WZ2}
0 = \fn_1^2 z_3 (z_3-1)^2 + z_1^2 (\fn_3 - \fn_4 z_3) \big( \fn_2 (1-z_3) + (\fn_3 - \fn_4)(1+z_3) \big) + \fn_1 z_1 (z_3-1) \big( \fn_3 (2z_3+1) - \fn_4 z_3 (z_3 +2) \big) \;.
\ee
However the combination $(\text{\ref{eq WZ2}}) - \fn_1 (z_3-1) (\text{\ref{eq WZ}})$ gives a linear equation in $z_1$:
\be
\label{expr W}
z_1 = \frac{2 \fn_1(\fn_4-\fn_3) z_3 (z_3-1)}{(\fn_3 - \fn_4z_3)\big[ (\fn_1-\fn_2)(z_3-1) + (\fn_3-\fn_4)(z_3+1) \big]} \;.
\ee
Finally, we substitute this back into (\ref{eq WZ}) obtaining a quadratic equation in $z_3$:
\begin{multline}
0 = (\fn_4 + \fn_1 -\fn_2 -\fn_3)( \fn_4 - \fn_1 + \fn_2 - \fn_3) (\fn_4 z_3^2 + \fn_3) \\
+ \Big( (\fn_3+\fn_4) \big[ (\fn_1-\fn_2)^2 + (\fn_3-\fn_4)^2 \big] - 2 (\fn_1+\fn_2)(\fn_3-\fn_4)^2\Big) z_3 \;.
\end{multline}
This gives two solutions for $z_3$, and substituting back into (\ref{expr W}) and (\ref{expr Y}) we find the values of the other scalars as well.

Hence, we find two solutions for the scalars:
\bea
\label{solution z's}
z_1 &= \frac{2(\fn_2 + \fn_3)(\fn_1-\fn_4)^2 - (\fn_1 + \fn_4) \big[ (\fn_2-\fn_3)^2 + (\fn_1-\fn_4)^2 \big] \pm 4 (\fn_4 - \fn_1) \sqrt{\Theta}}{2\fn_4 (\fn_4 - \fn_1 + \fn_2 - \fn_3)( \fn_4 - \fn_1 - \fn_2 + \fn_3)} \\
z_2 &=  \frac{2(\fn_1 + \fn_3)(\fn_2-\fn_4)^2 - (\fn_2 + \fn_4) \big[ (\fn_1 - \fn_3)^2 + (\fn_2 - \fn_4)^2 \big] \pm (\fn_4 - \fn_2) \sqrt{\Theta}}{2\fn_4 (\fn_4 + \fn_1 - \fn_2 - \fn_3)( \fn_4 - \fn_1 - \fn_2 + \fn_3)} \\
z_3 &= \frac{2 (\fn_1+\fn_2)(\fn_3-\fn_4)^2 - (\fn_3+\fn_4) \big[ (\fn_1-\fn_2)^2 + (\fn_3-\fn_4)^2 \big] \pm 4 (\fn_4 - \fn_3) \sqrt{\Theta}}{2 \fn_4 (\fn_4 + \fn_1 -\fn_2 -\fn_3)( \fn_4 - \fn_1 + \fn_2 - \fn_3) } \;,
\eea
where
\be
\Theta = \big( F_2 \big)^2 - 4 \fn_1 \fn_2 \fn_3 \fn_4 \;,\qquad
F_2 = \frac14 \bigg( \sum\nolimits_{a<b} \fn_a \fn_b  - \sum\nolimits_a \fn_a^2 \bigg) = \frac14 \Big( \sum\nolimits_a \fn_a \Big)^2 - \frac12 \sum\nolimits_a \fn_a^2  \;.
\ee
Let us also define
\be
\Pi = \frac18 \big( \fn_1 + \fn_2 - \fn_3 - \fn_4 \big)\big( \fn_1 - \fn_2 + \fn_3 - \fn_4 \big) \big( \fn_1 - \fn_2 - \fn_3 + \fn_4 \big)
\ee
as in the main text. The squares of the metric functions take the simple expressions
\bea
e^{4f} &= \frac{16 z_1 z_2 z_3}{(1+ z_1 + z_2 + z_3)^4} &&= \frac{\Pi^2}{2\Theta^2} \, \big( F_2 \pm \sqrt{\Theta} \big) \\
e^{4f_2} &= \frac{\big( \fn_1 z_2 z_3 + \fn_2 z_1 z_3 + \fn_3 z_1 z_2 - \fn_4 z_1 z_2 z_3 \big)^2}{4 z_1 z_2 z_3} &&= \frac12 \big( F_2 \pm \sqrt{\Theta} \big) \;.
\eea
As we show in Section \ref{sec: special n}, in the special case that the $\fn_a$'s are equal in pairs and $\kappa=-1$, a one-parameter family of solutions emerges.

To write down the metric functions directly, we first need to understand the positivity conditions on the fluxes $\fn_a$, such that a smooth regular horizon can exist. Such conditions are that $z_{1,2,3} >0$, $\Theta \geq 0$ and $\Upsilon>0$ where
\be
\Upsilon \,\equiv\,  \fn_4 z_1 z_2 z_3 - \fn_1 z_2 z_3 - \fn_2 z_1 z_3 - \fn_3 z_1 z_2 \;.
\ee
With a little bit of algebra one can prove the following equalities:
\bea
1+z_1 + z_2 + z_3 &= \pm \sqrt\Theta\, \frac{ \big[ F_2 + \fn_4 (\fn_4 - \fn_1 - \fn_2 - \fn_3) \mp \sqrt\Theta \big] }{ \fn_4 \Pi} \\
z_1 z_2 z_3 &= \frac{ \big[ F_2 + \fn_4 (\fn_4 - \fn_1 - \fn_2 - \fn_3) \mp \sqrt\Theta \big]^4 \big( F_2 \pm \sqrt\Theta \big) }{ 32 \fn_4^4 \Pi^2} \\
\Upsilon &= \frac{ \big[ F_2 + \fn_4 (\fn_4 - \fn_1 - \fn_2 - \fn_3) \mp \sqrt\Theta \big]^2 \big( F_2 \pm \sqrt\Theta \big) }{ 4 \fn_4^2 \Pi} \;.
\eea
The first one shows that, under the assumption that $\fn_4\neq0$ and $\Pi \neq 0$ (those special cases are analyzed in Section \ref{sec: special n}), $z_{1,2,3}>0$ guarantees that $\Theta\neq 0$ and the square bracket is non-vanishing. The second one then guarantees that $F_2 \pm \sqrt\Theta > 0$, and the third one shows that $\Pi$ and $\Upsilon$ have the same sign. Summarizing:
\be
z_{1,2,3} > 0 \;,\qquad \Theta>0 \;,\qquad \Pi > 0 \qquad\qquad\Rightarrow\qquad\qquad F_2 \pm \sqrt\Theta > 0 \;,\qquad \Upsilon > 0 \;.
\ee
Under those conditions, the metric functions are
\be
\label{general solution}
e^{2f} = \frac{\Pi}{\sqrt2\, \Theta} \, \big( F_2 \pm \sqrt\Theta \big)^{1/2} \;,\qquad\qquad e^{2f_2} = \frac1{\sqrt2} \, \big( F_2 \pm \sqrt\Theta \big)^{1/2} \;,
\ee
which give the radii of AdS$_2$ and $\Sigma_\fg$, respectively.

\subsubsection{Analysis of positivity}
\label{sec: positivity}

We want to precisely identify the region in the parameter space $\big\{\fn_a \,\big|\, \sum_a \fn_a = 2\kappa \big\}$ where the near-horizon solutions exist. First, let us impose the positivity constraints on the parameter space $\{\fn_a\}$, with no restriction on $\sum_a\fn_a$ and assuming $\fn_4, \Pi \neq 0$ (the special cases $\fn_4=0$ or $\Pi=0$ are analyzed in Section \ref{sec: special n}):
\be
\cD_\pm = \big\{ \fn_a \,\big|\, z_{1,2,3}^\pm > 0,\, \Theta > 0,\, \Pi >0 \} \;\subset \bR^4 \;,
\ee
where $z_{1,2,3}^\pm$ are the two solutions for the scalars in (\ref{solution z's}). It turns out that both domains are \emph{linear}, in the sense that they are bounded by hyperplanes.

The domain $\cD_-$ is easy to write:
\be
\cD_- = \big\{ \Pi>0,\, \fn_a < 0 \big\} \;.
\ee
This domain is unbounded. Actual solutions to the BPS equations follow from imposing the further constraint $\sum_a \fn_a = 2 \kappa$. On $S^2$ and $T^2$ clearly there are no solutions. On $H^2$ we can rewrite the region as
\be
\cD_-(H^2) = \Big\{ \big(\fn_1 + \fn_2 + 1 \big)\big(\fn_1 + \fn_3 +1 \big)\big( \fn_2 + \fn_3 +1 \big)<0 ,\quad \fn_{1,2,3}<0,\quad \fn_1+\fn_2+\fn_3 > -2 \Big\}
\ee
in terms of $\fn_{1,2,3}$. This domain in bounded.

The domain $\cD_+$ is
\be
\cD_+ = \big\{ \Pi>0,\, \text{three $\fn_a$'s } < 0 \big\} \;.
\ee
The domain is unbounded, and $\cD_- \subset \cD_+$. Now let us impose $\sum_a \fn_a = 2\kappa$. For $H^2$ we do not have further simplifications:
\be
\cD_+(H^2) = \big\{ \Pi>0 ,\, {\ts \sum_a \fn_a = -2} ,\, \text{three $\fn_a$'s } < 0 \big\} \;.
\ee
For $T^2$ we can write the resulting region as
\be
\cD_+(T^2) = \big\{ {\ts \sum_a \fn_a = 0} \,\big|\, \text{three $\fn_a$'s } < 0 \big\}
\ee
because the condition $\Pi>0$ is automatically satisfied, or equivalently as
\be
\cD_+(T^2) = \big\{ \fn_{1,2,3}<0 \big\} \cup \big\{ \fn_{1,2}<0,\; \fn_1+\fn_2+\fn_3>0 \} \cup \text{permutations}
\ee
in terms of $\fn_{1,2,3}$ only. For $S^2$ we can write
\be
\cD_+(S^2) = \big\{ {\ts \sum_a \fn_a = 2} \,\big|\, \text{three $\fn_a$'s } < 0 \big\} \;,
\ee
or equivalently
\be
\cD_+(S^2) = \big\{ \fn_{1,2,3} < 0 \big\} \cup \big\{ \fn_{1,2}<0,\; \fn_1+\fn_2+\fn_3> 2 \} \cup \text{permutations}
\ee
in terms of $\fn_{1,2,3}$ only.

\subsubsection{The special cases}
\label{sec: special n}

First, starting from the beginning, it is easy to see that if two, three or all four of the $\fn_a$'s are zero, then there are no regular solutions. These are precisely the cases with enhanced supersymmetry. The case $\fn_a = \kappa=0$ corresponds to M2-branes on $T^2$ preserving 1D $\cN=16$ supersymmetry. The case $\fn_1 = 2\kappa \neq 0$ and $\fn_{2,3,4}=0$ or permutations thereof corresponds to M2-branes on the (local) hyperk\"ahler space $T^*\Sigma_\fg$, preserving 1D $\cN=8$ supersymmetry. The case $\fn_{3,4}=0$ or permutations thereof corresponds to M2-branes on a local Calabi-Yau threefold, preserving 1D $\cN=4$ supersymmetry.

If one of the $\fn_a$'s vanishes, then $\Theta = F_2^2$ and it is clear that we should choose the upper sign. If one of $\fn_{1,2,3}$ vanishes, then the formul\ae{} above are directly applicable. If $\fn_4=0$ we do not expect anything special to happen, because the final result is symmetric under permutation of the $\fn_a$'s, however the formul\ae{} for the scalars are singular and one should either take the limit carefully, or repeat the computation from scratch. Either way, one obtains
\bea
z_1 &= \frac{(\fn_1 + \fn_2 - \fn_3)(\fn_1 - \fn_2 + \fn_3)}{4F_2} \;,\qquad\qquad z_2 = \frac{ (\fn_1 + \fn_2 - \fn_3)(-\fn_1 + \fn_2 + \fn_3)}{4F_2} \\
z_3 &= \frac{ (\fn_1 - \fn_2 + \fn_3)(-\fn_1 + \fn_2 + \fn_3)}{4F_2}
\eea
as well as $e^{2f} = \Pi/F_2^{1/2}$ and $e^{2f_2} = F_2^{1/2}$. These solutions only exist on $H^2$ and are already contained in $\cD_+(H^2)$.

More interesting is the case that $\Pi=0$. Suppose that only one of the three factors in $\Pi$ vanishes: this implies that the set $\{\fn_a\}$ contains at least three values---and we can assume that they are all non-vanishing, otherwise we are in one of the previous cases. In this case there are no regular solutions.

If two, but not three, of the factors in $\Pi$ vanish, then $\fn_1 = \fn_2 \neq \fn_3 = \fn_4$ (or permutations thereof). In this case on finds a one-parameter family of solutions in which the scalars are
\be
z_1 = \frac12 \Big( z_3 + 1 \pm \sqrt{ (z_3 + 1)^2 - \frac{4\fn_1}{\fn_3} z_3} \Big) \;,\qquad
z_2 = \frac12 \Big( z_3 + 1 \mp \sqrt{ (z_3 + 1)^2 - \frac{4\fn_1}{\fn_3} z_3} \Big)
\ee
in terms of the free value of $z_3$. The metric functions are
\be
e^{2f} = \sqrt{ \frac{\fn_1}{\fn_3}} \, \frac{z_3}{(z_3+1)^2} \;,\qquad\qquad e^{2f_2} = \sqrt{\fn_1 \fn_3} \;,
\ee
and the solutions exist for
\be
\fn_1 < 0 \;,\qquad \fn_3 < 0 \;,\qquad \frac{\fn_1}{\fn_3} \leq \frac{(z_3+1)^2}{4z_3}
\ee
because $\Upsilon = - \fn_1 z_3$. These solutions only exist on $H^2$. This one-parameter family of solutions should be thought of as a ``conformal manifold'' of exactly marginal deformations of the superconformal quantum mechanics. The entropy (as the central charges in higher dimensions) is constant on the conformal manifold.

If all three factors in $\Pi$ vanish, then $\fn_1 = \fn_2 = \fn_3 = \fn_4 = \frac\kappa2$ and we can assume that they are non-vanishing. Again we find a one-parameter family of solutions:
\be
z_1 = 1 \;,\qquad z_2 = z_3 \;,\qquad e^{2f} = \frac{z_3}{(z_3+1)^2} \;,\qquad e^{2f_2} = - \frac\kappa2
\ee
and permutations of the $z_{1,2,3}$. These solutions only exist on $H^2$, where $\kappa=-1$.

\subsubsection{The full analytic black hole solutions}

In the case of $S^2$ (\ie{} $\kappa=1$), the full analytic black hole solutions are in (\ref{BH_metric})-(\ref{rh and K(rh)}). Notice, however, that the radial coordinate $r$ used there is not the same radial coordinate used at the beginning of this appendix and, in particular, in (\ref{BPS equations full set}).

For $\fg>0$ the solutions are still written as in (\ref{BH_metric}), with $e^{-\cK(X)} = 8 \sqrt{X_1X_2X_3X_4}$ and
\be
\label{profile of scalars}
X_a = \frac14 - \frac{\beta_a}r \;,\qquad\qquad \sum\nolimits_a \beta_a = 0 \;,
\ee
however the constant $c$ related to the horizon radius $r_h$ is
\be
r_h^2 = c = 4 \sum\nolimits_a \beta_a^2 - \frac\kappa2
\ee
and the relation between the parameters $\beta_a$ and the fluxes $\fn_a$ is
\be
\fn_a - \frac\kappa2 = 16\beta_a^2 - 4 \sum\nolimits_b \beta_b^2 \;,
\ee
implying $\sum_a \fn_a = 2\kappa$. The latter can also be written as
\be
\fn_a = 16 \beta_a^2 - r_h^2 \;.
\ee
The inverse formula is
\be
\beta_a = \mp \frac{ 4 \big( \fn_a - \frac\kappa2 \big)^2 + \kappa^2 - \sum_b \fn_b^2 }{ 16 \sqrt\Pi} \;,
\ee
where $\Pi$ is the same as in (\ref{def Pi, F2, Theta}). The expressions in (\ref{rh and K(rh)}) remain valid.

From the radial profile of the scalars $X_a$ in (\ref{profile of scalars}) it is clear that whenever the near horizon solution is regular---in particular $X_a(r_h)>0$ and the horizon radius $r_h$ is positive---the full black hole solution is regular. Therefore the analysis of positivity we did in Section \ref{sec: positivity} gives the region in parameter space where smooth black hole solutions with regular horizon exist. In particular, for the case of $H^2$ (\ie{} $\kappa=-1$), when the parameters lie inside $\cD_-(H^2)$ one finds two black hole solutions with different entropy.

\section{$\boldsymbol{I}$-extremization: the example of a free chiral multiplet}
\label{app: freefield}

In this appendix we examine in details the ${\cal N}=2$ quantum mechanics of a free chiral multiplet. Although seemingly trivial,
the model contains some useful information. In particular, the index is extremized in correspondence with the exact R-symmetry
of the model.

\subsection{The massive case}

Consider an ${\cal N}=2$ quantum mechanics with $\mathfrak{u}(1|1)$ supersymmetry algebra
\be
\label{N2algebra}
\frac12 \{ \cQ, \wb\cQ \} = H - \sigma J \;,\qquad\qquad  \cQ^2 = \wb\cQ^2 = 0\;, \qquad\qquad  [\cQ,H ] = [\cQ, J] = 0 \;,
\ee
where $J$ is a flavor symmetry of the theory, $[J,H]=0$ and $\wb\cQ= \cQ^\dagger$. The Witten index
\be
\label{SUSY index app free}
\cI = \Tr \, (-1)^F \, e^{i \Delta J} \, e^{-\beta H}  =  \Tr \, (-1)^F \, e^{i (\Delta  + i \beta \sigma) J} \, e^{-\frac\beta2   \{ \cQ, \wb\cQ \}}
\ee
is independent of $\beta$ and it receives contributions only from ``chiral'' supersymmetric ground states that satisfy $H= \sigma J$.
As a result, it is a holomorphic function of the complex fugacity $y=e^{i (\Delta  + i \beta \sigma)} $ and it can be written as
\be
\cI = \Tr_{H = \sigma J} \, (-1)^F y^J \;.
\ee
Such an index for ${\cal N}=2$ quantum mechanics has been considered in \cite{Hori:2014tda} and evaluated by localization therein.
It is also related to the topologically twisted index of a three-dimensional theory  by  dimensional reduction on $S^2$ \cite{Benini:2015noa}.
In the three-dimensional language, $H$ is the Hamiltonian of the dimensionally reduced theory (and it depends on the magnetic fluxes), while $\Delta$ and $\sigma$ are expectation values for the background vector multiplet associated with the flavor symmetry $J$: $\Delta$ is a flat connection on $S^1$ and $\sigma$ is a real mass \cite{Benini:2015noa}.

We consider a model with a complex scalar $z=x_1 + i x_2$ and a complex fermion $\psi$ satisfying
\be
[x_j, p_k] = i \delta_{jk} \;,\qquad\qquad \{\psi, \wb\psi\} = 1 \;,\qquad\qquad \psi^2 = \wb\psi^2 = 0 \;.
\ee
The Hamiltonian and the flavor symmetry $J$ are
\be
H = \frac{\vec p^{\;2}}2 + \sigma^2 \,\frac{\vec x^{\,2}}2 - \frac\sigma2 [\wb\psi, \psi ] \;,\qquad\qquad J = x_1p_2 - x_2 p_1 + \frac12 [\wb\psi, \psi] \;.
\ee
The fields $z$ and $\wb\psi$ have charge $1$ with respect to $J$, and $\sigma$ plays the role of a real mass. The model can be obtained by
reducing on $S^2$  the topologically twisted theory of a  free three-dimensional chiral multiplet of R-charge 0 \cite{Benini:2015noa}.

We construct the spectrum using oscillators. The bosonic ones are
\be
\label{oscillators}
a_j = \sqrt{ \frac{|\sigma|}2}\, x_j + \frac{i p_j}{\sqrt{2|\sigma|}} \;,\qquad\qquad a_z = \frac{a_1 + i a_2}{\sqrt2} \;,\qquad\qquad a_{\bar z} = \frac{a_1 - i a_2}{\sqrt2} \;.
\ee
They satisfy $[a_j, a^\dag_k] = \delta_{jk}$, as well as $[a_z, a_z^\dag]= [ a_{\bar z} , a_{\bar z}^\dag] = 1$, $[a_z, a_{\bar z}^\dag]=0$ and conjugate. We have  a bosonic Fock space generated from $\ket0$ (defined such that $a_z \ket0= a_{\bar z} \ket0 = 0$) by the action of $a_z^\dag$ and $a_{\bar z}^\dag$.
The fermions give rise to a fermionic Fock space $\{ \ketup, \ketdown \}$. They are defined such that $\psi\ketdown = 0$ and $\ketup = \wb\psi \ketdown$. The  fermion number is $F = \psi \wb\psi$.

The Hamiltonian and the charge can be written as
\be\label{massivehamiltonian}
H = \big( a_{\bar z}^\dag a_{\bar z} + a_z^\dag a_z + 1 \big) |\sigma| - \frac\sigma2 [\wb\psi, \psi] \;,\qquad\qquad J = a_{\bar z}^\dag a_{\bar z} - a_z^\dag a_z + \frac12 [\wb\psi, \psi] \;.
\ee
Notice that $[H,J]=0$. The supercharges can be constructed as
\be
\begin{aligned} \cQ &= - 2i \sqrt\sigma\, a_z \psi \\ \wb\cQ &= 2i \sqrt\sigma\, a_z^\dag \wb\psi \end{aligned} \qquad \text{for } \sigma>0 \;,\qquad\qquad
\begin{aligned} \cQ &= 2i \sqrt{|\sigma|}\, a_{\bar z}^\dag \psi \\ \wb\cQ &= - 2i \sqrt{|\sigma|}\, a_{\bar z} \wb\psi \end{aligned} \qquad \text{for } \sigma<0
\ee
and satisfy the algebra (\ref{N2algebra}).

For $\sigma>0$, the ground state of the total Hamiltonian is $\ket0\otimes\ketup$: it is bosonic and has $H=\sigma/2$, $J = \frac12$. All excited states are obtained by acting with $\psi$, $a_z^\dag$, $a_{\bar z}^\dag$. Since all of them shift $H \to H + \sigma$, the first two shift $J \to J-1$ while the last one shifts $J \to J+1$, it turns out that all states satisfy $H \geq \sigma |J|$ and the only states with $H = \sigma J$ are $(a_{\bar z}^\dag)^n \ket{0,\uparrow}$. The normalized ``chiral'' states are
$$
\frac{(a_{\bar z}^\dag)^n}{\sqrt{n!}} \ket{0,\uparrow}
$$
and are annihilated by $\cQ$ and $\wb\cQ$.
The supersymmetric index $\cI$ in (\ref{SUSY index app free}) is then
\be
\cI = \Tr_{H=\sigma J} \, (-1)^F y^J = \sum\nolimits_{n=0}^\infty y^{\frac12 + n} = \frac{y^{1/2}}{1-y} \;.
\ee
Since $y = e^{i\Delta - \beta\sigma}$, for $\sigma>0$ the series is in fact convergent.

For $\sigma<0$, the ground state is $\ket0 \otimes \ketdown$: it is fermionic and has $H = |\sigma|/2$, $J = -\frac12$. All excited states are obtained by acting with $\wb\psi$, $a_z^\dag$, $a_{\bar z}^\dag$. They all shift $H \to H + |\sigma|$, but $\wb\psi$ and $a_{\bar z}^\dag$ shift $J \to J+1$ while $a_z^\dag$ shifts $J \to J-1$. Therefore all states satisfy $H \geq |\sigma J|$, while the normalized ``chiral'' states satisfying $H=\sigma J$ are
$$
\frac{(a_z^\dag)^n}{\sqrt{n!}} \ket{0,\downarrow}
$$
and are annihilated by $\cQ$ and $\wb\cQ$. The supersymmetric index is
\be
\cI = \Tr_{H = \sigma J} \, (-1)^F y^J = - \sum\nolimits_{n=0}^\infty y^{-\frac12 - n} = \frac{y^{1/2}}{1-y}
\ee
as before. This series is convergent for $\sigma<0$ as it should.

From the example, it appears that the index is only defined in the massive theory. The states counted by the index do not have a well-defined limit as $\sigma\to0$ (this is manifest in the Schr\"odinger representation), and the two series would not converge for $\sigma=0$. Thus the index for zero real mass is defined as the limit of the index with $\sigma \ne 0$. However, for the free chiral case we can still make sense of the index in the massless $\sigma=0$ case if we use generalized states $\ket{\vec x}$ (or $\ket{\vec p}$) in the Schr\"odinger representation. Let us compute
\be
\cI = \Tr \, (-1)^F e^{i\Delta J} e^{-\beta H} = \sum_{\alpha \,=\, \uparrow, \downarrow} \, \int d^2p\, \langle \vec p\, \alpha | (-1)^F e^{i\Delta J} e^{-\beta H} |\vec p\, \alpha \rangle
\ee
with $H = \vec p^{\,2}/2$ and $y=e^{i\Delta}$. The trace factorizes:
\be
\cI = \sum_{\alpha \,=\, \uparrow, \downarrow} \langle \alpha| (-1)^F e^{\frac{i\Delta}2 [\wb\psi, \psi]} |\alpha \rangle \;\cdot\; \int d^2p\, \langle \vec p\, | e^{i\Delta J_\text{bos}} e^{-\beta H} |\vec p\, \rangle \;.
\ee
The fermionic trace is easily  computed to be $y^{1/2} - y^{-1/2}$. To compute the bosonic trace, we notice that the operator $e^{i\Delta J_\text{bos}}$ rotates the $x$-plane and $p$-plane by an angle $\Delta$. We thus find
$$
\int d^2p\; \langle R_{-\Delta}\, \vec p\,| e^{-\beta H} |\vec p\,\rangle = \int d^2p\; e^{-\beta \vec p^{\,2}/2} \; \delta^{(2)} (R_{-\Delta} \vec p - \vec p) = \frac1{\big| \det( R_{-\Delta} - \unit) \big|} = \frac1{2(1-\cos\Delta)} \;.
$$
Notice that, eventually, only the ground state $|\vec p = 0\rangle$ with $H=0$ contributes, however the correct contribution depends crucially on the density of states in a neighborhood of $H=0$.
Finally
\be\label{indexchiral}
\cI = \frac{y^{1/2} - y^{-1/2}}{2(1 - \cos\Delta)} = \frac{y^{1/2}}{1-y}
\ee
as before.

Notice that the expression  (\ref{indexchiral}) is not a single-valued function of $y$ due to an anomaly for the flavor symmetry. In three dimensions this is due to a parity anomaly and it can be cured by adding a Chern-Simons term for the background flavor field \cite{Benini:2015noa}. In quantum mechanics we should add a Wilson line \cite{Hori:2014tda}.

\subsection{The massless case}

The case of interest for this paper is the massless case. By setting $y=e^{i\Delta}$ we have
\be
\cI = \frac{i}{2 \sin \frac{\Delta}{2} } \;.
\ee
This is extremized at $\Delta=(2 k+1) \pi$ with integer $k$, which corresponds to $y=-1$. The value of the index is $|\cI( \Delta = \pi) | =1/2$ which is not an integer.
We can understand this value as a zeta-function regularization
$$
\cI(\Delta= \pi) = \frac{i}{2} = i \, \big( 1 -1 +1 -1 + \ldots \big) \;,
$$
which is consistent with the geometric series expansion of (\ref{indexchiral}) for $y=-1$. As we  have  seen, unfortunately,  the index in the massless case
should be suitably regularized. Only if we define it as a limit of the massive case we can make sense of it as a sum over states of the discrete spectrum.

On the other hand, we also have good news. In the massless case there can be a superconformal algebra and the free chiral theory provides an example of that.
In the massless limit $\sigma\to0$ we find
\be
\cQ \to p \, \psi \;,\qquad\qquad \wb\cQ \to \bar p \, \wb\psi \;,\qquad\qquad H = \frac12 \{\cQ, \wb\cQ\} = \frac{\vec p^{\,2}}2 \;,
\ee
where we defined the holomorphic momentum $p = p_1 + i p_2$. The theory also gains conformal symmetry. We can define the operators
\be
K = \frac{\vec x^{\,2}}2 \;,\qquad\qquad\qquad D = \frac{\vec x \cdot \vec p + \vec p \cdot \vec x}2 \;,
\ee
that satisfy  the $\sl(2,\bR) \simeq \so(2,1) \simeq \su(1,1)$ conformal algebra
\be
[D,H] = 2iH \;,\qquad\qquad [D,K] = -2iK \;,\qquad\qquad [H,K] = -iD \;.
\ee
Here $D$ is the generator of dilations, and $K$ of special conformal transformations.
We can also define the conformal supersymmetries
\be
\cS = z \, \psi \;,\qquad\qquad\qquad \wb\cS = \bar z \, \wb\psi
\ee
with $\wb\cS = \cS^\dagger$, and the R-symmetry current
\be
\label{exactR}
R = \frac{\bar p \, z - p \, \bar z}{2i} + [\psi, \wb\psi] = x_2 p_1 - x_1 p_2 + [\psi, \wb\psi] = - J - \frac12 [\wb\psi, \psi] = -J + F - \frac12 \;,
\ee
which satisfy
\be
[\cQ, K] =-i \cS \;, \qquad \{ \cQ, \cS \} = \{\wb\cQ, \wb\cS\} = 0 \;,\qquad  \{\cQ, \wb\cS\} = D - i R \, .
\ee

All together these operators satisfy an ${\cal N}=2$ superconformal algebra \cite{Ademollo:1975an, Coles:1990hr, Michelson:1999zf, BrittoPacumio:2000sv, Britto-Pacumio:2001zma}.   We can write it  in a compact way
by defining the operators
\bea
L_0 &= \frac{H+K}2 \;, \qquad\qquad\qquad\qquad  & \cG_{\pm\frac12} &= \frac{\cQ \mp  i \cS}{\sqrt2} \\
L_{\pm1} &=  \frac{H-K \mp iD}2 \;, & \wb\cG_{\pm\frac12} &= \frac{\wb\cQ \mp  i \wb\cS}{\sqrt2} \;.
\eea
We find indeed
\bea
\label{virasoro}
[L_m, L_n] &= (m-n) L_{m+n}							& [L_m, \cG_r] &= \frac{m-2r}2 \cG_{m+r}	\qquad	& [R, \cG_r] &= \cG_r \\
\{ \cG_r, \wb\cG_s \} &= 2L_{r+s} + (r-s) \delta_{r,-s} R \qquad	& [L_m, \wb\cG_r] &= \frac{m-2r}2 \wb\cG_{m+r}	& [R, \wb\cG_r] &= - \wb\cG_r \\
\{ \cG_r, \cG_s\} &= \{ \wb \cG_r, \wb \cG_s \} = 0
\eea
with $m,n=0,\pm1$ and $r,s = \pm\frac12$, where we recognize the $\su(1,1|1)$ superalgebra in the Virasoro form. The Hermiticity properties are $L_0^\dag = L_0$, $L_{\pm1}^\dag = L_{\mp1}$, $\cG_{\pm\frac12}^\dag = \wb\cG_{\mp\frac12}$ and $R^\dag = R$.

The R-symmetry operator $R$ is uniquely singled out by the superconformal algebra. This is the {\it exact} R-symmetry of the superconformal quantum mechanics.  We can relate it to the extremization of the index as follows. The critical point of $\big| \cI(\Delta) \big|$ is at $\Delta = \pi$. At that point, using (\ref{exactR}), we have
\be
\cI (\Delta=\pi) = \Tr \, (-1)^F (-1)^J e^{-\beta H} = i \Tr \, (-1)^R e^{-\beta H} \;.
\ee
In other words, the extremization precisely singles out the exact R-symmetry!

We may ask if there is some symmetry at work in this simple example behind the selection of $R$ by extremization. The index $\cI$ is purely imaginary and its extremization $\partial_\Delta |\cI|=0$ is equivalent to
\be
-i \, \partial_\Delta \cI = \Tr \, (-1)^F J \, e^{i\Delta J} e^{-\beta H} =0 \;.
\ee
Then $\Delta=\pi$ is an extremum if
\be
\Tr \, (-1)^R J \, e^{-\beta H} = 0  \;.
\ee
But the theory is invariant under time-reversal: $H$ is invariant while the currents $J$, $R$ change sign. Also $(-1)^R$ in invariant since $R$, as defined in (\ref{exactR}),
has integer spectrum: the bosonic part is the generator of a rotation in the $x$-plane (the component $J_z$ of angular momentum with eigenvalues $m\in \mathbb{N}$) and $[\psi, \wb\psi]$ is integer-valued on the fermionic states $\{ \ketup, \ketdown \}$.
Being odd under time-reversal, $\Tr \, (-1)^R J \, e^{-\beta H}$ must be zero.

Notice that, since $H$ has a continuum spectrum,  all the previous traces must be regularized, using the generalized eigenstates of the momentum or by taking a suitable limit of the massive theory.

\subsection{The alternative superconformal index}
\label{sec: superconformal index}

Using the superconformal algebra we can define an alternative superconformal index making use of the operator $L_0$ that has integral spectrum \cite{Michelson:1999zf, BrittoPacumio:2000sv, Britto-Pacumio:2001zma}. From $\{ \cG_{-\frac12}, \wb\cG_{\frac12} \} = 2 L_0 - R$ we see that
\be
\cI_c = \Tr \, (-1)^R e^{-\beta(2 L_0 -R)}
\ee
is an index, independent of $\beta$ and which takes contribution only from states annihilated by $ \cG_{-\frac12}$ and $\wb\cG_{\frac12}$.
By analyzing the representation theory of the superconformal algebra, one can show that $\cI_c$ gets contributions only from singlets and chiral primaries in (short) chiral representations (annihilated by $\cG_{\pm\frac12}$ and $\wb\cG_{\frac12}$).

In the case of a free chiral,
\be
2 L_0 = \frac{1}{2} \, \vec p^{\,2} + \frac{1}{2} \, \vec x^{\,2}
\ee
is the Hamiltonian of a harmonic oscillator. In fact we can formally map the massless problem to that of a {\it massive} chiral field
with $\sigma = -1$. By explicitly computation we find
\be
2 L_0 - R = H_{\sigma=-1} + J
\ee
and
\bea
\cG_{\frac12} & = - i \sqrt2 \, a_z \, \psi \;, \qquad\qquad\qquad & \wb\cG_{\frac12} &= - i \sqrt2 \, a_{ \bar z} \, \wb\psi \;, \\
\cG_{-\frac12} & = i \sqrt2 \, a_{\bar z}^\dag \, \psi \;, & \wb\cG_{-\frac12} &=  i \sqrt2 \, a_z^\dag\, \wb\psi \;,
\eea
where $H_{\sigma=-1}$ is the Hamiltonian (\ref{massivehamiltonian}) for $\sigma=-1$ while $a_z$ and $a_{\bar z}$ are the oscillators (\ref{oscillators}) for $\sigma=-1$: $a_z = \frac{z + i p}{2}$, $a_{\bar z} = \frac{\bar z + i \bar p}{2}$. The chiral primary states of the superconformal algebra are $(a_z^\dag)^n \ket{0,\downarrow}$ with R-charge $1-n$, so that the superconformal index is
\be
\cI_c = \sum\nolimits_{n=0}^\infty (-1)^{1-n} = - \frac{1}{2} \;.
\ee
We  see that even the superconformal index requires a regularization, since there are infinitely many chiral primaries, and it coincides (up to a phase) with the regularized Witten index  that we have computed above: $-i \, \cI = \Tr \, (-1)^R e^{-\beta H}$.

\section{Attractor mechanism for half-BPS horizons in $\boldsymbol{\cN=2}$ supergravity}
\label{sec: attractorsAPP}

Here we derive a particularly useful identity for half-BPS near-horizon solutions in gauged supergravity that clarifies the attractor mechanism,%
\footnote{The discussion in this section includes all attractors of asymptotically AdS$_4$ black holes, but it is not restricted to them: examples of the same type of attractor behavior can be found in extremal non-BPS black holes in Minkowski space, and in other BPS black holes with more exotic asymptotics such as hyperscaling-violating Lifshitz.}
following the standard ${\cal N}=2$ supergravity conventions \cite{Andrianopoli:1996cm}. In view of our results in the main text, we rewrite in a particularly useful way the known attractor equations, with the goal to provide a clearer holographic picture of the topologically twisted index.

The attractor mechanism for AdS$_4$ black holes in gauged supergravity was studied in details \eg{} in \cite{Cacciatori:2009iz, Dall'Agata:2010gj, Hristov:2010ri, Katmadas:2014faa, Halmagyi:2014qza}. Here we follow \cite{Dall'Agata:2010gj} as it provides a general picture with both electric and magnetic charges, but we make a particular choice for the sections as in \cite{Hristov:2010ri}. Let us introduce the main quantities we deal with. The ``central charge'' is
\be
{\cal Z} = e^{{\cal K}/2} \left(F_{\Lambda} p^{\Lambda} - X^{\Lambda} q_{\Lambda}\right) \,\equiv\, e^{{\cal K}/2} {\cal R} \;,
\ee
where the last equality serves as a definition for the quantity $\cal{R}$. The electric and magnetic charges $q_{\Lambda}$, $p^{\Lambda}$ are defined by the corresponding fluxes through a sphere at any point of spacetime and are conserved via the Maxwell equations and Bianchi identities, respectively. The ``central charge of the gaugings'' is
\be
{\cal L} = e^{{\cal K}/2} \left(g F_{\Lambda} \xi^{\Lambda}  - g X^{\Lambda} \xi_{\Lambda}\right) = - e^{{\cal K}/2} g \xi_{\Lambda} X^{\Lambda} \;,
\ee
where in the second equality we set $\xi^{\Lambda}=0$ since we do not consider magnetic gaugings.%
\footnote{One can always symplectically rotate a given gauged theory to this choice of gauging frame, so there is no loss of generality in this choice.}
Now let us focus on the BPS equations that hold at the black hole horizon, as derived in \cite{Dall'Agata:2010gj} (eqns. (3.9) and (3.5) respectively),
\be
\label{attractor_eqs}
{\cal Z} = i \, R_{S^2}^2 \, \cL \;,\qquad\qquad\qquad D_j {\cal Z} = i \, R^2_{S^2} \, D_j {\cal L} \qquad \forall j \;,
\ee
where the derivatives are with respect to the complex scalars $z_j$, $D_j = \partial_j + {\cal K}_j/2$ and ${\cal K}_j = \partial_j {\cal K}$. These are the BPS attractor equations for AdS$_4$ black holes that are written in a completely general symplectic-invariant way. In particular, there is still a scaling symmetry for the choice of symplectic sections $X^{\Lambda}$ since the number of physical scalars is one less. This scaling symmetry is a remnant of the conformal symmetry in off-shell supergravity and one can always make a gauge choice for it, if needed. Here we decide to make the particular gauge choice
\be
\label{gauge_choice}
2 \xi_{\Lambda} X^{\Lambda} = 1 \;.
\ee
This choice was already implicitly made in the main text, and it was built in the ``ansatz'' for the solutions in \cite{Hristov:2010ri}. One can further see that the choice \eqref{gauge_choice} leads to the explicit appearance of the function $e^{\cal K}$ in the warp factor, which follows from the extra BPS flow equation we are not considering here.\footnote{The additional BPS equation (2.33) of \cite{Dall'Agata:2010gj} fixes $2 R_\text{AdS$_2$}^2 = e^{- \cK}$ on the horizon in accordance with the solution we presented in the main text.}
This choice is made at the level of the theory, and it holds everywhere in spacetime, not just at the horizon. Such a choice does not lead to any physical observable, as the metric, scalars and gauge fields are gauge invariant. However it does change their functional dependence on the sections, and choosing \eqref{gauge_choice} we put the physical solution in a form that is most convenient for us.

Another reason for choosing \eqref{gauge_choice} is the simplification in the holographic dictionary. As we saw in the dual field theory, the chemical potentials $\Delta_a$ obey a similar relation and can be identified with $X_a$ up to a proportionality constant. A different gauge choice would have led to a different identification and a more cumbersome notation. In this sense what we derive below for ${\cal R}$ is not a gauge-invariant statement, but this does not change the underlying physical picture. One can always refer back to \eqref{attractor_eqs} for the scale-invariant equations.

With the gauge choice \eqref{gauge_choice}, the first attractor equation in \eqref{attractor_eqs} gives at the horizon:
\be
{\cal R} = - \frac{ig}{2} \, R_{S^2}^2 \qquad\qquad\Rightarrow\qquad\qquad  |{\cal R}| \,\propto\, S_{BH} \;,
\end{equation}
meaning that $|{\cal R}|$ is equal to the entropy up to a proportionality constant. This result is valid in two-derivative supergravity, and will generically change with higher derivative corrections. Keeping in mind that ${\cal R}$ is a function of the sections, the second equation in \eqref{attractor_eqs} gives
\be
0 = \partial_j {\cal R} + {\cal K}_j  \Big( {\cal R} + \frac{ig}{2} \, R_{S^2}^2 \Big) \qquad\qquad \Rightarrow\qquad\qquad \partial_j \cR = 0 \;.
\ee
This is valid in the gauge \eqref{gauge_choice} that determines, say, $X^0$ in terms of the other sections. Therefore the derivative with respect to the physical scalars $z_j$ can be traded for a derivative with respect to the sections, if we impose \eqref{gauge_choice}. We finally find
\be
\label{R-extremization-sugra}
\frac{\partial {\cal R}}{\partial X^{\Lambda}} \Big|_\text{horizon} = 0 \;,
\ee
that the function ${\cal R}$ is extremized at the horizon. This fixes the values of the complex scalars, and it can be thought of as an attractor equation. Furthermore the value of ${\cal R}$ at the extremum is proportional to the black hole entropy. This is valid for all supersymmetric asymptotically AdS$_4$ black holes, with a general choice of electric and magnetic charges and complex sections $X^{\Lambda}$ under the constraint \eqref{gauge_choice}.

Due to the exact match between the twisted index and the quantity ${\cal R}$ in the particular case considered in the main text, it is natural to expect that this continues to hold for all AdS$_4$ black holes with a field theory dual (note that \eqref{R-extremization-sugra} holds for other BPS horizons as well). It is then tempting to speculate about a more general correspondence between ${\cal R}$ and the Witten index of the dual 1D superconformal quantum mechanics also in cases without AdS$_4$ asymptotics.

\bibliographystyle{JHEP}
\bibliography{BHEntropy}

\end{document}